\documentclass{emulateapj}
\usepackage{natbib}
\usepackage{epsfig}
\usepackage{amsmath}

\newcommand{\xcounits}{\mbox{cm$^{-2}$ (K km s$^{-1}$)$^{-1}$}}
\newcommand{\acounits}{\mbox{M$_\odot$ pc$^{-2}$ (K km s$^{-1}$)$^{-1}$}}

\shorttitle{The Multiphase Fountain in M82}
\shortauthors{Leroy et al.}

\begin{document}

\slugcomment{Accepted for publication in the Astrophysical Journal} 
\title{The Multi-Phase Cold Fountain in M82 Revealed by a Wide, Sensitive Map of the Molecular ISM}

\author{
Adam K. Leroy\altaffilmark{1,2}, 
Fabian Walter\altaffilmark{3}, 
Paul Martini\altaffilmark{1,4}, 
H\'el\`ene Roussel\altaffilmark{5}, 
Karin Sandstrom\altaffilmark{6,7}, 
J\"urgen Ott\altaffilmark{8}, 
Axel Weiss\altaffilmark{9}, 
Alberto D. Bolatto\altaffilmark{10}, 
Karl Schuster\altaffilmark{11}, 
Miroslava Dessauges-Zavadsky\altaffilmark{12}
}

\altaffiltext{1}{Department of Astronomy, The Ohio State University, 140 West 18th Avenue, Columbus, OH 43210}
\altaffiltext{2}{National Radio Astronomy Observatory, 520 Edgemont Road, Charlottesville, VA 22903, USA}
\altaffiltext{3}{Max Planck Institute f\"ur Astronomie, K\"onigstuhl 17, 69117, Heidelberg, Germany}
\altaffiltext{4}{Center for Cosmology and AstroParticle Physics (CCAPP), The Ohio State University, 191 W. Woodruff Ave, Columbus, OH 43210, USA}
\altaffiltext{5}{Institut d'Astrophysique de Paris, Sorbonne Universit\'es, UPMC (Univ. Paris 06), CNRS (UMR 7095), 75014 Paris, France }
\altaffiltext{6}{Steward Observatory, University of Arizona, 933 North Cherry Avenue, Tucson, AZ 85721, USA}
\altaffiltext{7}{Center for Astrophysics and Space Sciences, University of California, San Diego, 9500 Gilman Drive, Mail Code 0424, La Jolla, CA 92093, USA}
\altaffiltext{8}{National Radio Astronomy Observatory, PO Box O, 1003 Lopezville Road, Socorro, New Mexico 87801, USA}
\altaffiltext{9}{Max-Planck-Institut f\"ur Radioastronomie, Auf dem Hügel 69, D-53121 Bonn, Germany}
\altaffiltext{10}{Department of Astronomy, University of Maryland, College Park, MD, USA}
\altaffiltext{11}{IRAM, 300 rue de la Piscine, 38406 St. Martin d\textquoteright H\`{e}res, France}
\altaffiltext{12}{Observatoire de Geneve, 1290 Sauverny, Switzerland}

\begin{abstract}
We present a wide area ($\approx 8 \times 8$~kpc), sensitive map of CO (2--1) emission around the nearby starburst galaxy M82. Molecular gas extends far beyond the stellar disk, including emission associated with the well-known  outflow as far as 3~kpc from M82's midplane. Kinematic signatures of the outflow are visible in both the CO and {\sc Hi} emission: both tracers show a minor axis velocity gradient and together they show double peaked profiles, consistent with a hot outflow bounded by a cone made of a mix of atomic and molecular gas. Combining our CO and {\sc Hi} data with observations of the dust  continuum, we study the changing properties of the cold outflow as it leaves the disk. While H$_2$ dominates the ISM near the disk, the dominant phase of the cool medium changes as it leaves the galaxy and becomes mostly atomic after about a kpc.  Several arguments suggest that regardless of phase, the mass in the cold outflow does not make it far from the disk; the mass flux through surfaces above the disk appears to decline with a projected scale length of  $\approx 1$--$2$~kpc. The cool material must also end up distributed over a much wider angle than the hot outflow based on the nearly circular isophotes of dust and CO at low intensity and the declining rotation velocities as a function of height from the plane. The minor axis of M82 appears so striking at many wavelengths because the interface between the hot wind cavity and the cool gas produces H$\alpha$, hot dust, PAH emission, and scattered UV light. We also show the level at which a face-on version of M82 would be detectable as an outflow based on unresolved spectroscopy. Finally, we consider multiple  constraints on the CO-to-H$_2$ conversion factor, which must change across the galaxy but appears to be only a factor of $\approx 2$ lower than the Galactic value in the outflow.
\end{abstract}

\keywords{}

\section{Introduction}
\label{sec:intro}

Galactic winds \citep[][]{CHEVALIER85,HECKMAN90} may  enrich the circumgalactic and intergalactic medium and affect the evolution of galaxies by moving fuel for future star formation from the disk to the inter- or circum-galactic medium. Indeed, feedback --- often attributed to active galactic nuclei but potentially also due to intense star formation--- is often cited as a necessary element to produce the luminosity function observed in the present day galaxy population \citep[e.g.,][]{CROTON06}. Feedback due to star formation is often specifically invoked as the mechanism by which galactic disks self-regulate, making it essential to produce commonly observed scaling relations \citep[e.g.,][among many others]{ANDREWS11,HOPKINS12}. Enormous abundances of dust and gas, much of it enriched, are now observed outside of galaxies via redenning and absorption \citep[e.g.,][]{MENARD10,WERK13} and this reservoir of gas is so large \citep[e.g.,][]{WERK14} that the interplay of the galaxy disk with circumgalactic gas and dust must represent an essential aspect of galaxy evolution. Even in the case where material is driven out of a galaxy disk but does not escape the galaxy \citep[a ``Galactic fountain'' scenario,][]{SHAPIRO76}, the redistribution of baryons may shape the future evolution of the galaxy.

As a major mode of feedback from the disk into the halo, Galactic winds driven by star formation represent a key aspect of this problem \citep[see the review by][]{VEILLEUX05}. Indeed, there is observational evidence that outflows driven by star formation are pervasive at both low \citep[e.g.,][]{ARMUS95,LEHNERT95,RUPKE05} and  high redshift \citep[e.g.,][]{NEWMAN12}. Based on newer observations, there are hints that outflows may often entrain large masses of molecular gas and dust from the star-forming disk, though  note that molecular winds are not a common result of simulations. These starburst-driven molecular winds have been observed in emission in stacks of nearby ULIRGs \citep{CHUNG11} and resolved near the disks of the local starbursts including M82 \citep{WALTER02}, NGC 2146 \citep{TSAI09,KRECKEL14},  NGC 3256 \citep{SAKAMOTO06B}, and NGC 253 \citep{BOLATTO13A}. Molecular winds apparently driven by active galactic nuclei have also been seen in many systems \citep[e.g.,][]{CICONE14}, including Markarian 231 \citep{FERUGLIO10} and NGC 1266 \citep{ALATALO11}. An observationally-motivated picture is emerging for these AGN-driven molecular winds; here we focus on the less well-understood phenomenon of starburst-driven winds by studying the most famous example.

Given the potential importance of this phenomenon, we know relatively little about the physics of molecular winds in starburst galaxies on large scales. In particular, the evolution of gas as it moves away from the starburst, including the interplay of different components of the interstellar medium remains poorly explored. We know that the gas must change phases: most of the gas in the starbursts that drive the wind is in the molecular phase, but almost all of the circumgalactic gas is ionized \citep[][and references therein]{WERK14}, so that the cold, neutral component must either change phases or fall back to the disk.  However, this phase change has not been resolved observationally. The fate of molecular gas associated with a wind also remains unclear: is gas merely removed from the starburst region and then cycled back to a new part of the disk (``a fountain'') or is it expelled to spend significant time in the halo (a true ``outflow''), perhaps even escaping the galaxy?

Because of its proximity, the prototype edge-on starburst and galactic wind galaxy M82 \citep{LYNDS63,OCONNELL78} represents an ideal target to study a galactic outflow in several phases of gas and dust simultaneously. Indeed, for a long time M82 hosted the only known example of a molecular outflow driven by  star formation \citep{WALTER02}. In this paper, we present a new sensitive, wide area map of CO $J=2\rightarrow1$ emission around M82 (\S \ref{sec:data}).  The map shows that molecular gas extends for kiloparsecs in every direction beyond the central disk, including CO emission coincident with the tidal {\sc Hi} streamers connected to M81 \citep{YUN93}, the extended dust halo \citep{ENGELBRACHT06,KANEDA10,ROUSSEL10}, and the outflow visible in H$\alpha$, scattered UV, and X-ray emission. It also includes kinematic information that allows us to show broad, multi-component line profiles in the outflow, slowing rotation with increasing vertical displacement, and a clear velocity gradient along the outflow axis. Combining all of this information, we argue that the cold gas around M82 can be best described by a ``fountain''-type model. The cold material confines and funnels the outflow, with the interface giving rise to some of the most spectacular emission features seen in M82. However, the cold material cast out of the disk does not itself travel more than a few kpc from the galaxy, on average, instead forming a spheroidal halo with characteristic scale length $\approx 1$--$2$~kpc. The cold gas around M82 is extended compared to the central starburst, which is approximately 300~pc in size, and the stellar disk, which has vertical scale length $\approx 0.4$~kpc. However, the distribution of cold material is confined compared to the extent of the circumgalactic medium observed in other galaxies via absorption line or reddening measurements, $\sim 10$--$100$~kpc.

Our starting framework for the geometry of the M82 outflow is the paper by \citet{SHOPBELL98}. They consider the hot and ionized gas in detail and conclude that the H$\alpha$ arises from two cones with base a few hundred pc and slightly ajar from one another. The H$\alpha$ arises mostly from the walls of this cone, which presumably indicates an encasing medium of cold material. Given the relatively low resolution of our data, we emphasize the fate of the outflow once it is launched. For studies of the molecular outflow near its base, see the study of M82 by \citet{WALTER02} and recent demonstration of dense gas associated with the base of the outflow by \citet{KEPLEY14}.

Throughout the paper we adopt a distance of $3.6$~Mpc \citep[][]{FREEDMAN94} so that  $1\arcsec \approx 17.5$~pc and $1\arcmin \approx 1$~kpc. The inclination of the galaxy will often be relevant to our analysis and following \citet{COKER13}, \citet{WESTMOQUETTE07}, and \citet{GOTZ90}, we will assume that M82's disk is inclined at $80\arcdeg$ and the outflow to be perpendicular to the disk, inclined at $10\arcdeg$ relative to the plane of the sky. Because this inclination is uncertain and even small changes can have large impact on the derived outflow velocities, we also show results for $\pm 5^\circ$. For a thorough review of M82 that is beyond the scope of this paper, we refer the reader to \citet{WESTMOQUETTE07}. 

In the appendices, we discuss related issues not appropriate for the main text. The first appendix quantifies the prospect of finding systems like M82 from unresolved spectroscopy by artificially inclining the galaxy. The second appendix investigates the CO-to-H$_2$ conversion factor and conditions within the molecular gas in detail. The third appendix explores some systematics related to our treatment of the dust spectral energy distribution, including the role of collisional heating. The final appendix discusses an algorithmic approach for correcting single dish maps for contamination by stray light picked up the in the ``error beam'' of the telescope.

\section{Observations}
\label{sec:hera}

\begin{figure}
\epsscale{1.2}
\plotone{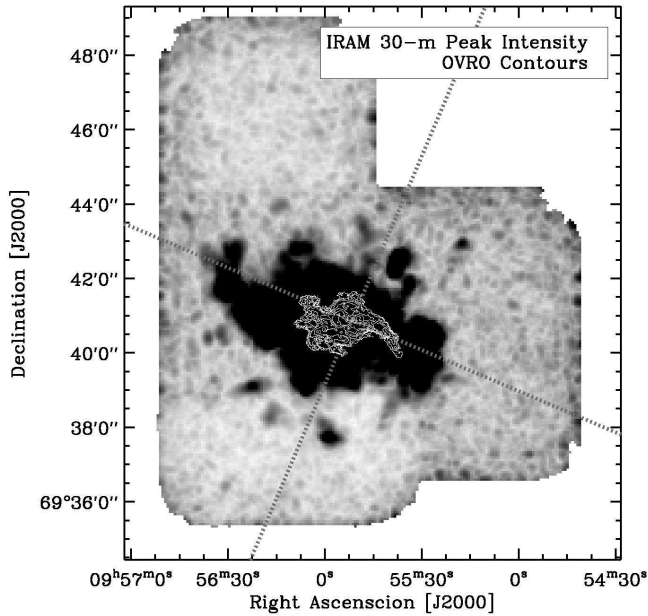}
\caption{Peak CO $J=2\rightarrow1$ emission from our IRAM 30-m observations of M82 on a high stretch (saturating at 50~mK)  and $20\arcsec$ resolution (grayscale). The figure shows the full area surveyed, chosen to cover the outflow, stellar disk, and the {\sc Hi} streamer induced by the ongoing interaction with M81 \citep{YUN93}. For comparison, we plot the CO $J=1\rightarrow0$ interferometer map from \citet[][in contour]{WALTER02}, which initially demonstrated the molecular outflow in M82. Gray lines indicate our adopted major and minor axes.
\label{fig:peaki}}
\end{figure}

\begin{figure*}
\epsscale{1.2}
\plotone{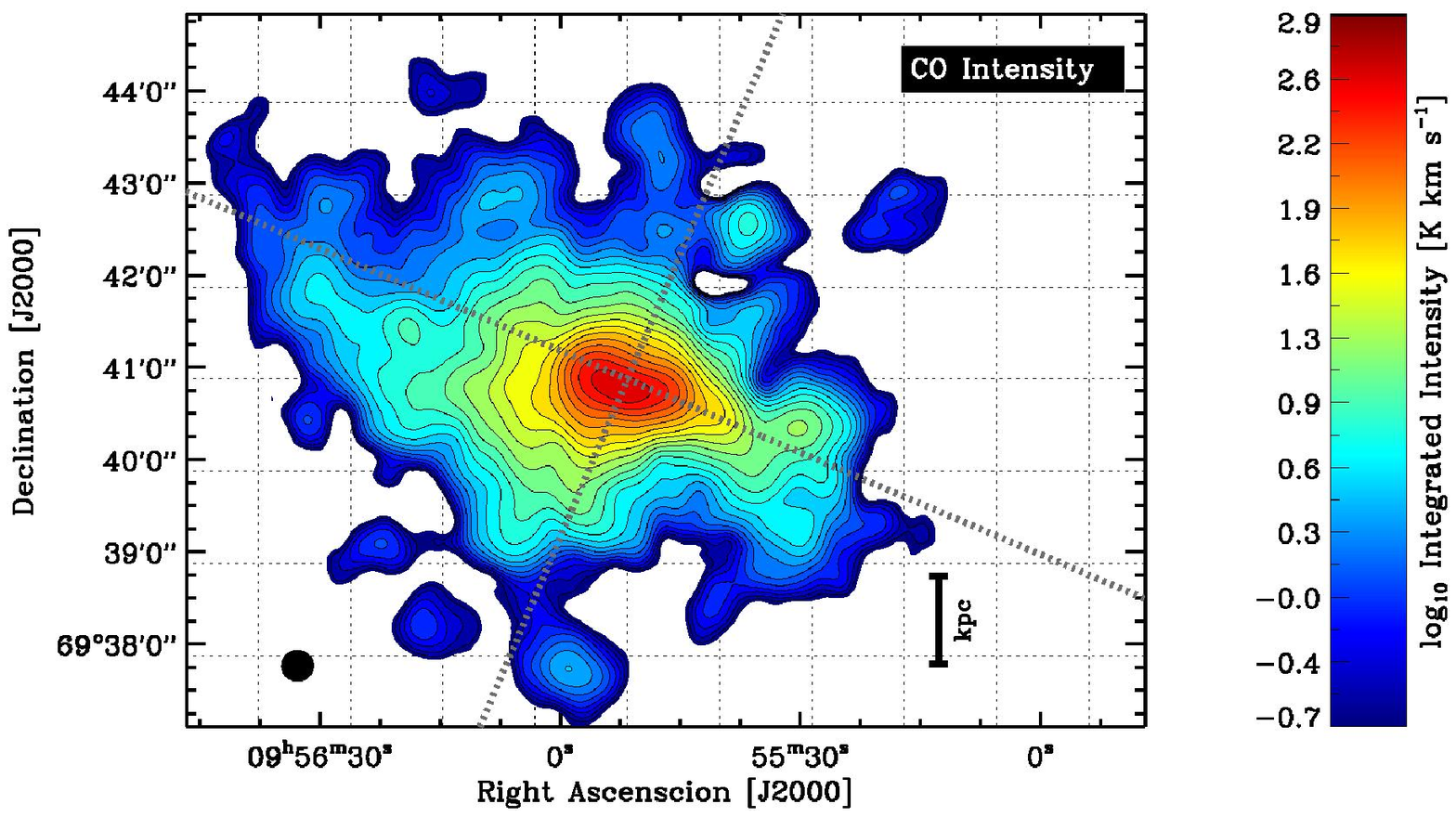}
\plotone{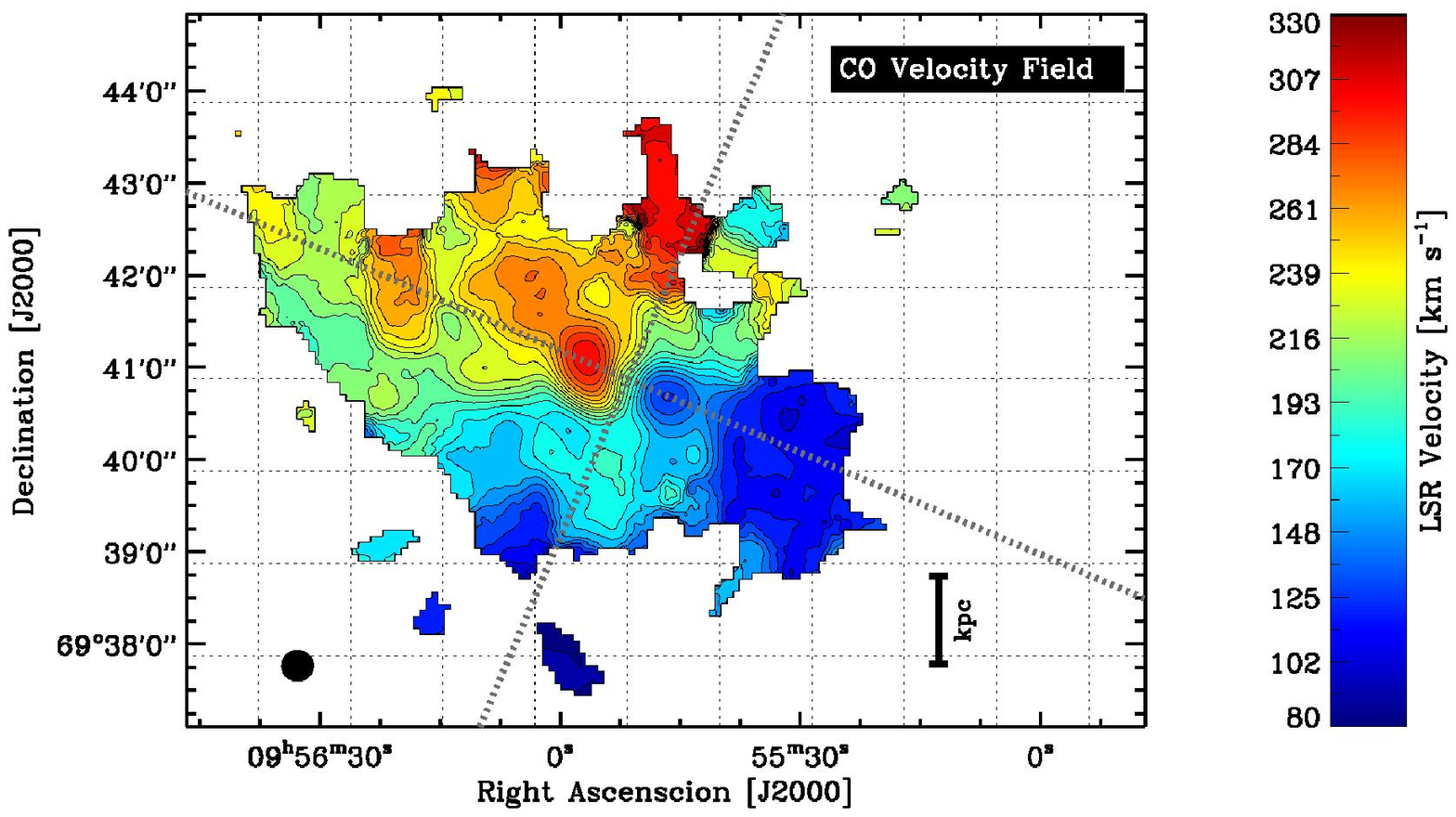}
\caption{CO $J=2\rightarrow1$ emission from M82 at $20\arcsec$ resolution. The top panel shows integrated intensity, with contours beginning at 0.2 K~km~s$^{-1}$ and increasing by factors of $\sqrt{2}$. The bottom panel shows the emission-weighted mean velocity, on a linear stretch from 80~km~s$^{-1}$ to 330~km~s$^{-1}$. To indicate scale, we plot a black bar of size 1~kpc at the distance of M82. Both panels show the bright, rotating central disk surrounded by a region of CO emission that extends for many kpc in each direction. Some of this extended emission may be associated with an extended disk, some of it appears associated with the tidal streamers that link M82 to M81 in {\sc Hi}, and some of the material is coincident with the starburst-driven outflow visible in H$\alpha$ and X--ray emission.
\label{fig:cube}}
\end{figure*}

We used the Heterodyne Receiver Array \citep[HERA,][]{SCHUSTER04} on the IRAM 30-m telescope to observe a wide footprint around M82 in CO (2--1) emission. The data were obtained between January and March 2007 and then reduced using the HERACLES pipeline \citep[][]{LEROY09}. The reduction included baseline fitting using polynomials up to fifth order at both the individual spectrum and cube stage, rejection of pathological spectra based on comparing their rms noise to expectations from the radiometer equation, and construction and subtraction of OFF spectra from the ends of the scan legs. We gridded the data onto a cube with a $4\arcsec$ pixel scale using a $17\arcsec$ gridding kernel; combined with the native $10.5\arcsec$ beam size of the IRAM 30-m at 230~GHz, this yielded a map with effective beam size $20\arcsec$. This is coarser than other HERA maps of nearby galaxies \citep[e.g.,][]{LEROY13} because in order to cover a wide area with good baseline stability, the observations did not critically sample the 30-m beam. The channel spacing of the  WILMA backend used to obtain the data is 2.6~km~s$^{-1}$. In the Appendix, we describe an iterative correction scheme to account for the effects of the 30-m error beam (Kramer, Pe\~nalver, \& Greve 2013, IRAM memo). We apply the error beam correction to the cube. The integrated correction to the flux is $\approx 10\%$ and locally the effect seldom exceeds 20--30\% of the flux. We use a Hann kernel to smooth the cubes and work with a final velocity resolution of 5.2~km~s$^{-1}$. Finally, we apply the estimated main beam (0.58) and forward (0.92) efficiencies of the 30-m at 230~GHz to convert from the measured antenna temperature to main beam temperature. At this resolution, the cube has rms noise per 5.2~km~s$^{-1}$ channel of $7$~mK in units of main beam temperature. For reference, our beam size and observing frequency imply $\approx 17.4$~Jy per Kelvin.

To highlight the extent and sensitivity of the maps, Figure \ref{fig:peaki} shows the peak intensity of this CO $J=2\rightarrow1$ emission on a high stretch over the full survey area. We plot the OVRO interferometer map by \citet{WALTER02} in white contour for reference. Figure \ref{fig:cube} shows the CO $J=2\rightarrow1$ integrated intensity and intensity weighted mean velocity maps over the region where we detect CO. This map has dramatically increased field of view relative to the OVRO map of \citet{WALTER02} but much coarser resolution. It has greatly improved resolution and sensitivity compared to the FCRAO single-dish map by \citet{TAYLOR01}. It complements the CO $J=1\rightarrow0$ map of \citet{SALAK13}, which has comparable resolution but much lower sensitivity than our HERA map, with $\approx 10\times$ higher rms noise in $T_{\rm mb}$ units for matched velocity channels. Similarly, it improves on the extent and sensitivity of the CO $J=3\rightarrow2$ map presented by \citet{WILSON12}. We will see below that comparing the CO $J=3\rightarrow2$, $J=2\rightarrow1$, and $J=1\rightarrow0$ lines yields interesting constraints on physical conditions.

\section{Data At Other Wavelengths}
\label{sec:data}

In order to place the molecular superstructure of the galaxy in context,  we compare our new CO map to observations of M82 at other wavelengths. As tracers of the ionized wind, we use a narrow-band image of H$\alpha$ emission from the Local Volume Legacy survey \citep{DALE09,LEE09}; H$\alpha$ velocity information from longslit spectra by \citet{MCKEITH95} and a Fabry-Perot cube by \citet{SHOPBELL98}. We also use far-ultraviolet (FUV) emission from GALEX \citep{HOOPES05,GILDEPAZ07}, and X--ray emission from {\em Chandra} \citep{STRICKLAND04,STRICKLAND07}.  To trace starlight, we use the 2MASS $K$ band image from \citet[][]{JARRETT03}; the more sensitive IRAC 3.6$\mu$m map from LVL is contaminated by PAH emission in the region of the outflow.

To study the distribution of dust in the halo of M82, we use infrared observations from {\em Herschel} at 70, 160, 250, and 350$\mu$m \citep{ROUSSEL10} and {\em Spitzer} maps at 3.6, 8, and 70$\mu$m from the Local Volume Legacy survey \citep{ENGELBRACHT06,DALE09,LEE09}. As discussed in \citet{ROUSSEL10}, the presence of the bright starburst in the field means that studying M82's dusty halo in the infrared requires high dynamic range imaging. Following \citet{ROUSSEL10}the {\em Herschel} data have been ``cleaned'' in a manner similar to radio interferometry data, though in detail the procedure is refined from that paper. The algorithm reconstructs the image as a series of point sources, in each case replacing the complex {\em Herschel} point spread function, which has substantial extended structure, with a Gaussian. To estimate the {\em Herschel} PSF, we combined observations of point sources, bright sources for the wings and fainter sources for the core to avoid saturation. These models took into account the rotation of the telescope. The IR images were then cleaned by subtracting a scaled version of the PSF, corrected for orientation of the telescope from the brightest point source in the image and replacing it with a Gaussian containing equivalent flux. We cleaned to a limiting brightness of $0.1$, $0.04$, $0.6$, and $0.2$ Jy~arcsec$^{-2}$ at 70, 160, 250, and 350$\mu$m. The beam of the Gaussian used to replace the PSF at each wavelength was: $4.2\arcsec$ (70$\mu$m), $8.6\arcsec$ ($160\mu$m), $13.5\arcsec$ (250$\mu$m), and 19$\arcsec$ ($350\mu$m), in each case corresponding to about 75\% of the FWHM of the {\em Herschel} PSF. The final ``cleaned'' {\em Herschel} data have approximately Gaussian beams for their bright emission and we convolve these to a common resolution of $\approx 24\arcsec$  for most analysis. For most of this paper the key point of this analysis is we are able to distinguish faint emission even in the presence of M82's bright starburst. The final SPIRE images had a small positive zero-level offset, which we fit and subtracted away from the galaxy. The effect of this subtraction is shown in the appendix.

Using the matched-resolution, cleaned infrared intensity maps, we build a point-by-point spectral energy distribution, which we fit using the models of \citet{DRAINE07A} following the parameterization of \citet{DRAINE07B} and \citet{ANIANO12}. The resulting fits include estimates of the dust surface density, $\Sigma_{\rm dust}$, and the mean radiation field $\left< U\right>$ for each location\footnote{The models distinguish a PDR-like component and minimum radiation field. In this paper we consider only their combination, $\left< U \right>$.} Other  parameters in the model, for example the PAH mass fraction, $q_{\rm PAH}$, and the breakdown of the radiation field distribution, are less certain and we do not consider them in this paper \citep[see][for discussion of the PAH distribution]{ENGELBRACHT06,KANEDA10,BEIRAO15}. The \citet{DRAINE07A} models consider only radiative heating of dust. Shocks and processing of small dust grains are known to be important around M82 \citep[e.g.][]{KANEDA10,BEIRAO15}, so collisional heating of dust might become important in some areas around the galaxy. However, in the appendix we show that for reasonable limits to the bulk gas temperature and even a modest radiation field we do not expect collisional heating to be dominant over most of the area around M82. However, more detailed modeling of the interface between the very hot, rare wind and the cold material is certainly needed.  The appendix also shows that the choice of the \citet{DRAINE07B} models does not substantially bias our results; most of our analysis requires a linear tracer of dust mass and we show that within a factor of $\approx 2$, the \citet{DRAINE07B} models match the results of fitting a modified blackbody modulo a linear scaling. The appendix also compares the $\left< U \right>$ from the model fit to the best-fit dust temperature.

In order to help constrain the physical conditions in the molecular gas, we use maps of other CO transitions. We take CO $J=1\rightarrow0$ from \citet{WALTER02}, who combined OVRO and IRAM 30-m data and maps of the CO $J=3\rightarrow2$ line from the JCMT Nearby Galaxies Legacy Survey \citep{WILSON12}. The appendix presents a detailed comparison of the three maps, while the main text summarizes the results.

We also compare to {\sc Hi} 21-cm emission, for which we use several data sets. The D-configuration data set of \citet{YUN93} and \citet{YUN94} provides a low resolution ($\approx 70\arcsec$), high surface brightness sensitivity view. Our own reduction of the \citet{YUN93} C-configuration data provides a higher resolution view (which we convolve to 20$\arcsec$ resolution to match the CO data). Finally the best matched resolution column density estimate of {\sc Hi} comes from our combination of the archival VLA data, Effelsberg single dish data, and modeling to convert the {\sc Hi} absorption into emission. This best-estimate column density map has resolution $\approx 15\arcsec$ and we convolve it to the $20\arcsec$ resolution of the CO data for comparison. We  derive intensity weighted mean velocity fields and velocity dispersions at 20$\arcsec$ and $\approx 70\arcsec$ resolution. We treat these velocity measurements as undefined near the disk of the galaxy where {\sc Hi} goes into absorption against the continuum.

Collectively, these data give us information on the excitation of CO, mostly near the disk; the location and kinematics of the outflow as seen in ionized gas (H$\alpha$); the distribution of small dust grains, likely polycyclic aromatic hydrocarbons (PAHs, $8\mu$m); the distribution of larger dust grains (70--350$\mu$m); the interstellar radiation field; the distribution of stars, and the location and kinematics of atomic gas (21-cm). We mention additional processing as we present individual comparisons.

\section{A Wide, Sensitive CO $J=2\rightarrow1$ Map}
\label{sec:thecube}

\begin{figure}
\plotone{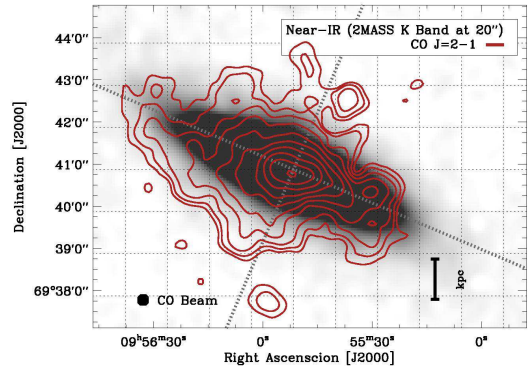}
\caption{
\label{fig:stars} 
Our CO (2-1) map (red contours) relative to a high-stretch plot of the $K$-band light from stars (grayscale) with major and minor axes shown by gray lines. CO emission extends well beyond the stellar distribution along the minor axis of the galaxy.
}
\end{figure}

\begin{figure*}
\epsscale{1.0}
\plotone{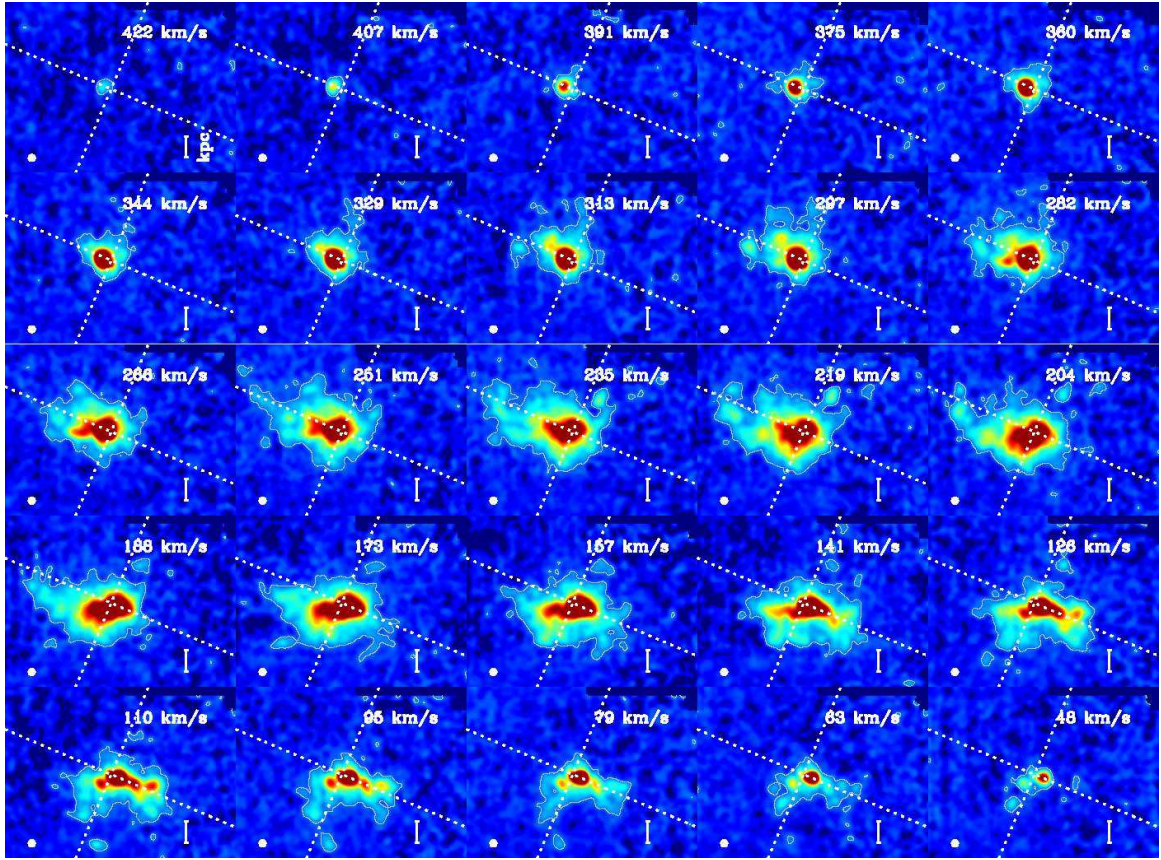}
\caption{
\label{fig:chan} 
Individual channel maps of CO emission over the same area as Figure \ref{fig:cube}. The figure shows every third channel over the velocity range of significant emission with the numbers in each map indicating the mean LSR velocity for that 5.2~km~s$^{-1}$ wide channel. A gray contour shows emission at $\approx 21$~mK, or  $S/N = 3$ in an individual channel. Dotted lines show the major and minor axis. A bar in each panel shows a length corresponding to 1~kpc for reference and a white circle shows the beam.
}
\end{figure*}

Our CO $J=2\rightarrow1$ observations, shown in Figure \ref{fig:cube}, offer the most complete view of the molecular medium  around M82 to date. Together, the integrated intensity map and velocity field show a bright, quickly rotating disk embedded in a  distribution of fainter CO emission that stretches for kiloparsecs. Indeed, it is striking that the columnated morphology familiar from UV, X--ray, and H$\alpha$ images of the M82 wind would be hard to identify from the  integrated CO intensity map alone. Instead, CO emission extends in every direction, similar to results from infrared imaging of dust emission \citep{ENGELBRACHT06,ROUSSEL10,KANEDA10}. 

The picture becomes clearer when we compare the CO to the stellar disk in Figure \ref{fig:stars}. CO extends far above the stellar disk, with  the largest extensions above and below the central starburst region along the minor axis of the  galaxy. This is suggestive of M82's familiar outflow, especially as seen in H$\alpha$. The molecular emission above and below the disk is not confined to the minor axis, however, with CO emission extending above and below all along the major axis of the galaxy.

Even though the integrated intensity is surprisingly amorphous, the velocity field shows an impressive degree of structure. The central disk, which is by far the brightest feature in the integrated intensity image (note the logarithmic stretch in Figure \ref{fig:cube}), shows strong, clear rotation. Above and below this disk, a velocity gradient is visible along the minor axis, a point we will will return to below (\S \ref{sec:kin}). Even above and below the main disk, rotation is still visible as a velocity gradient along the major axis, but the strength of the rotation seems to weaken as one moves away from the main disk.

\begin{figure*}
\epsscale{1.0}
\plotone{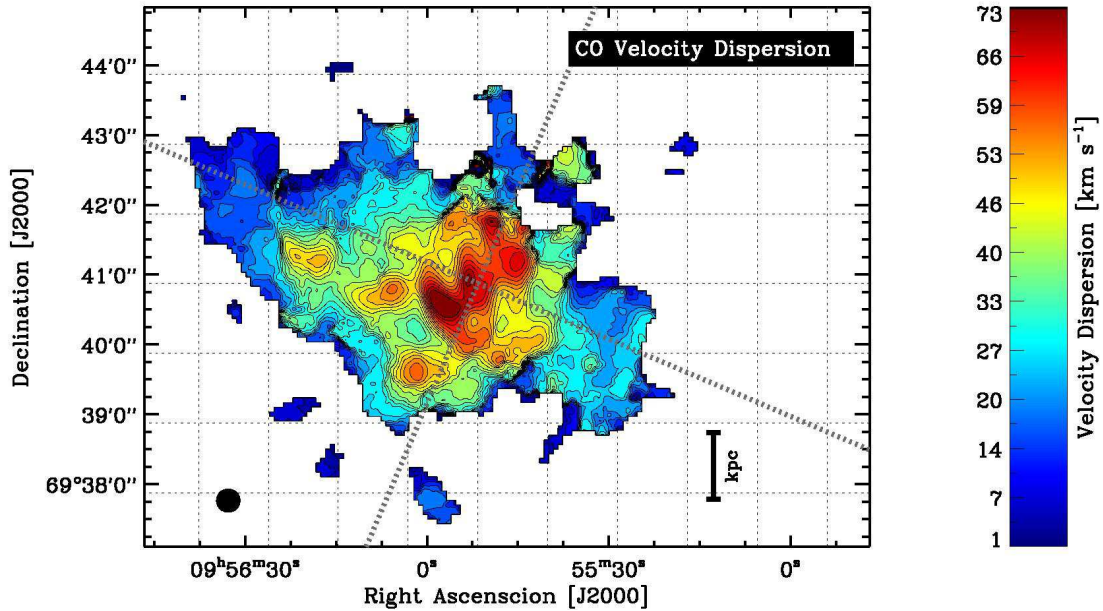}
\caption{
\label{fig:vdisp} 
Line of sight velocity dispersion calculated from our CO map. The broadest line widths are concentrated in, above, and below the starburst near the minor axis of the galaxy. Some of these broad features show evidence for multiple components, signatures of gas surrounding the conical outflow.}
\end{figure*}

This more detailed structure is also visible in the individual channel maps of CO emission (Figure \ref{fig:chan}), where is it possible to pick out clumps of CO emission far above and below the main disk, filamentary structures that extend up and down in individual channels (e.g., see $v_{\rm LSR} = 235$, $110$, and $79$~km~s$^{-1}$), and an extended low surface brightness region around the main disk that includes an eastward extension of the galaxy. The channel maps also show that the emission can have a very broad velocity distribution. Features often persist across several channels and the step between channels is $\Delta v \approx 15$~km~s$^{-1}$. Some of the broad line widths reflect rotation in an edge-on disk. However, the broadest features tend to be concentrated along the minor axis. Figure \ref{fig:vdisp} shows this by plotting the line of sight velocity dispersion calculated from the second moment.

\section{Association With the Superwind and Decomposition into Regions}

\begin{figure*}
\plottwo{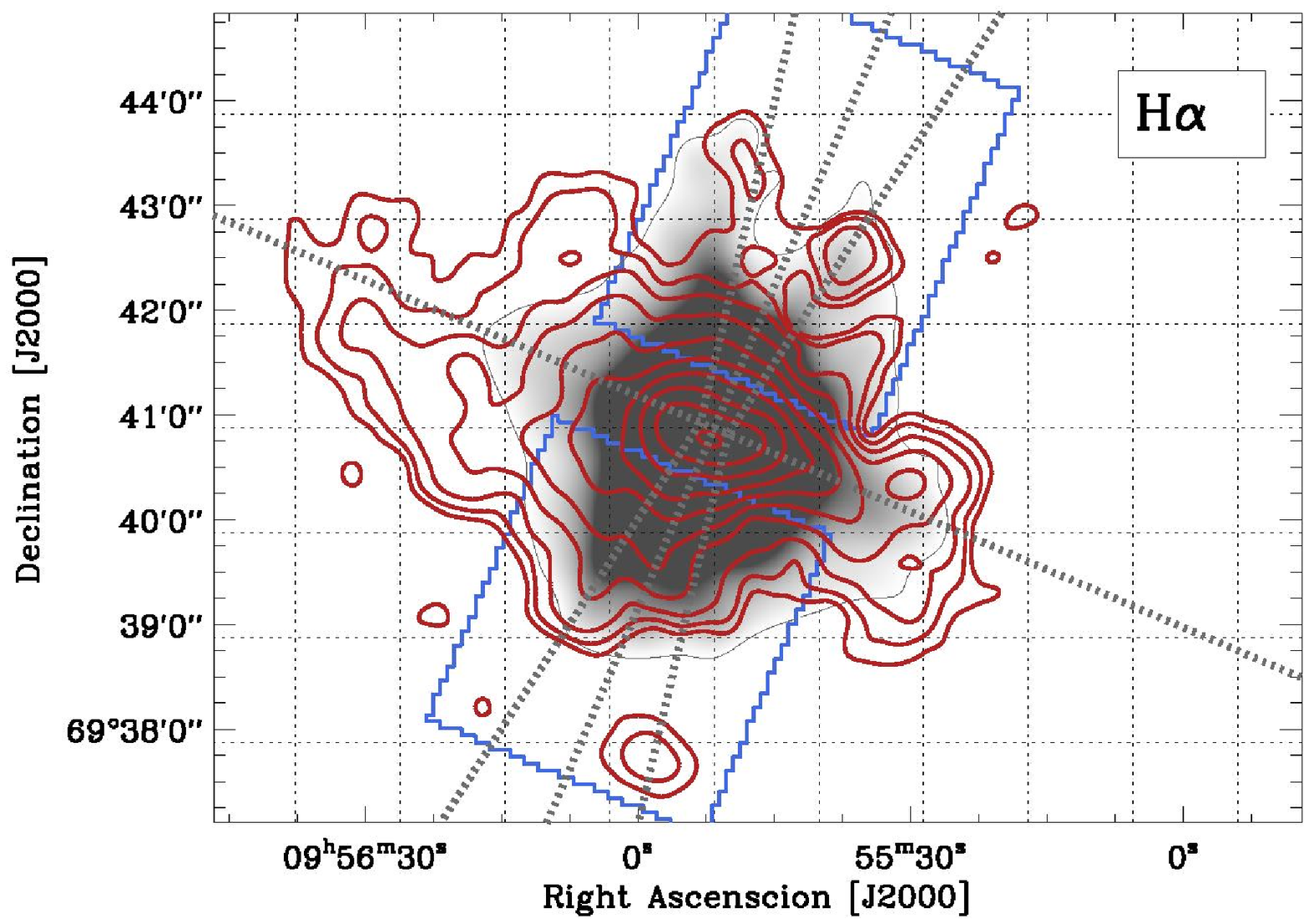}{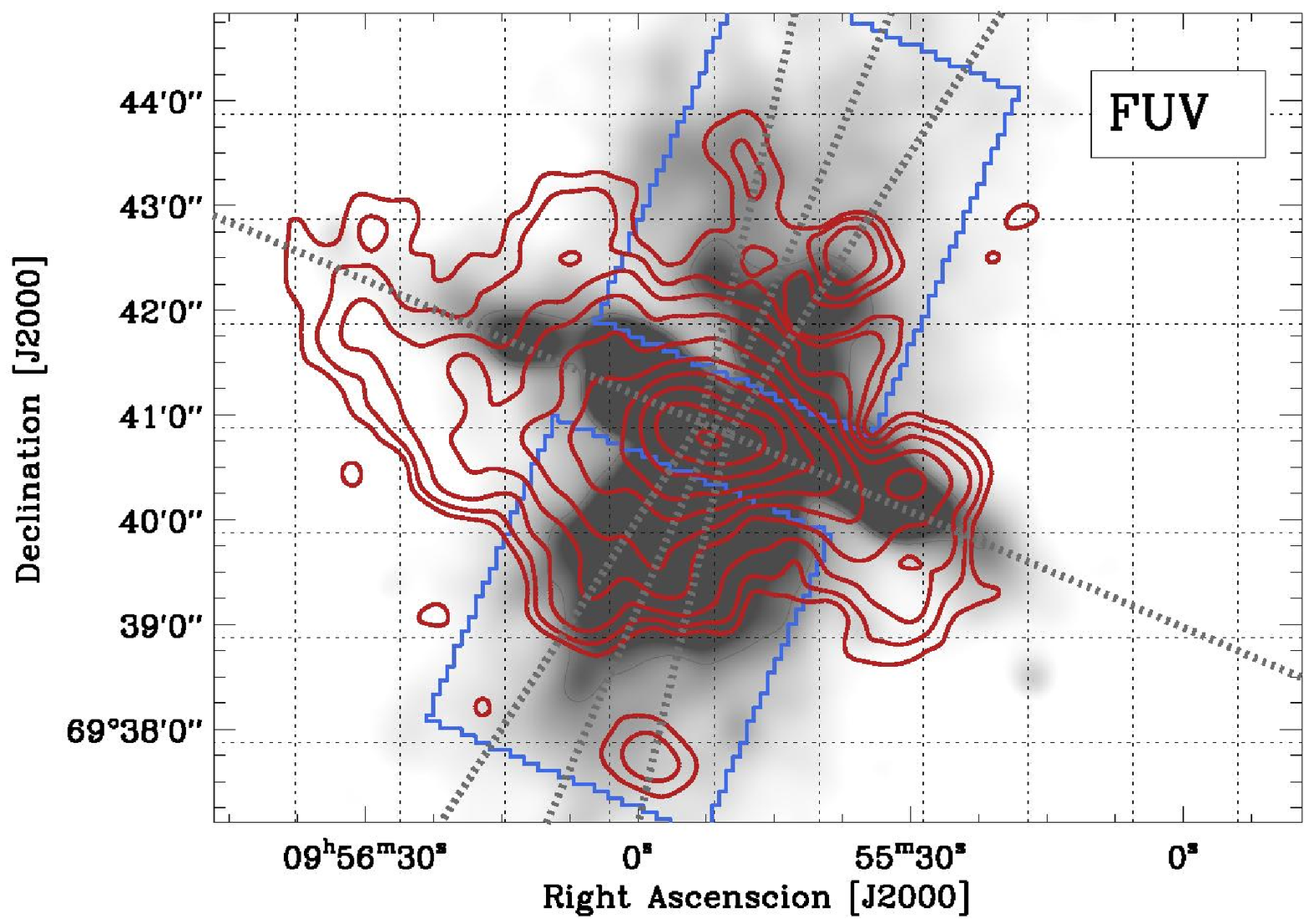}
\plottwo{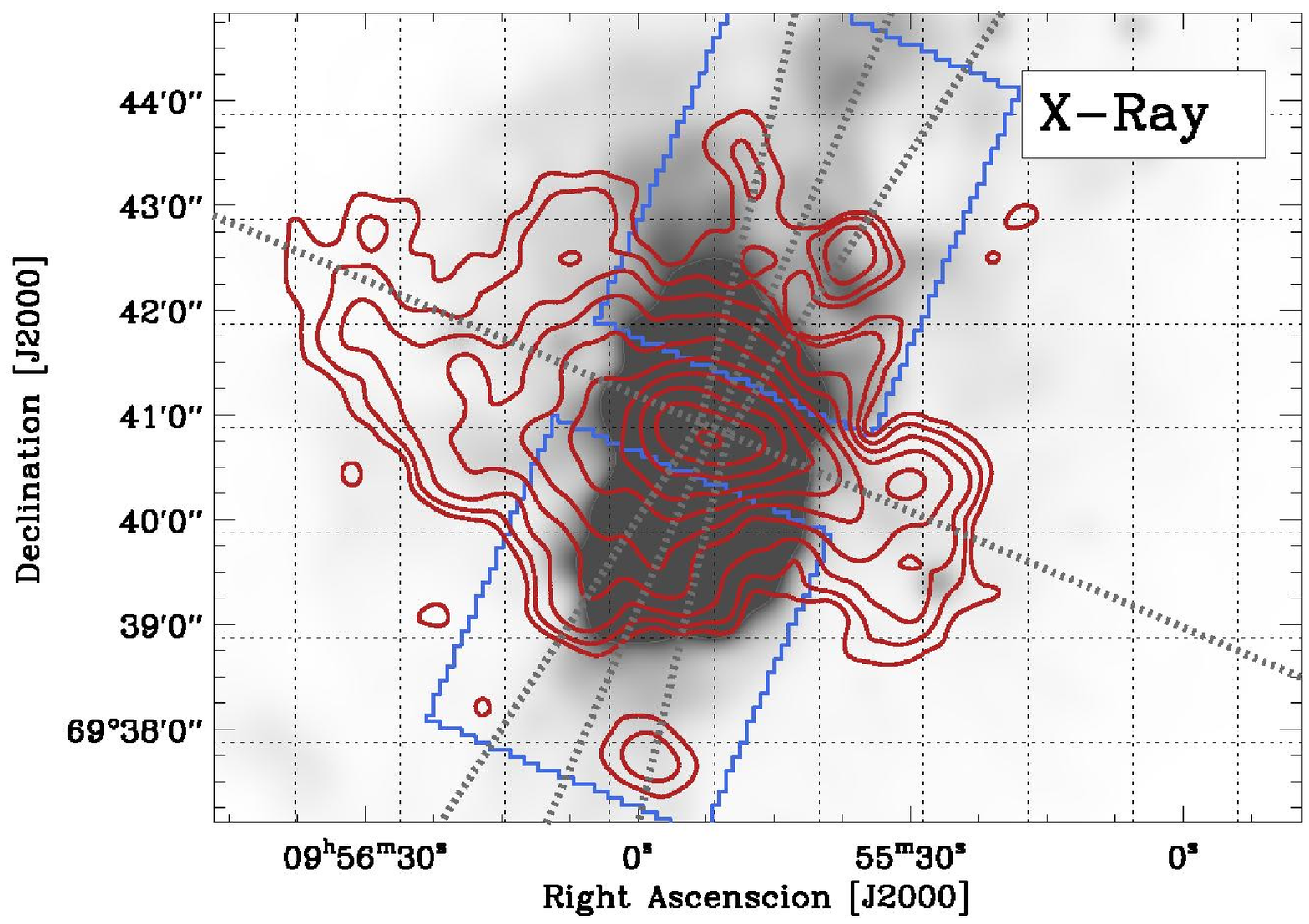}{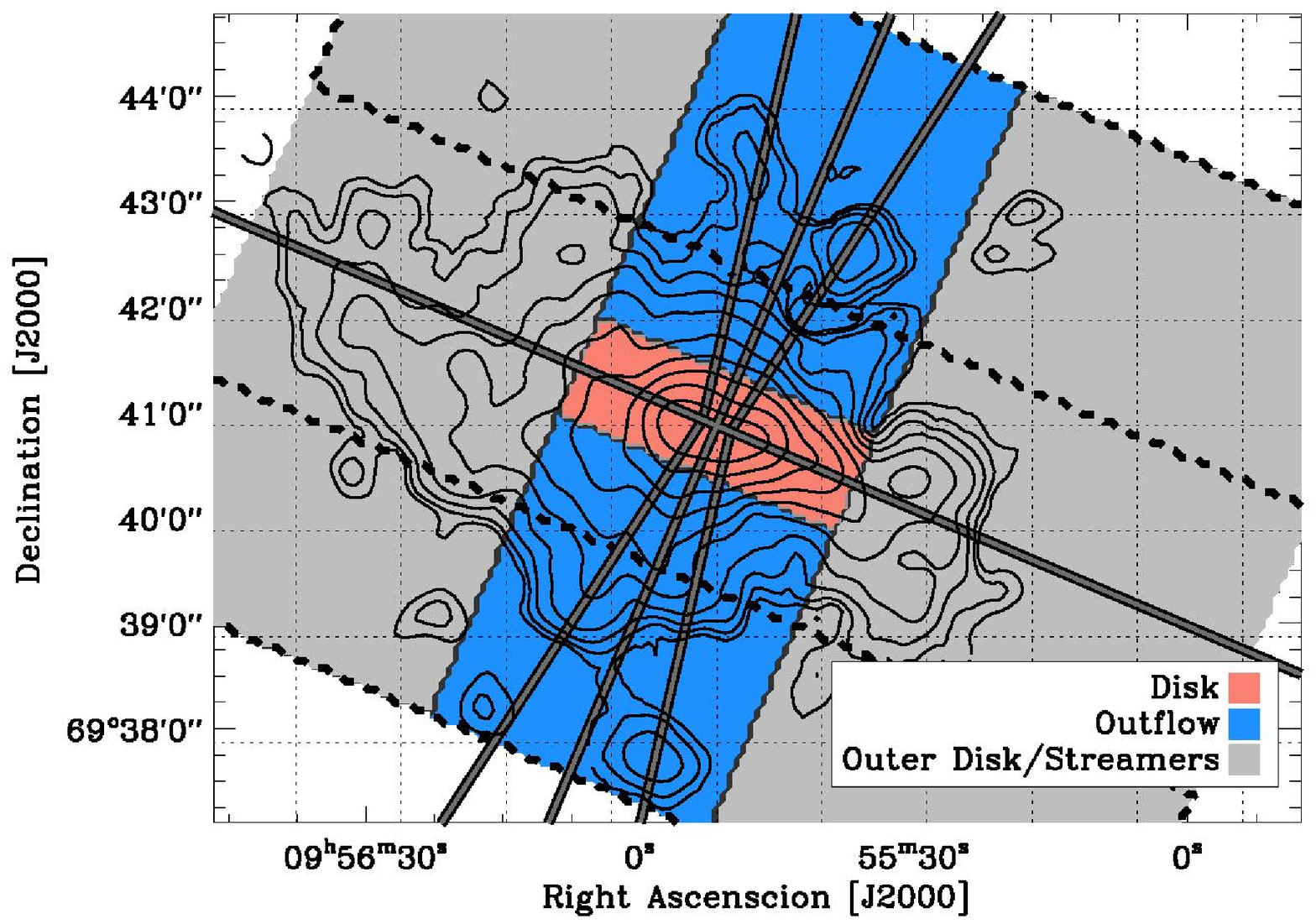}
\caption{
\label{fig:wind} 
Our CO (2-1) map (red contours) plotted over tracers of the M82 outflow in grayscale: ({\em top left}) H$\alpha$, ({\em top right}) far ultraviolet, and ({\em bottom left}) soft X-ray emission. The bottom right panel shows the two-dimensional region assignments that we use to break the map into emission mainly from M82's disk (pink), emission associated with the outflow (blue), and emission not clearly associated with either --- mostly the extended disk and molecular gas associated with the tidal streamers (gray). The  outflow region is also shown in as a rectangular blue outline in the other three panels. We use this region, which has width $\pm 1.5$~kpc about the minor axis, in a number of subsequent calculations. The final panel shows the major and minor axes and includes lines $\pm 1.5$~kpc above and below the major axis. The final panel also shows a cone with opening angle $20^\circ$, which is the opening angle implied by the line splitting that we observe.}
\end{figure*}

Figure \ref{fig:wind} shows the CO distribution in contours relative to signatures of the well-known hot outflow from M82: H$\alpha$ (a mixture of in situ and scattered emission), scattered far-ultraviolet (FUV) emission, and soft X--ray emission from hot gas. Above and below the disk, we see qualitative correspondence between the CO and the outflow tracers, particularly the H$\alpha$.  The coincidence might be expected given that the FUV light shows the location of dust grains  scattering emission from the  starburst and some of that dust is likely to be mixed with molecular gas. Similarly, H$\alpha$ emission traces recombinations, which will occur in regions of high gas density, often along the surface of neutral clouds.

In the next sections, we show a kinematic association of {\sc Hi} and CO. \citet{WALTER02} have shown a similar association of CO and H$\alpha$. This suggests that the ionized, atomic, and molecular phases are mixed along the minor axis. We also see that the neutral (atomic and molecular combined) gas matches the morphology of dust surface density very well. There is also a good correspondence of CO emission with the filaments of PAH emission seen in the mid-infrared. Thus  the three ``cold'' phases of the ISM --- H$_2$, {\sc Hi}, and dust --- appear closely associated.  Modeling the dust spectral energy distribution shows evidence of high interstellar radiation fields along the minor axis --- i.e., the dust temperature is high along the outflow --- providing additional evidence that this cold gas is indeed interfacing with the outflow (though shocks or collisional heating at the interface with the hot gas might also contribute some to the heating). Moreover, individual CO and the {\sc Hi} spectra along the minor axis appear broad and sometimes double-peaked, indicative of a multi-directional outflow almost aligned with the plane of the sky. 

This evidence suggests that the extended CO emission along the minor axis is associated with M82's superwind. Our maps also include the bright starburst region and  show a large molecular disk that extends along the major axis of the galaxy out into the tidal streamers created by the galaxy's interaction with M81 \citep{YUN94}. In order to study the molecular component of the  outflow quantitatively, we have decomposed the area around M82 into regions dominated by the outflow, the bright  central disk, and the extended disk.

The last panel in Figure \ref{fig:wind} shows our region definition. Based on the CO intensity and velocity, we define a bright disk region (red). We define the outflow region to cover areas of bright H$\alpha$, FUV, and X--ray emission outside this region (blue; shown in all panels). The remaining area corresponds to the extended disk and tidal streamers (gray). The extent of the ``outflow'' (blue) region parallel to the major axis is 3~kpc (i.e., $\pm 1.5$~kpc about the major axis). For simplicity, we do not treat the north and south differently, although \citet{SHOPBELL98} have shown asymmetry between the two in both orientation and opening angle.  Figure \ref{fig:wind} and subsequent figures show that our simple region definition captures signatures of hot gas, hot dust, the minor axis gradient, and wide line profiles. We also indicate a cone with a base of $300$~pc and an opening angle of $20\arcdeg$ ($\pm 10\arcdeg$); we will find some evidence for this opening angle from line splitting later in the paper and the geometry corresponds approximately to the \citet{SHOPBELL98} picture of the outflow.

A caveat to this decomposition is that some emission above and below the midplane in the outflow region may still come from an extended disk. While it is highly inclined ($i \approx 80\arcdeg$), M82 is not perfectly edge on. We do observe an extended disk of CO  along the major axis. At $i=80\arcdeg$, a 5~kpc disk would be foreshortened to have extent of about a kpc along the minor axis, which may contaminate measurements of the outflow. We will see below that material beyond about 1.5~kpc from the midplane (the two dashed lines in Figure \ref{fig:wind}) is very securely associated with the outflow. Closer to the disk we show evidence that the outflow appears important to the CO emission, but expect contamination from a foreshortened extended disk.

Table \ref{tab:lco} reports the integrated emission for our whole map and individual regions. We report both the flux, from summing  $I_{\rm CO} \times \Omega$, and the luminosity, $I_{\rm CO} \times \Omega \times d^2$. We also note the flux above a height of $1.5$~kpc in the outflow region, which we can securely assign to the outflow.

\begin{deluxetable}{lcc}[h]
\tabletypesize{\scriptsize}
\tablecaption{Integrated CO Emission \label{tab:lco}}
\tablewidth{0pt}
\tablehead{
\colhead{Region} & 
\colhead{CO Flux} & 
\colhead{CO Luminosity} 
\\
\colhead{} & 
\colhead{(K km s$^{-1}$ $\arcsec^2$)} & 
\colhead{(K km s$^{-1}$ pc$^2$)}
}
\startdata
Entire HERA Map & $4.3 \times 10^5$ & $5.3 \times 10^8$ \\
$\ldots$ Central Disk & $3.1 \times 10^5$ & $3.8 \times 10^8$ \\
$\ldots$ Outflow Region & $0.8 \times 10^5$ & $1.0 \times 10^8$ \\
$\ldots$ Outflow $|z| > 1.5$~kpc & $0.12 \times 10^5$ & $0.15 \times 10^8$ \\
$\ldots$ Disk \& Streamers & $0.4 \times 10^5$ & $0.5 \times 10^8$ 
\enddata
\tablecomments{CO $J=2\rightarrow1$ emission. Luminosities assume $d=3.6$~Mpc.
For region definition see Figure \ref{fig:wind}.}
\end{deluxetable}

\section{CO and Other Phases of the Cool ISM}
\label{sec:atomicdust}

\begin{figure*}
\epsscale{1.0}
\plottwo{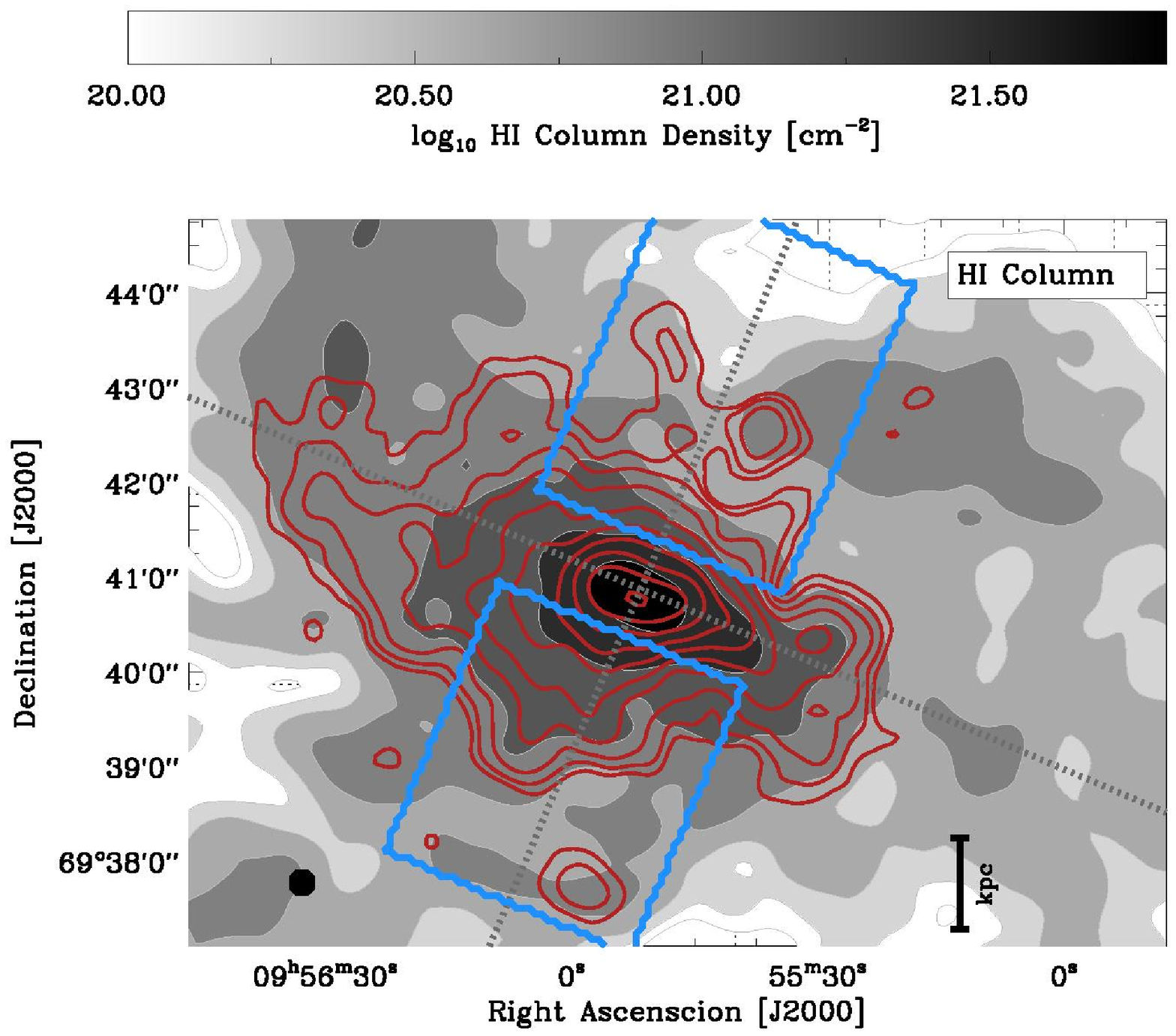}{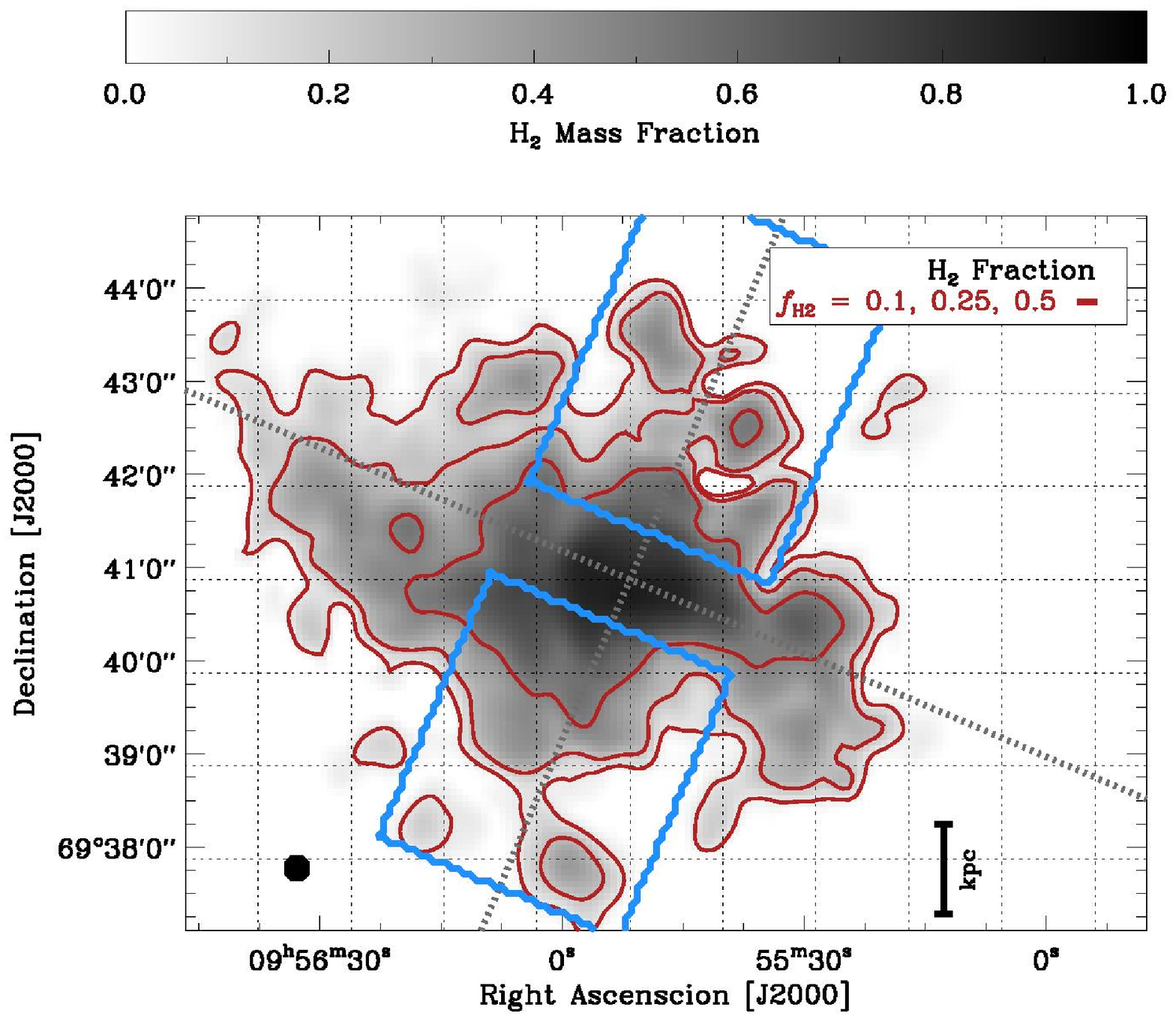}
\plottwo{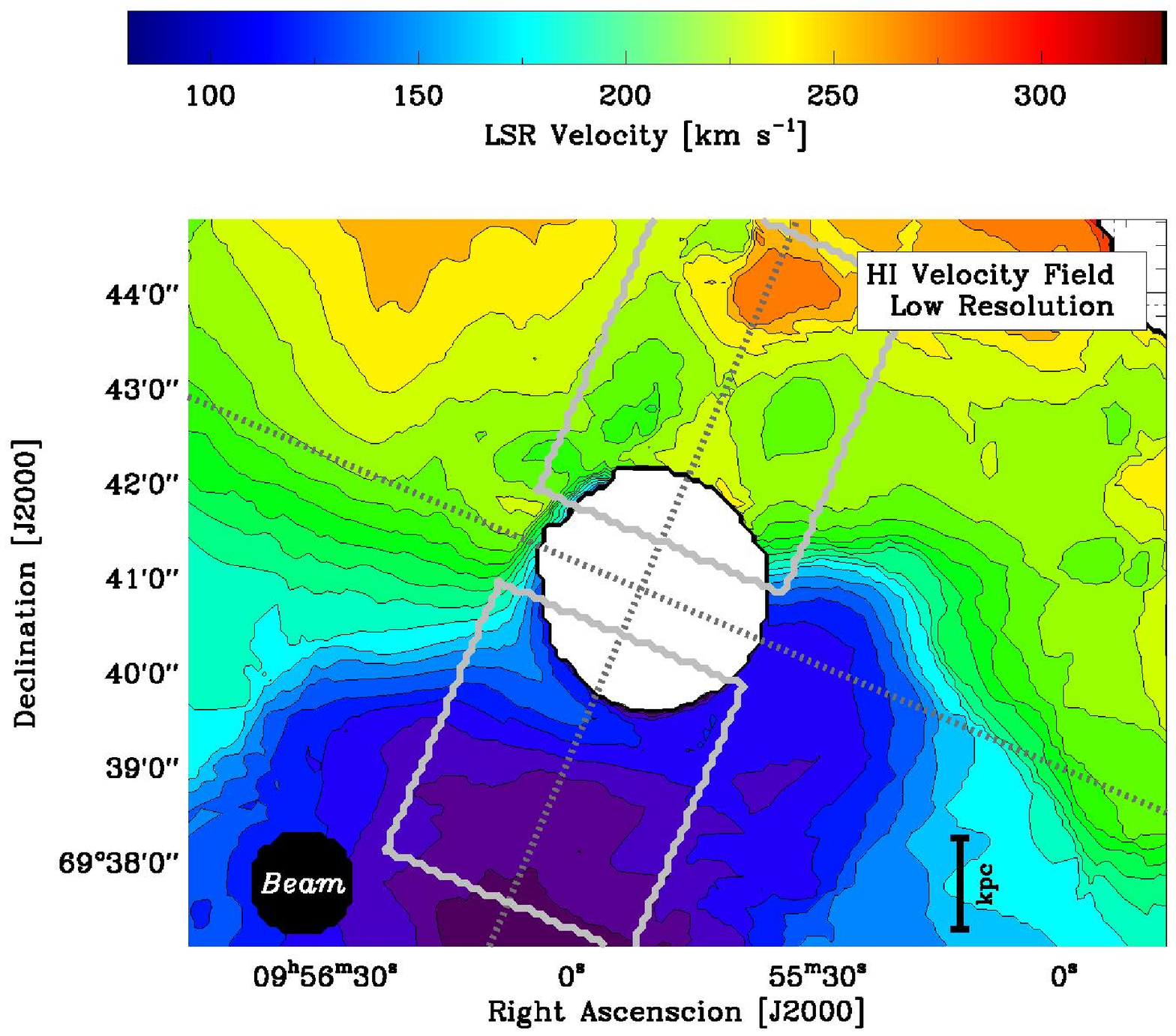}{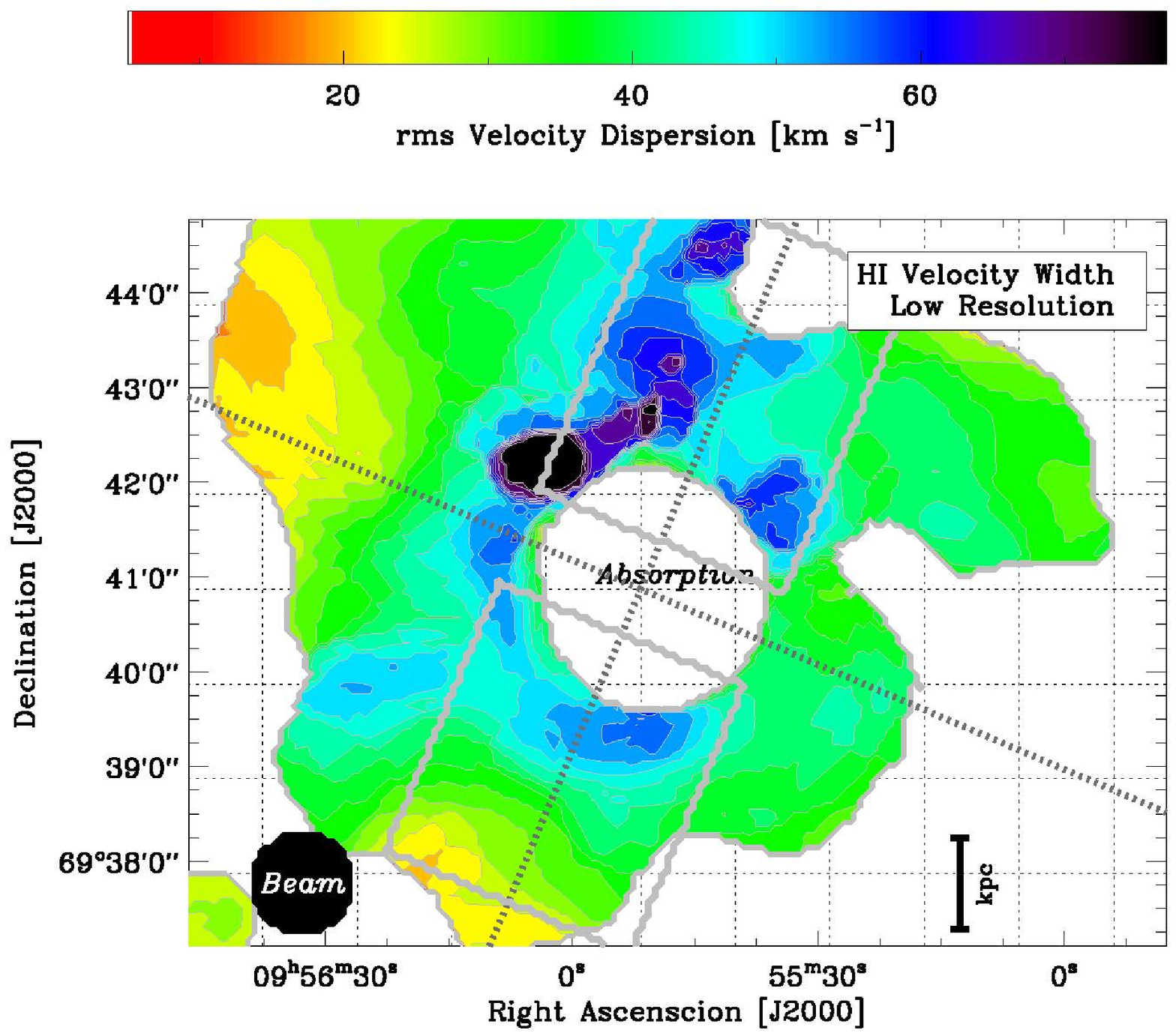}
\caption{
\label{fig:hi} 
Atomic gas over the area of our CO survey. ({\em top left}) {\sc Hi} column density (grayscale) at matched resolution with CO contours (red contours). Blue rectangles indicate the area of the outflow (same as Figure \ref{fig:wind}). ({\em top right}) Approximate H$_2$ mass fraction, $f_{\rm H2} = M_{\rm H2} / (M_{\rm H2} + M_{\rm HI})$, with contours indicating where the gas is 10, 25, and 50\% molecular. ({\em bottom left}) {\sc Hi} velocity field at lower resolution (beam in the lower left corner) to take advantage of increased signal to noise.  Again the area of the outflow is indicated. Note the minor axis gradient \citep{COTTRELL77}, similar to what we observe in the CO (Figure \ref{fig:cube}). ({\em bottom right}) {\sc Hi} velocity dispersion, also at low resolution, tracing the line width over regions of high {\sc Hi} column. Regions associated with the outflow appear to have higher velocity dispersions, reflecting more complex, multi-component line profiles.}
\end{figure*}

Neutral atomic hydrogen ({\sc Hi}) and dust both also extend to great distances around M82 and represent key points of comparison to understand the evolution of the molecular phase of the superwind.

\subsection{Atomic Gas}

Although most of the gas in the disk of M82 appears to be molecular, the extended tidal features are mostly atomic, so that an enormous {\sc Hi} superstructure surrounds M82 \citep{COTTRELL77,YUN93,YUN94}.  Figure \ref{fig:hi} compares the column density and kinematics of atomic gas to our CO map. In the top left panel, we see that the shapes of CO and {\sc Hi} features near the starburst are similar, but that the {\sc Hi} features extend well beyond CO. The high column {\sc Hi} arcs to the northeast and northwest are the well-known tidal features induced by the interaction with M81 \citep{YUN94}. CO is present in these tidal features, but as they extend progressively further from the disk the amount of molecular gas diminishes. The top right panel in Figure \ref{fig:hi} shows that directly by plotting $f_{\rm H2}$, the fraction of gas mass that is molecular (as opposed to atomic); here we use $\alpha_{\rm CO} = 2$~\acounits\ for simplicity. Gas near the starburst and in the stellar disk is mostly molecular but becomes progressively less so with increasing distance from the burst and disk.

Figure \ref{fig:hi} shows that the {\sc Hi} exhibits the same minor axis velocity gradient as the CO. The {\sc Hi} minor axis velocity gradient was first recognized by \citet{COTTRELL77}, who placed a lower limit of 23~km~s$^{-1}$~kpc$^{-1}$ on the gradient and noted the link to the minor axis  H$\alpha$ filaments. In general, the CO and {\sc Hi} velocity fields agree. Comparing CO and {\sc Hi} velocities across the survey area, we find no mean offset, though there is a substantial 20~km~s$^{-1}$ rms scatter between individual local {\sc Hi} and CO mean velocities. We defer a more detailed investigation to future work focused on improved {\sc Hi} data. Thus from Figure \ref{fig:hi}, we conclude that the {\sc Hi} and {\sc H$_2$} appear mixed where both are  present, showing similar morphology and kinematics, but that {\sc Hi} is much more extended than H$_2$ traced by CO.

Finally, the bottom right panel in Figure \ref{fig:hi} shows that broad {\sc Hi} lines, indicated by a high rms line of sight velocity dispersion, tend to be associated with the minor axis of the galaxy, a signature of the complex line profiles associated with the outflow.

\subsection{Dust}

\begin{figure*}
\epsscale{1.1}
\plotone{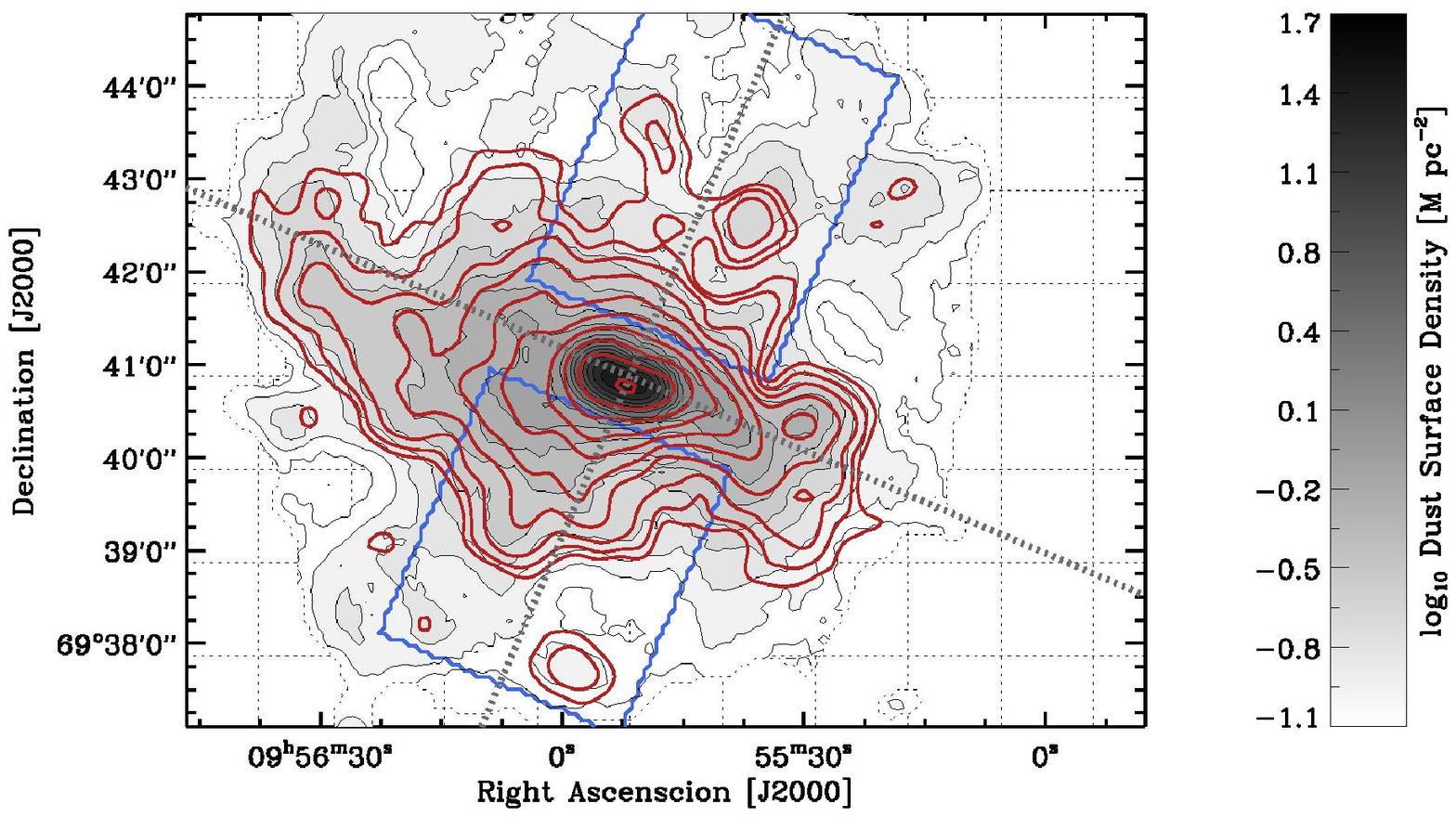}
\plotone{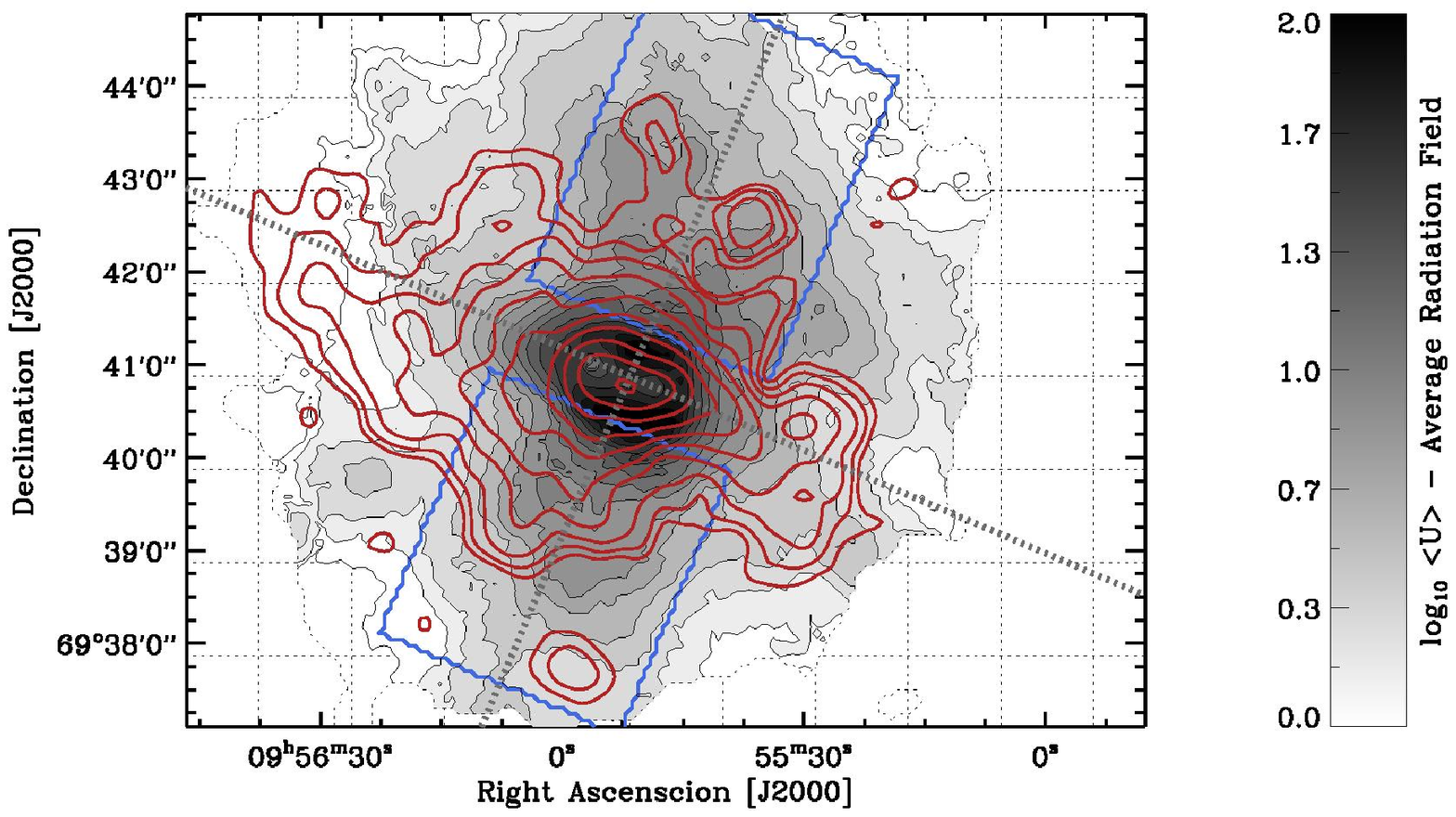}
\caption{
\label{fig:dustmaps} Dust mass surface density ({\em top}) and the mean interstellar radiation field illuminating the dust ({\em bottom}), derived from point-by-point fitting of the infrared spectral energy distribution following \citet{DRAINE07A,DRAINE07B}. Before fitting, all data were convolved to the limiting resolution of {\em Herschel} at 350$\mu$m. In the left panel the red contours show CO integrated intensity contours. In both panels, blue outlines show our ``outflow'' region (Figure \ref{fig:wind}). The dust mass contours begin at $\log_{10} \Sigma_{\rm dust} \left[ {\rm M}_\odot~{\rm pc}^{-2} \right] = -1$ and step by $0.15$. The $\left< U \right>$ contours begin at $1$ (a Solar Neighborhood field) and step by $\sqrt{2}$. The dashed region shows where we carry out the SED fit, which corresponds to an area above a limiting surface brightness of $I_{250} = 15$~MJy~sr$^{-1}$. The dust mass distribution shows a remarkable match to the CO emission. The faint extensions of $\Sigma_{\rm dust}$ beyond the CO  are real and correspond to features seen in the {\sc Hi} map, so that dust around M82 appears mixed with a combination of atomic and molecular gas (see the appendix for an analysis of the  dust-to-gas ratio and conversion factor based on these data). The right panel shows that the radiation field (equivalently dust temperature) returned by our SED fitting shows hotter dust in the region of the outflow. This argues for a direct association between the cool gas and the superwind and agrees with the observation of FUV and H$\alpha$ light scattered off dust grains along the minor axis \citep[e.g., see][]{HOOPES05,COKER13}.}
\end{figure*}

\begin{figure*}
\epsscale{1.2}
\plotone{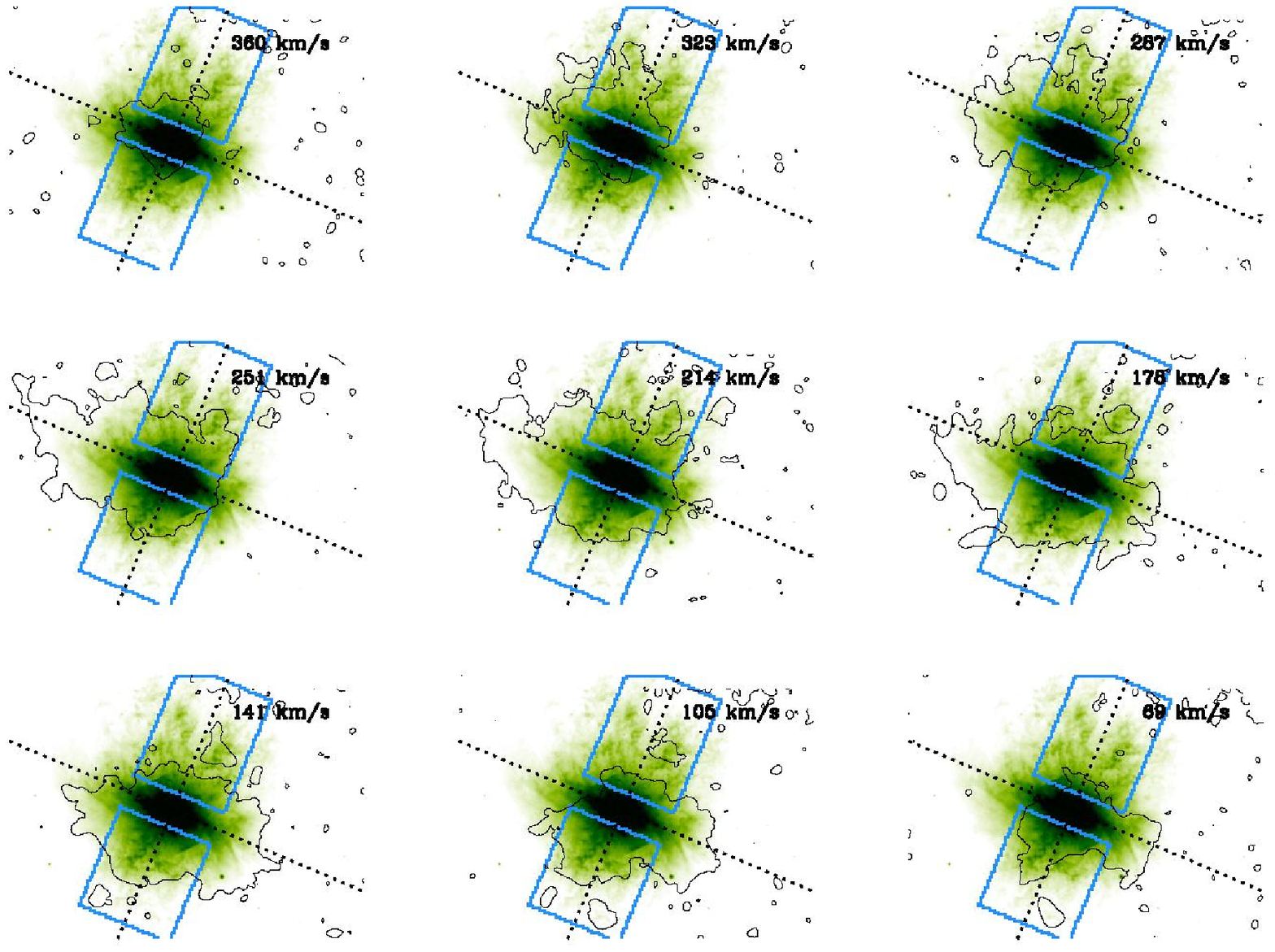}
\caption{
\label{fig:chan_pah} 
Contours for individual channel maps of CO emission in black contour over the $8\mu$m emission \citep[mostly from PAHs,][]{ENGELBRACHT06,DALE09} in the the same area as Figure \ref{fig:cube}. A blue outline shows our two dimensional ``outflow'' region. Individual CO features exhibit a good match to the filamentary structures seen in the mid-IR dust map at much higher resolution. The black contours show where peak CO emission across a seven channel range exceeds $\approx 21$~mK, or $S/N = 3$ in an individual channel.}
\end{figure*}

We expect dust to be mixed with both molecular and atomic gas. We also expect that most of the dust associated with the M82 outflow will have been entrained along with the cooler material because the hot portion of the superwind will quickly destroy dust. \citet{SHOPBELL98} estimate the timescale for grain destruction by sputtering in the very hot (X--ray) portion of the wind to be only $\sim 10^5$~yr \citep[see][]{OSTRIKER73,DRAINE79}. Dust associated with the cooler phase can survive much longer. \citet{SHOPBELL98} also point out that dust should even survive for a long time ($\sim 10^7$~yr) in the lower temperature ionized gas associated with H$\alpha$ emission or a galactic halo, $T \sim 10^{5}$~K (instead of the $10^6$~K associated with the hot X--ray emitting gas). Therefore although dust directly associated with the hot wind will be destroyed, if material ejected in a cool phase becomes ionized as it leaves the galaxy we do not expect its dust to be immediately destroyed. Dust therefore offers the potential to uniformly measure gas mass in all but the hottest phase of the outflow.

Figure \ref{fig:dustmaps} shows the results of our IR SED fitting, the distribution of dust mass surface density, $\Sigma_{\rm dust}$, and the mean interstellar radiation field heating the dust, $\left< U \right>$ (equivalently dust temperature, which might include some contribution from collisional heating; see the appendix). The dust mass distribution shows an excellent morphological match to the CO, though low surface density dust extends further from the galaxy than CO emission. These extensions all match features in the {\sc Hi} maps (Figure \ref{fig:hi}). Together, the three maps show the neutral gas becoming more  atomic and less molecular at greater distances from M82 and both phases mixed with dust. Most of the visible dust emission can be associated with either {\sc Hi} or CO, so Figure \ref{fig:dustmaps} does not show evidence for a substantial component of dust associated with ionized or hot gas except perhaps above $\approx 3$~kpc to the north of the galaxy.

The outflow is particularly evident in $\left< U \right>$ (right panel). As found by \citet{ROUSSEL10}, the infrared colors in the wind region indicate hotter dust, and thus a more intense interstellar radiation field along the outflow. UV emission scattered off dust shows that radiation from the starburst escapes along the cavity blown by the superwind \citep{HOOPES05,COKER13}. This same radiation heats the dust in or along the edge of the cavity. Dust in the extended disk is shielded by large amounts of intervening material. Inasmuch as dust appears well-mixed with gas, this  provides further evidence that the cool gas along the minor axis is associated with the outflow.

The dusty outflow from the galaxy is particularly visible in the 8$\mu$m emission, which is dominated by PAH emission \citep[see the spectral mapping of][]{BEIRAO15}. The $8\mu$m emission \citep{ENGELBRACHT06} shows a filamentary structure similar to the H$\alpha$ \citep[e.g., see][]{KANEDA10}, extending several kpc above and below the galaxy. We show the comparison in Figure \ref{fig:chan_pah}, where we plot CO contours from individual channel maps over the 8$\mu$m emission. CO features in individual channel maps show good correspondence to the $8\mu$m filaments, which in turn show detailed structure down to the resolution limit of IRAC, $\sim 2\arcsec \approx 35$~pc at the distance of M82. Until high resolution CO data become available over a wide field, the IRAC emission may give the best detailed view of cold  ISM structure in the outflow \citep[e.g., see][]{VEILLEUX09B}.

\subsection{Molecular Mass Estimates and Dust-to-Gas Ratio}

We wish to translate CO intensity into molecular mass, which requires an estimate of the CO-to-H$_2$ conversion factor. \citet{BOLATTO13B} review this topic and \citet{BOLATTO13A} have considered the CO-to-H$_2$ conversion factor for the case of the molecular wind in NGC 253. In the appendix, we constrain  $\alpha_{\rm CO}$ in M82 by comparing low-$J$ CO transitions and combining $\Sigma_{\rm dust}$, CO, and {\sc Hi} emission. Following \citet{WEISS01} and \citet{WEISS05} we conclude that the observed line ratios do not require optically thin gas and indeed appear difficult to reproduce for optically thin CO emission. The three lowest $J$ $^{12}$CO transitions could arise from either hotter, low density ($T \gtrsim 30$~K, $n \lesssim 10^3$~cm$^{-3}$) gas or cooler dense gas ($T \lesssim 20$~K, $n \sim 10^5$~cm$^{-3}$). Previous results using also using CO isotopologues by \citet{WEISS01} and \citet{WEISS05} argue for the former case. This hot, low density gas implies $\alpha_{\rm CO} \sim 2$--$3$ times lower than Galactic, but with significant uncertainties. This agrees with analysis of the virial masses of resolved clouds by \citet{KETO05}.

We use dust emission as an independent tracer of the total cool gas to derive another constraint on $\alpha_{\rm CO}$. In the appendix, we show that bright CO emitting regions in M82 show a high ratio of  CO (2--1) to 350$\mu$m continuum emission when compared to a large reference sample compiled from the HERACLES and KINGFISH surveys.  This argues for $\alpha_{\rm CO}$ $\approx 2.5$ times lower in M82 than in a typical disk galaxy. This is more than an order of magnitude higher than the optically thin value.

We also use the modeled dust surface density, $\Sigma_{\rm dust}$ and follow a modified version of \citet{SANDSTROM13} and \citet{LEROY11} to compare dust, CO, and {\sc Hi}. This approach describes regions of a galaxy using a fixed gas-to-dust ratio, $\delta_{\rm GDR}$, and a fixed CO-to-H$_2$ conversion factor, $\alpha_{\rm CO}$, so that

\begin{equation}
\delta_{\rm GDR} \times \Sigma_{\rm dust} = \alpha_{\rm CO} \times I_{\rm CO} + \Sigma_{\rm HI}~.
\end{equation}

\noindent Here $\Sigma_{\rm dust}$ and $\Sigma_{\rm HI}$ are the mass surface densities of dust and atomic gas inferred from IR and 21-cm observations. $I_{\rm CO}$ is the CO intensity. Observations of a wide range of CO, {\sc Hi},  and dust values simultaneously constrain the gas-to-dust ratio and CO-to-H$_2$ conversion factor  \citep[see][for more details]{LEROY11,SANDSTROM13}. Adopting this approach, we simultaneously solve for the best-fit $\alpha_{\rm CO}$ and $\delta_{\rm GDR}$ in the extended disk of M82. We find the same result for $\alpha_{\rm CO}$  implied by the simple $I_{\rm CO 2-1}/I_{\rm 350}$ treatment and a dust to gas mass ratio of $\delta_{\rm GDR}^{-1} \approx 0.009$. 

Our M82 dust-to-gas ratio is $\approx 35\%$ lower than the value found by \citet{SANDSTROM13}. They studied the CO-bright parts of many disk galaxies using the same  CO line and dust models and found a dust-to-gas ratio of $0.014$. Our value does resemble the $0.01$ found by \citet{DRAINE07B} considering integrated galaxy SEDs using the same  dust models. Thus M82 appears somewhat dust-poor compared to the inner part of a large disk galaxy. This could agree with its small mass and the mass-metallicity relation \citep{TREMONTI04} and indeed there is some evidence of subsolar metallicity in M82 \citep{ORIGLIA04,NAGAO11} despite its apparently normal abundance patterns \citep{FORSTERSCHREIBER01}. Note that we  emphasize comparative statements when discussing the dust-to-gas ratio because the \citet{DRAINE07A} dust models  appear to have a normalization that is $\approx 2$ times too high. Thus our best-estimate true dust-to-gas mass ratio would be  $\approx 0.005$ ($\approx 1$-to-$200$).

Synthesizing $\alpha_{\rm CO}$ estimates from $\Sigma_{\rm dust}$, IR intensity, and CO lines, our best estimate $\alpha_{\rm CO}$ is given in Equation \ref{eq:brokenaco}. It is $\alpha_{\rm CO} = 2.5$~\acounits in the extended disk and faint parts of the outflow and $\alpha_{\rm CO} = 1$~\acounits in the bright part of the outflow and the starburst. We also report results for the simpler case of a single $\alpha_{\rm CO}^{2-1} \approx 2$~\acounits . Given that our estimates span $\alpha_{\rm CO}^{2-1} \approx 1$--$4$~\acounits , a factor of $\approx 2$ represents a reasonable, realistic systematic uncertainty on the conversion to molecular mass.

These conversion factors, temperatures, and densities differ from the Milky Way but are not truly extreme. Based on this analysis, one might speculate that the extended CO emission comes from a close analog of the resolved cloud population seen by \citet{KETO05} in the starburst region of M82. Our knowledge of CO substructure is circumstantial, but the small size of the bright PAH filaments, the ``normal'' CO line ratios, and the similar $I_{\rm CO 2-1}/I_{350}$ ratios all argue that physical conditions inside the gas may not change overwhelmingly between the inner disk and the brighter parts of the outflow. Reinforcing this idea, \citet{KEPLEY14} showed some evidence of dense gas traced by HCO$^+$ extending off the starburst disk into the outflow. This would support the idea that the material in the M82 outflow consists of dense, opaque clouds of molecular gas.

\section{Minor Axis Mass Profile: Changing Phase and Falling Profiles}
\label{sec:massprof}

\begin{figure*}
\plotone{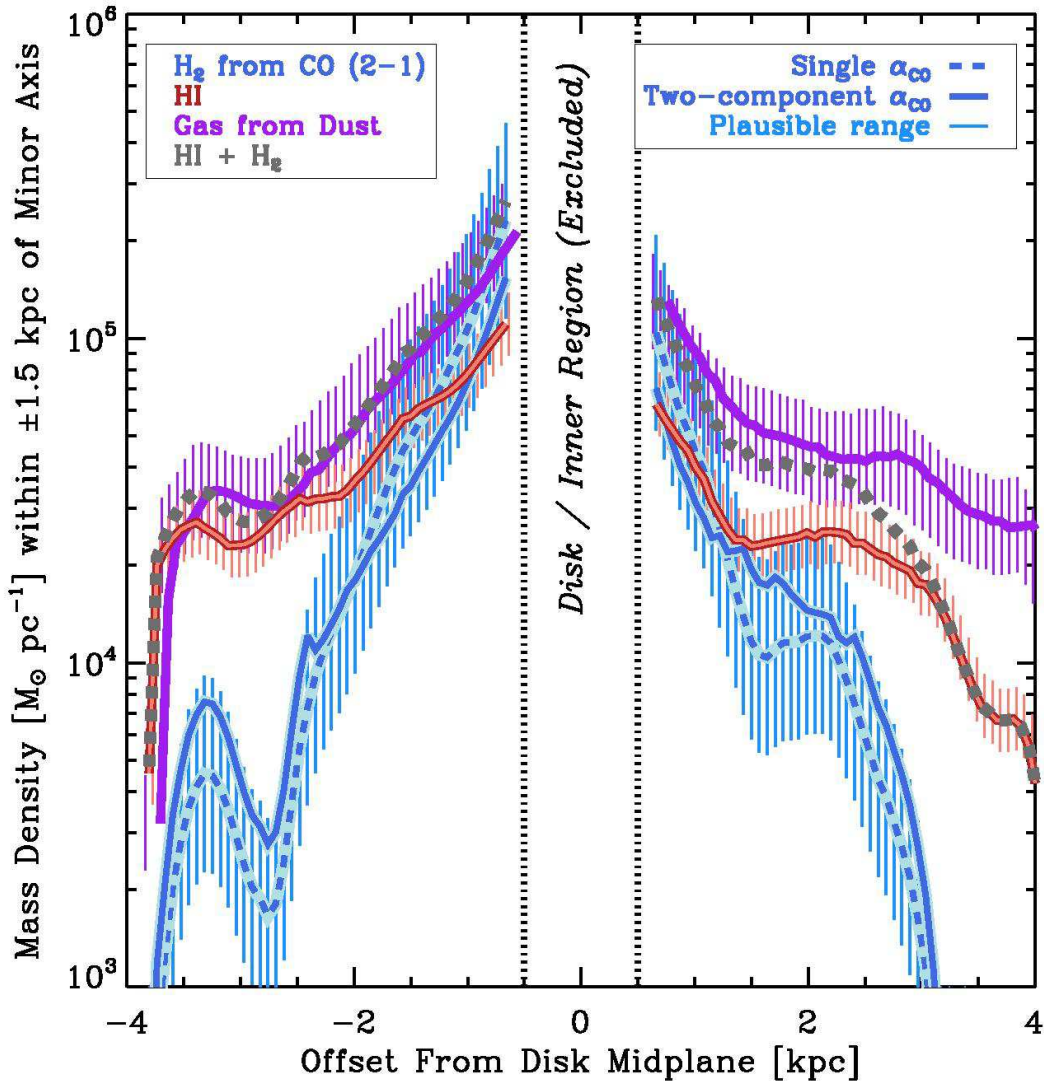}
\caption{\label{fig:vertprof}
Linear mass density profile along the direction of the outflow of molecular gas (blue), atomic gas (red), and total gas estimated from dust (purple). The $y$-axis  reports the mass per unit length integrated within $\pm 1.5$~kpc of the minor axis (the blue rectangle in Figure \ref{fig:wind}) as a function of height above the plane. That is, the curves are proportional to average projected mass surface surface density and the value can be multiplied by  length along the minor axis to recover a mass. Here ``gas from dust'' indicates dust surface density multiplied by the estimated ratio of gas mass to dust mass, $\Sigma_{\rm dust} \times \delta_{\rm DGR}^{-1}$ with $\delta_{\rm DGR} = 0.009$ (see appendix). Error bars indicate systematic uncertainties, e.g., from the CO-to-H$_2$ conversion factor, dust-to-gas ratio, or calibration. Although we show vertical displacements as low as $\approx 500$~pc, we caution that close to the  disk, the association of material with outflow is less straightforward.}
\end{figure*}

\begin{figure*}
\plotone{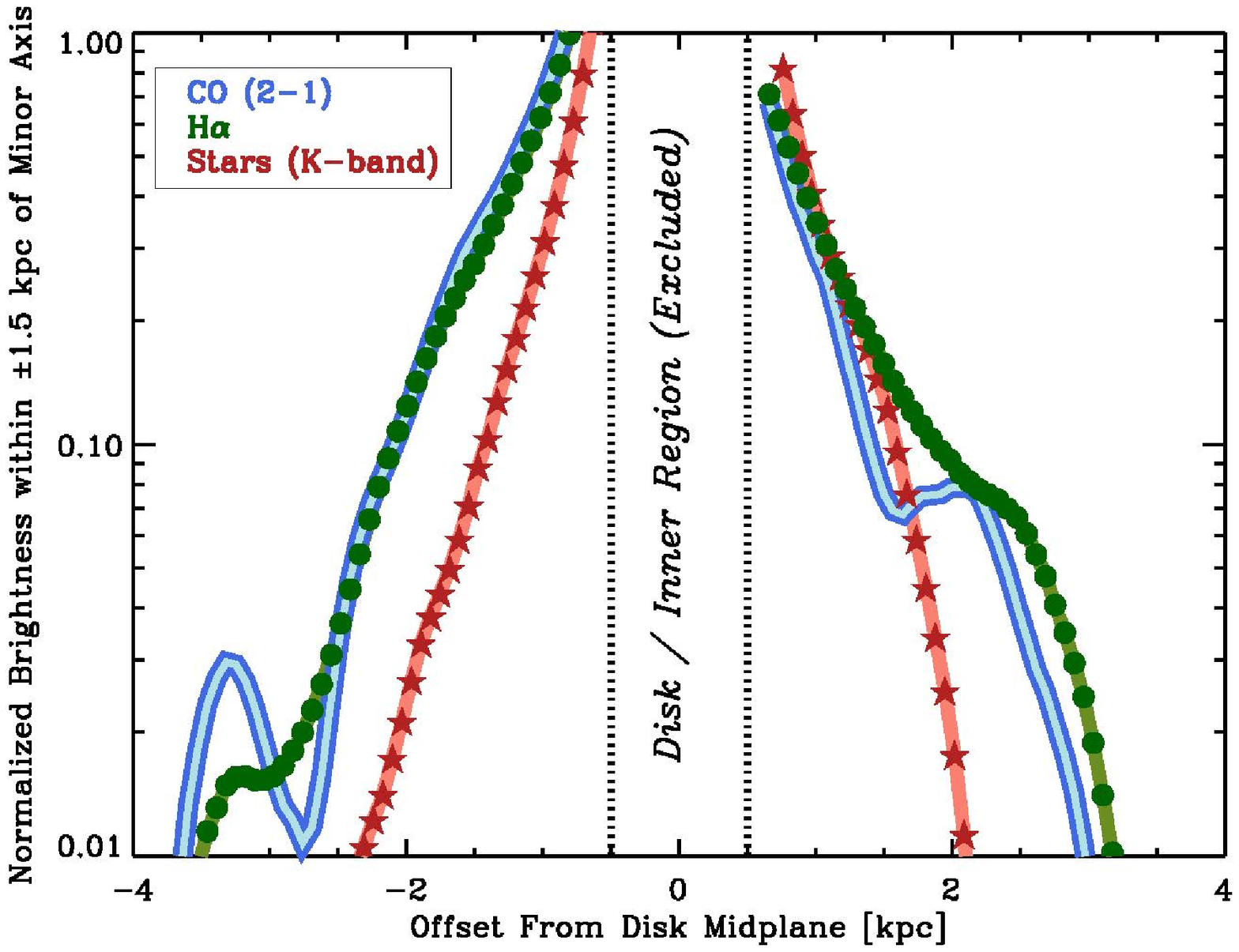}
\caption{\label{fig:hastarprof} Intensity of $K$-band light (starlight), H$\alpha$ emission, and CO along the minor (outflow) axis. All three profiles are normalized to have an average value of unity at $\approx \pm 0.5$~kpc. Note the slight misalignment between the CO midplane and the stellar disk midplane (also visible in Figure \ref{fig:stars}). H$\alpha$ and CO are clearly more extended than the stellar disk and decline at a very similar rate at greater offsets from the disk midplane.}
\end{figure*}

With H$_2$, {\sc Hi}, and dust estimates we can examine the composition and total amount of cold interstellar material along the outflow of M82. Figure \ref{fig:vertprof} shows our best estimate of this profile along the outflow region. The $y$-axis shows mass per unit vertical length integrated over a width $\pm 1.5$~kpc around the major axis. The $x$-axis shows vertical displacement from the disk with positive indicating north. We show profiles for H$_2$, {\sc Hi}, and total gas estimated from dust assuming a dust to gas ratio $\delta_{\rm DGR} = 0.009$. The vertical error bars indicate the systematic uncertainty in the mass profile. For H$_2$ this is dominated by the CO-to-H$_2$ conversion factor. For dust this is set by our adopted dust to gas ratio.

Figure \ref{fig:hastarprof} shows the corresponding plot for H$\alpha$ emission and near-infrared light, which tracks the distribution of stellar mass. Here we do not estimate mass, but plot intensity averaged as a function of vertical displacement. Again we average within $\pm 1.5$~kpc of the minor axis (the blue ``outflow'') region. We normalize the profiles to unity near $\pm 0.5$~kpc to show the relative decline of CO, starlight, and H$\alpha$.

\begin{deluxetable}{lc}[h]
\tabletypesize{\scriptsize}
\tablecaption{Scale Lengths Along the Minor (Outflow) Axis \label{tab:mass}}
\tablewidth{0pt}
\tablehead{
\colhead{Quantity} & 
\colhead{Scale Length (kpc)} 
}
\startdata
molecular gas & $0.5 \pm 0.1$ \\
H$\alpha$ intensity & $0.6 \pm 0.1$ \\
atomic gas & $1.7 \pm 0.2$ \\
$f_{\rm H2} \equiv {\rm H}_2 / ({\rm HI} + {\rm H}_2)$ & $\approx 0.9$  \\
dust mass & $1.4 \pm 0.3$ \\
starlight & $0.4 \pm 0.1$ \\
\enddata
\tablecomments{Over the range 0.5--4~kpc from the disk. See Figure \ref{fig:vertprof}. The $\pm$ values indicate the spread between north and south. For an outflow velocity of $450$~km~s$^{-1}$, it takes about $2.2$~Myr to travel 1~kpc.}
\end{deluxetable}

Figure \ref{fig:vertprof} shows that {\sc Hi} and H$_2$ both contribute a large fraction of the gas mass around M82 between $0.5$ and $1.5$~kpc from the plane. H$_2$ likely makes up most of the gas within $\approx 1$~kpc, though this depends on the adopted $\alpha_{\rm CO}$. In this same regime, the gas implied by dust is a good match to the total {\sc Hi}+H$_2$ gas. This is formally by construction (see appendix) but our results would be largely the same if we simply adopted the \citet{DRAINE07B} dust-to-gas ratio.

At larger heights, the surface density of all phases declines with increasing distance from the disk. The strongest decline is in the molecular component, which declines sharply with increasing distance to both the north and south of the galaxy\footnote{This is not a sensitivity effect; we detect CO over most lines of sight out to at least $\pm 2$~kpc, by which time the decline is clearly evident.}. The associated exponential  scale length is $\approx 0.5$~kpc and almost the same to the north ($\approx 0.42$~kpc) and south ($\approx 0.53$~kpc) of the galaxy.

The fall-off in CO closely matches the decline in H$\alpha$ intensities (Figure \ref{fig:hastarprof}). Thus, Figure \ref{fig:wind} and the match in the profiles in Figure \ref{fig:hastarprof} support the idea that bright H$\alpha$ emission arises from the interface of the cavity carved by the hot wind with cold, denser material \citep[e.g.,][]{SHOPBELL98,WESTMOQUETTE07}. As the dense, confining material falls off, the H$\alpha$ intensity also falls off almost in lock-step.

Figure \ref{fig:hastarprof} also shows the decline in the stellar profile at large heights. The slight vertical  misalignment between the bright CO disk and the stellar disk appears in the asymmetry of the north and south profiles (this is also visible in Figure \ref{fig:stars}). With a scale length of $\approx 0.4$~kpc, the stars have the sharpest decline of any component, emphasizing the degree to which the gas --- even the H$_2$ --- really has been  ejected from the central region. In particular, CO and H$\alpha$ emission outside $\sim 1.5$~kpc is very clearly extraplanar and we shall see that this is the emission that can best be linked directly to the hot outflow.

The other two neutral gas tracers decline more slowly than the CO. For $|z| < 3.5$~kpc, the {\sc Hi} shows a scale length $\approx 1.7$~kpc, that is roughly three times longer than the H$_2$. This longer scale length of {\sc Hi} compared to H$_2$ means that the phase of the neutral ISM is changing systematically as one  moves away from the disk. The molecular fraction $f_{\rm H2} \equiv {\rm H}_2 / ({\rm HI} + {\rm H}_2)$ declines almost exponentially with a scale length $\approx 0.9$~kpc as a function  of vertical displacement from the disk. As a result, molecular gas contributes substantially to the ISM around M82 within about 1.5~kpc of the disk but is subdominant compared to {\sc Hi} by a vertical displacement of 3~kpc. For the most part, these statements are robust to the conversion factor prescription adopted, though near the disk this matters to the exact accounting of where H$_2$ exceeds {\sc Hi}.

Dust shows a scale length intermediate between the CO and {\sc Hi}, $\approx 1.4$~kpc --- steeper to the south and somewhat shallower to the north. The dust profile combined with $\delta_{\rm GDR} \approx 0.009$ is a reasonable match to the total gas inferred from adding the {\sc Hi} and H$_2$ estimates (the gray line). There is not clear evidence for a large amount of ionized gas associated with dust around M82. If present, this would appear as dust without associated gas traced by either CO or {\sc Hi}. The best candidate for such emission would be above $3$~kpc to the north of the galaxy, but this is near the edge of our maps and faint, so we hesitate to ascribe much significance to it.

We have argued based on dust temperatures, {\sc Hi} velocity dispersions, and the apparent kinematic association of CO and {\sc Hi} that the {\sc Hi} in Figure \ref{fig:vertprof} is probably linked to the disk and outflow rather than superimposed high latitude material.  In this case, given an outflow velocity a displacement along the minor axis translates linearly to the time since gas was expelled from the starburst and the $x$ axis in the profiles can be thought of as a timeline. If we make the assumption that the starburst has continuously expelled gas at a constant rate then we can consider this profile as a plot of the time evolution of gas as it leaves the starburst. Thus our first-order interpretation of the vertical profiles is that the phase of the gas changes  as it moves away from the galaxy and that gas and dust disappear from the minor axis outflow as they reach larger and larger distances from the disk. The relevant timescales are all of order a Myr; for an outflow velocity of $450$~km~s$^{-1}$ (Section \ref{sec:pv}) it takes gas $\approx 2.3$~Myr to travel a kpc. Therefore roughly every Myr, there is an $e$-folding in the amount of H$_2$ associated with the outflow and roughly every $2$~Myr the gas becomes a factor of $e$ more atomic. Meanwhile the overall amount of gas in the outflow diminishes with an $e$-folding time of $\sim 3$~Myr.

For the phase change, the logical explanation is that photodissociation by light from the starburst and dissociation in shocks outpace H$_2$ formation as clouds move away from the galaxy. It seems fair to assume that the density and shielding of molecular gas will decline compared to the starburst if a cloud is placed in the lower ambient pressure of the outflow with less nearby material to shield it and no larger molecular medium to replenish the cloud mass. Meanwhile the dust modeling and scattered UV light both clearly show that much of the cool ISM along the outflow is being bathed in a strong radiation field that presumably includes dissociating photons. Shocks also appear to play an important role in the outflow, at the very least processing small dust grains in the outflow \citep[e.g.,][]{BEIRAO15} and heating a small fraction of the H$_2$ to very high temperatures \citet{VEILLEUX09B}. The expected rate at which $f_{\rm H2}$ will change depends on cloud properties and geometry (combined shielding and density structure and incident  radiation field), but we note that in simulations treating the opposite problem (molecule formation), \citet{GLOVER12} see dramatic changes in the molecular fraction of initially  atomic clouds in roughly a Myr. It seems reasonable to consider that if one tipped the balance in the other direction, a cloud could be dissociated  on a similar timescale.

\section{The Wide-Field View of Dust and Gas: the Finite Extent of the Cool Material}

\begin{figure*}
\plotone{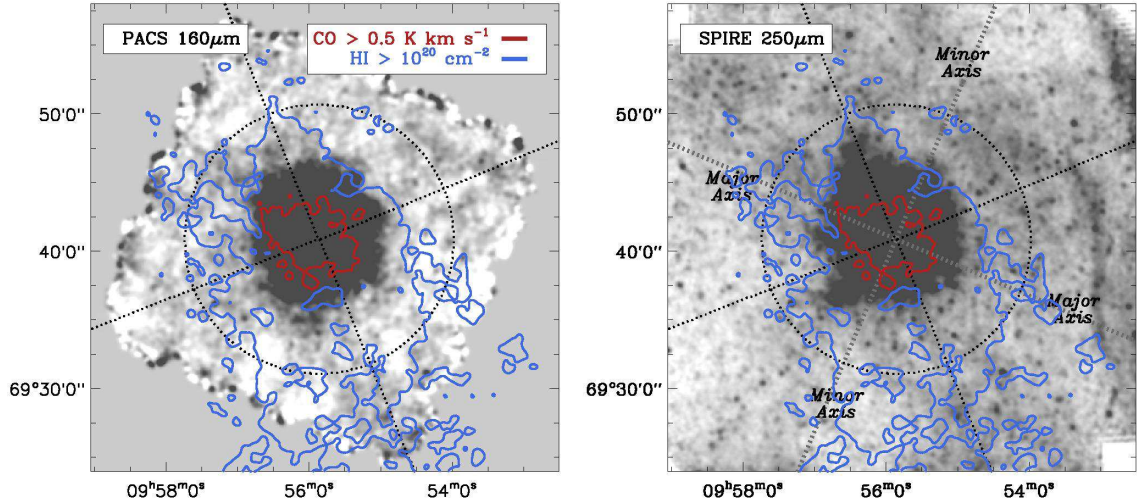}
\caption{\label{fig:wideir} Infrared images at 160$\mu$m (left) and 250$\mu$m (right) shown at high saturation with CO (red) and {\sc Hi} (blue) contours showing the extent of bright  molecular and atomic gas. Dotted black lines show the major and minor axis definitions used for the profiles in Figure \ref{fig:wideprof}, which are also labeled in the right panel. The circle indicates a radius of 10~kpc.
}
\end{figure*}

\begin{figure*}
\plotone{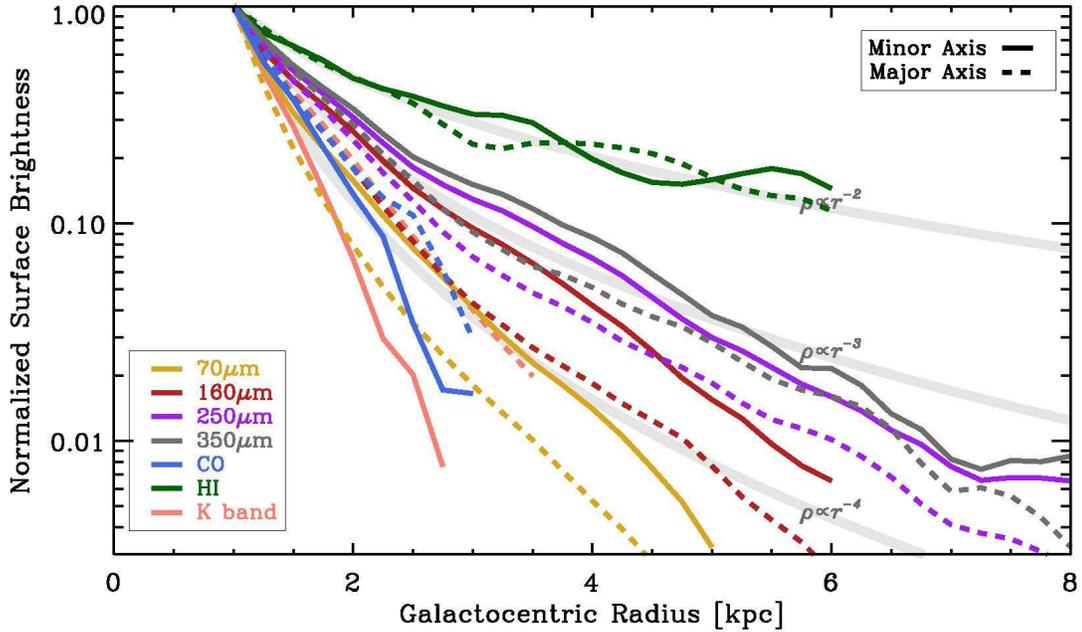}
\caption{\label{fig:wideprof} Radial profiles of infrared intensity and gas tracers over a wide area around M82. The $x$-axis shows displacement on the sky from the galaxy center. The $y$-axis shows intensity normalized to unity at $r = 1$~kpc. We build separate profiles along the major axis (dotted lines) and along the minor axis (solid lines). Gray lines show the projected surface brightness profiles of several density distributions. The profiles show that: (1) CO emission declines more rapidly than any other ISM tracer, indicating a changing phase of the gas around M82; and unlike any other tracer it is more extended along the major axis, reflecting M82's extended gas disk (2) dust emission is slightly more elongated along the minor axis, likely as a result of the outflow, but this effect is not large; (3) dust temperature, indicated by infrared color, declines steadily moving away from the galaxy; (4) {\sc Hi} is the most extended circumgalactic cool component, likely because it reflects the tidal interaction in addition to the outflow; and (5) aside from {\sc Hi}, all of the other tracers are concentrated towards the galaxy. That is, their profiles decline much faster than a projected $\rho \propto r^{-2}$ density profile and so there is less emission (and thus mass) in each successive shell. Table \ref{tab:mass} reports scale lengths associated with these profiles.
}
\end{figure*}

\begin{deluxetable}{lcc}[h]
\tabletypesize{\scriptsize}
\tablecaption{Scale Lengths Over a Wide Field}
\tablewidth{0pt}
\tablehead{
\colhead{Band} & 
\colhead{Minor Axis} &
\colhead{Major Axis} \\
\colhead{} & 
\colhead{[kpc]} &
\colhead{[kpc]} 
}
\startdata
70$\mu$m & $0.8$ & $0.7$ \\
160$\mu$m & $1.1$ & $0.9$ \\
250$\mu$m & $1.7$ & $1.5$ \\
350$\mu$m & $1.6$ & $1.4$ \\
CO & $0.5$ & $0.6$ \\
{\sc Hi} & $3.0$ & $2.6$ \\
$K$-band & $0.4$ & $0.7$
\enddata
\tablecomments{Fits to data outside 1~kpc. ``Minor axis'' refers to an opening angle of $90\arcdeg$ about the north and south outflow axis. ``Major axis'' refers to the rest of the area.
Figure \ref{fig:wideir} shows the relevant geometry and the Figure \ref{fig:wideprof} shows the profiles used for the fit.}
\end{deluxetable}

To help quantify the ultimate fate of cold material that is ejected from M82, we also calculate profiles of gas and dust emission over all azimuthal angles out to large scales. Following the logic of the previous section, if there is less material in each successive shell around the galaxy then continuity arguments suggest that material must stall or fall back. The minor axis profile suggests that this stalling or fallback happens. By considering all azimuthal angles in this analysis, we account for the possibility that emission may move out of the outflow region that we have defined. Recall that even if the material becomes ionized, we would expect it to remain visible in dust emission unless the ionized gas is very hot. 

Because we work with faint emission, we use the IR images themselves for these profiles rather than the dust SED fits. The SED fitting requires significant detections over individual lines of sight in many bands, while the images allow us to average data into radial profiles and so even recover signal at radii where individual lines of sight are not detected. This also means that our results are sensitive to the zero point of the map, but Figure \ref{fig:wideir} and the profiles do not show evidence for either faint extended ``negative'' emission or a plateau to a background level; we show the effect of uncertainty in the zero point in the appendix.

We moved all maps to the resolution and astrometry of the cleaned SPIRE 350$\mu$m map. We then  divide the galaxy into quadrants oriented along the major and minor axis. The dotted lines in Figure \ref{fig:wideir} illustrate the geometry, showing the division into quadrants with the circle indicating a radius of 10~kpc. We constructed major and minor axis profiles out to very large radius, which we plot in Figure \ref{fig:wideprof}. The figure shows average profiles for the major (dotted) and minor (outflow, solid) axis for each IR band, CO, {\sc Hi}, and $K$ band. For comparison, we plot the profiles expected for spherical, optically thin distributions with density profiles $\rho \propto r^{-\alpha}$ and $\alpha = -2$, $-3$, and $-4$. The $\alpha = -2$ profile offers a key point of comparison because it shows the borderline between a diverging and a converging mass profile.

Figures \ref{fig:wideir} and \ref{fig:wideprof} show that the cold material is concentrated around M82. The observed IR profiles decline steeply compared to a $\rho \propto r^{-2}$ profile. Even extrapolating the observed distribution to infinity beyond the sensitivity limits of the data, most of the material would still be concentrated near M82. Only the {\sc Hi} appears to follow a profile that implies equal mass in successive shells ($\rho \propto r^{-2}$). Figure \ref{fig:wideir} shows that this is because progressively larger circles encompass more and  more of the extended tidal material. A secondary effect may be that most of the gas near the galaxy itself is molecular rather than atomic.

Thus, high-stretch infrared images of a wide field around M82 (Figure \ref{fig:wideir}) do not suggest that material flows indefinitely out into the halo. Instead, Figure \ref{fig:wideprof} argues that the deep {\em Herschel} images by \citet{ROUSSEL10} already capture most of the cold material present around M82 and that this material is mainly confined to a spherical region a few kpc in size around the galaxy.

In detail, Figures \ref{fig:wideir} and \ref{fig:wideprof} show a number of other points important to interpret the cold outflow. First, the dust temperature is hotter along the outflow than outside the outflow. It also becomes cooler as one moves further from the galaxy \citep[as shown in][]{ROUSSEL10}, both along the outflow and outside the outflow. This can be seen in the quicker decline of the $70\mu$m and even the $160\mu$m bands relative to the SPIRE bands. However, while the dust becomes cooler, it does not become so cold that the $250\mu$m-to-$350\mu$m color changes in an extreme way (see the very similar decline of the gray and purple lines in Figure \ref{fig:wideprof}). Taking this to mean simply that the peak of the blackbody spectrum (in $F_{\nu}$) drops below $\lambda = 160\mu$m but not $\lambda = 250\mu$m, then the dust cools but not too far below $\sim 20$~K. Dust is not vanishing from the IR images because it becomes too cool to see \citep[see also the comparison to mm observations in][]{ROUSSEL10}.

Second, the dust distribution is round on large scales. Again following \citet{ROUSSEL10}, who noted a large amount of dust not associated with the superwind, note that while the IR  scale lengths do tend to be slightly shorter along the major axis (into the plane of the galaxy), this effect is modest, a difference of usually only $\sim 0.1$~kpc or so. This gives rise to the round appearance of the highly saturated IR images in Figure \ref{fig:wideir}. We experimented with fitting ellipses to the 160$\mu$m and 250$\mu$m isophotes and found axis ratios of $\approx 1.1$ (ellipticity $\approx 0.1$) for the low intensity contours that encompass a large part of the dust distribution.

Third, the differences between the extended {\sc Hi} distribution and the dust emission imply a varying dust-to-gas ratio in the cool gas around M82. Much of the  extended {\sc Hi} emission at the relatively high contour of $N_{\rm HI} = 10^{20}$~cm$^{-2}$ ($\approx 1$~M$_\odot$~pc$^{-2}$) does not have a clear  analog in the high-stretch SPIRE maps. Some spatial filtering may be expected from the data processing \citep{ROUSSEL13}, but a basic comparison shows that we would expect to detect 250$\mu$m emission from much of the extended {\sc Hi}, especially after averaging, if the same dust were mixed with {\sc Hi} throughout. Much of  the {\sc Hi} superstructure may thus represent gas from the extended {\sc Hi} disk of M82 that was originally poor in dust.

Fourth, these figures show the confined nature of the CO very clearly. Molecular gas is, as we have seen, extended compared to the stellar disk along the minor axis. However, it is confined compared to the other tracers of the cool gas. In contrast to the other gas tracers but in agreement with the stars, the molecular gas extends more along the major axis than the minor axis.

Finally, the figures shows a quantitative meaning for the frequently quoted extent of $\sim 6$~kpc for M82's dusty halo \citep{ENGELBRACHT06,ROUSSEL10}:  by this distance the cool dust intensity drops by two orders of magnitude relative to the intensity at $r \approx 1$~kpc. At the same time, this distance represents  about the point at which the intensity of dust emission approaches the sensitivity of {\em Herschel}. If the decline is exponential, then all but a few percent of the mass outside $1$~kpc lies inside $6$~kpc. If decline is closer to $\rho \propto r^{-3}$, then there will be equal mass per logarithmic interval of radius, so that we expect the same amount of material between 10 and 100~kpc as between 1 and 10~kpc. In either case, the present-day flow of material through successive surfaces around the galaxy must diminish with increasing distance.

A major caveat to these mass continuity arguments, particularly at large scale, is the assumption that the starburst has a continuous, fixed mass outflow rate. Mass continuity must apply, but the outflow rate from the starburst itself may vary with the star formation history plus some delay time for the relevant feedback processes to play out. A natural next step to understand the M82 outflow will be an analysis that combines the star formation history of the burst with the radial distribution of outflowing cold material.

\section{Kinematics of the Outflow: Velocity Gradient and Line Splitting}
\label{sec:kin}

\begin{deluxetable}{lc}[h]
\tabletypesize{\scriptsize}
\tablecaption{Minor Axis (Outflow) Velocity Profile \label{tab:vel}}
\tablewidth{0pt}
\tablehead{
\colhead{Quantity} & 
\colhead{Value} 
\\
}
\startdata
Mean velocity & \\
\ldots systemic & 211~km~s$^{-1}$ \\
\ldots $z < -1.5$~kpc & 118~km~s$^{-1}$ \\
\ldots $z > 1.5$~kpc & 269~km~s$^{-1}$ \\
Mean outflow velocity & \\
\ldots $\left<v_{\rm out}\right> \sin i$ & 75~km~s$^{-1}$ \\
\ldots $\left<v_{\rm out}\right>$ for $\sin i = 10\arcdeg$ & 430~km~s$^{-1}$\\
\ldots $\left<v_{\rm out}\right>$ for $\sin i = 5\arcdeg$ & 860~km~s$^{-1}$ \\
\ldots $\left<v_{\rm out}\right>$ for $\sin i = 15\arcdeg$ & 290~km~s$^{-1}$ \\
95\% envelope (relative to $v_{\rm sys}$) & \\
\ldots for $z < -1.5$~kpc & 138~km~s$^{-1}$ \\
\ldots for $z > 1.5$~kpc & 132~km~s$^{-1}$ \\
Minor axis gradient (on sky) & \\
\ldots within $\pm 1.5$~kpc of minor axis & 32~km~s$^{-1}$~kpc$^{-1}$ \\
Typical line width & \\
\ldots at peak for $z < -1.5$~kpc & 110~km~s$^{-1}$ \\
\ldots at peak for $z > 1.5$~kpc & 140~km~s$^{-1}$ \\
\ldots integrated for $z < -1.5$~kpc & 160~km~s$^{-1}$ \\
\ldots integrated for $z > 1.5$~kpc & 230~km~s$^{-1}$ \\
\enddata
\tablecomments{See Figures \ref{fig:cube} and \ref{fig:posvel}}
\end{deluxetable}

\begin{figure*}
\plotone{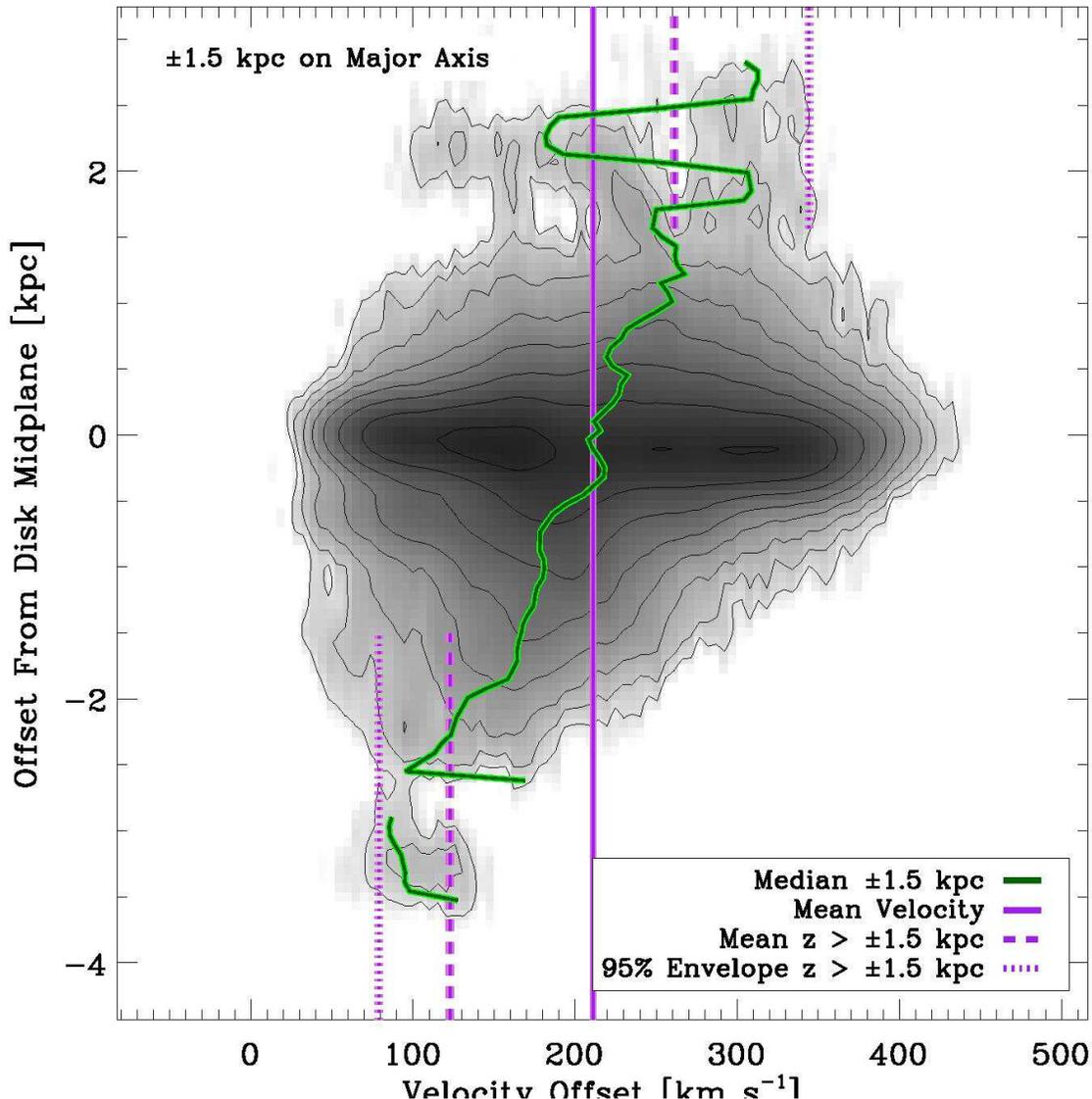}
\caption{
\label{fig:posvel} 
Integrated CO (2--1) position-velocity cut along M82's outflow. We plot intensity integrated over $\pm 1.5$~kpc along the major axis as a function of displacement along the minor axis. Intensity in the image shows intensity of emission with contours stepping by factors of two. Green lines show the median velocity for each vertical displacement, revealing increasing velocity with displacement from the plane. This signifies either acceleration or a wide range of ejecta velocities. The solid purple line shows the mean velocity while the dashed lines show the mean velocity above 1~kpc vertical displacement. The dashed purple lines show the 95\% envelope of emission over the same range, an indicator of the maximum ejecta velocity. Note the very clear line splitting to the north of the galaxy (discussed in terms of a conical geometry, below), but this is less clear to the south, where the lines are simply very broad.}
\end{figure*}

So far we have examined the distribution of different mass components around M82. Our observations also yield high spectral resolution measurements of the kinematics of the gas. These offer a powerful tool to investigate the mechanics of the outflow. Most obviously, the minor axis of M82 shows a  velocity gradient \citep[noticed in the molecular gas as early as][]{NAKAI87}, which is strikingly evident in our velocity field (Figure \ref{fig:cube}) and data cube. This gradient, which is also present in the {\sc Hi} \citep{COTTRELL77}, lets us estimate the magnitude of the outflow velocity and explore the possibility of gas accelerating away from the plane. The line profile along individual lines of sight also offers key information. Its shape has often been invoked as a way to explore the three dimensional  geometry of an outflow. Specifically, the appearance of two components along an indivdiual line of sight (``line splitting'') has been taken as  evidence for a tilted conical geometry \citep[e.g.,][]{MCKEITH95,SHOPBELL98}. 

\subsection{Position-Velocity Diagram, Minor Axis Gradient, and CO Outflow Velocity}
\label{sec:pv}

Figure \ref{fig:posvel} shows the minor axis position-velocity diagram, with north corresponding to positive offsets. We use this to study the minor axis gradient, the outflow velocity, and line splitting. Particularly at large vertical displacements, the CO is patchy with incomplete coverage, so to build Figure \ref{fig:posvel} we integrate over a width $\pm 1.5$~kpc  around the minor axis of the galaxy. That is, this is the integrated position velocity diagram for the outflow region in Figure \ref{fig:wind}.

Purple lines in Figure \ref{fig:posvel} show the systemic velocity of the galaxy (solid), the mean velocity for all emission with $|z| > 1$~kpc above or below the disk (dashed), and the 95\% envelope of emission $> 1$~kpc above the disk (dotted lines). The outflow is expected to be launched from a small central region \citep{SHOPBELL98}, so that, naively, the difference between the average velocity above and below the disk and the systemic  velocity of the galaxy should yield the projected speed of the  outflow, $\left<v_{\rm outflow}\right>~\sin i = (\left<v_{|z|>1~{\rm kpc}}\right> - v_{\rm sys})$.  Table \ref{tab:vel} summarizes these velocities. The projected mean velocity offset between the high latitude emission and the systemic velocity is $\approx 78$~km~s$^{-1}$. The inclination, $i$, of the outflow is uncertain  and because $i$ is small, this value exerts a large influence on the derived velocity. For $i = 10\arcdeg$, $v_{\rm outflow} \approx 450$~km~s$^{-1}$, while for a moderately higher $i = 15\arcdeg$, this drops to $v_{\rm outflow} \approx 300$~km~s$^{-1}$. Thus for our working inclination of $i=10\arcdeg$, the CO outflow speed is $450$~km~s$^{-1}$. For comparison, M82's maximum rotation velocity  is just under 200~km~s$^{-1}$ \citep[see][]{SOFUE98,GRECO12} and the H$\alpha$ outflow velocity is 525 to 625 km~s$^{-1}$ \citep{SHOPBELL98}.

The velocity gradient along the minor axis appears as a green line in Figure \ref{fig:posvel}. Southern positions show more blueshifted emission while northern positions show more redshifted emission, on average. This has the same sense as the H$\alpha$ velocity gradient \citep{MCKEITH95} and the {\sc Hi} gradient \citep{COTTRELL77}, which is increasingly blueshifted emission below the disk and increasingly redshifted emission above the disk. The magnitude of the gradient is $\approx 32$~km~s$^{-1}$~kpc$^{-1}$ in the plane of the sky.

We interpret Figure \ref{fig:posvel} as a blend of two phenomena, rotation along the major axis and outflow along the minor axis. In a moment, we will see that rotation slows with increasing height. Meanwhile, a conical outflow morphology produces a double peaked line centered on the mean velocity of the outflow,  with the two components reflecting the different tilts of the edge \citep[for more, see][]{SHOPBELL98} . A blend of these two components will produce Figure \ref{fig:posvel} and will naturally create the gradient. Rotation  dominates within $\pm 0.5$~kpc of the disk, producing the broad, almost contribute most emission. This is also the regime where Figure \ref{fig:hastarprof} show a clear excess of emission in the CO and H$\alpha$ relative to starlight. In the north the line splitting is stunningly evident in this range, while in the south the dual-peak morphology is present from about 2 to 2.5~kpc  but then only one component persists beyond 2.5~kpc. To the south, the cold material traced by dust and gas is mostly to the east of the region of hot dust temperatures  (see Figure \ref{fig:dustmap}), so this may indicate only partial covering of the outflow cone by high column cold material.

Figure \ref{fig:posvel} also plots the envelope of emission, calculated as the 95$^{th}$ percentile of emission sorted by velocity. This envelope describes emission at a wide range of vertical offsets well. The fact that the envelope does not curve substantially in or out argues that acceleration does not dominate the CO kinematics. That is, even though we observe a velocity gradient, we interpret this to reflect a changing blend of components rather than an acceleration of material as it moves away from the disk.

\subsection{Line Splitting and Conical Geometry}

\begin{figure*}
\plottwo{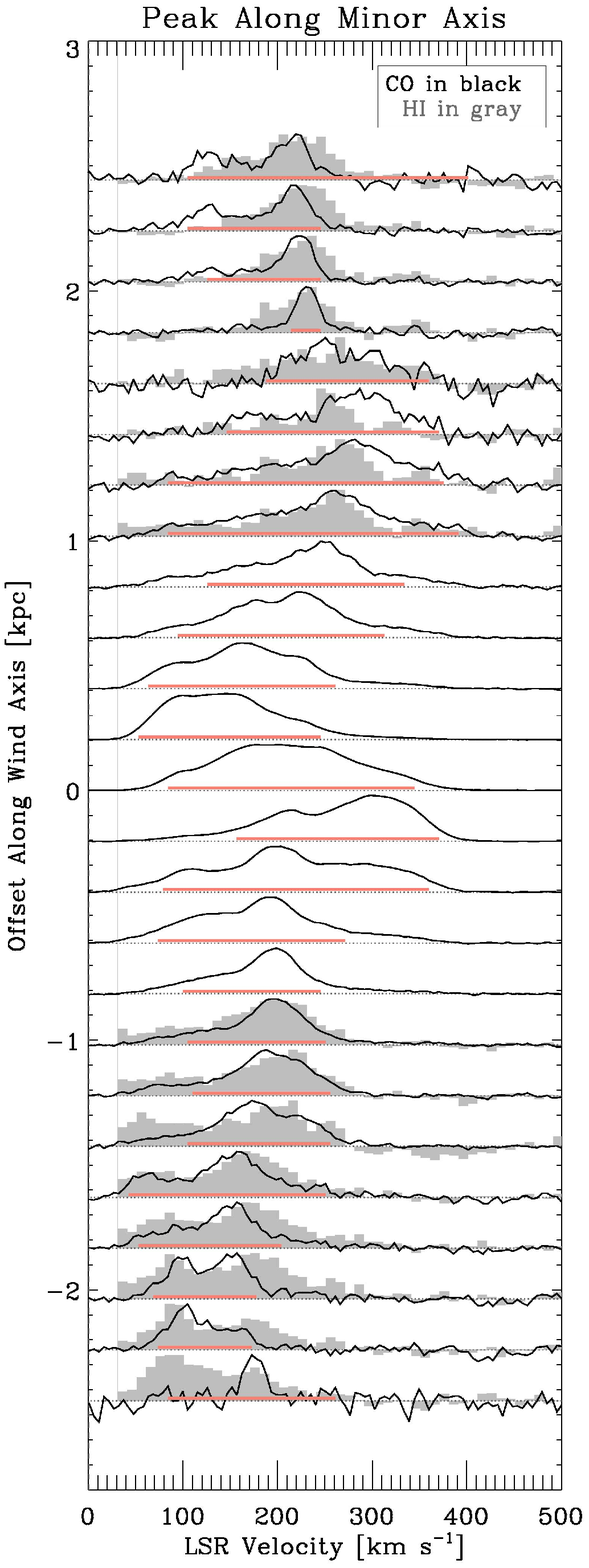}{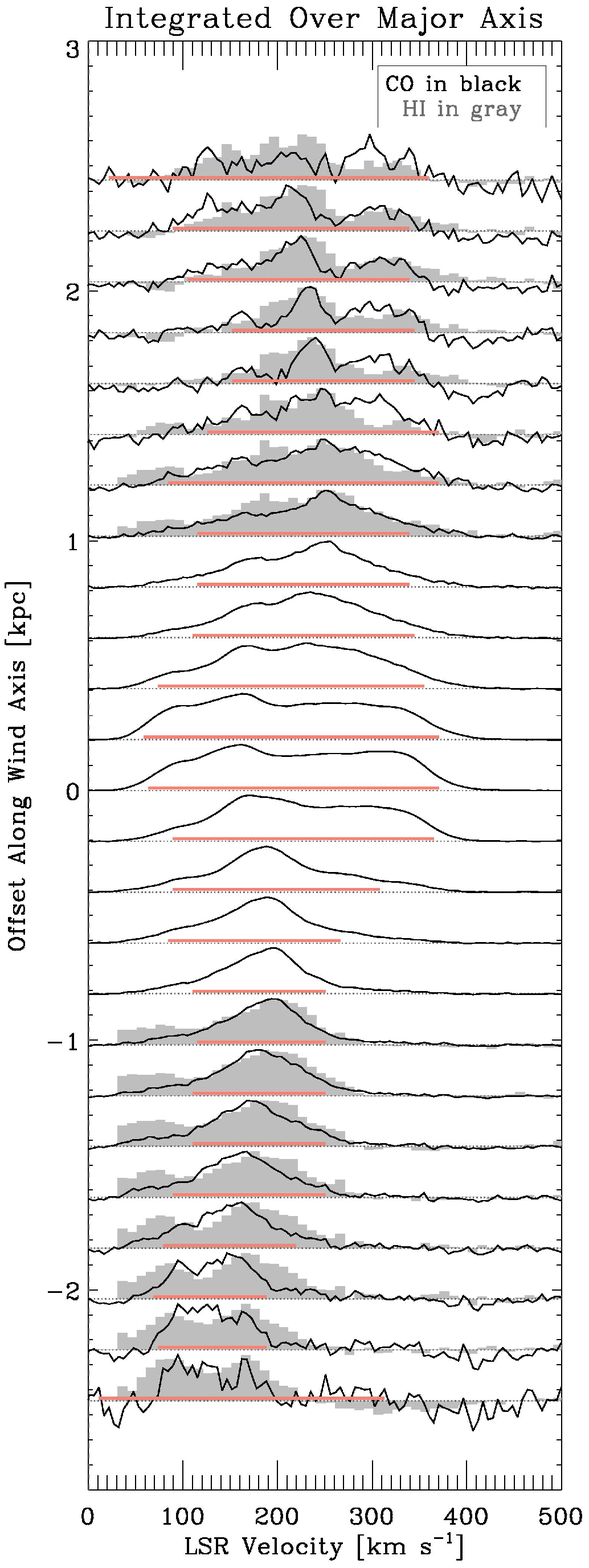}
\caption{
\label{fig:spec} CO (black) spectra as a function of vertical displacement from the major axis. ({\em left}) The spectrum at the location of peak brightness (i.e., an individual line of sight) at each vertical displacement and {\em right} integrated over $\pm 1.5$~kpc along the major axis. Note the shift of the peak, the line splitting at high vertical displacements in the integrated spectrum (mostly to the north), and the significant detections of  very broad lines even to very large vertical displacements. Light red horizontal lines show quarter-maximum line widths, with the average high latitude velocity widths reported in Table \ref{tab:vel}. Under simple assumptions, the width of the lines combined with an estimate of the outflow speed suggests an opening angle of $\approx 20\arcdeg$. In gray, we plot {\sc Hi} spectra, which have lower quality, for displacements that are not contaminated by absorption.
}
\end{figure*}

\begin{figure*}
\plotone{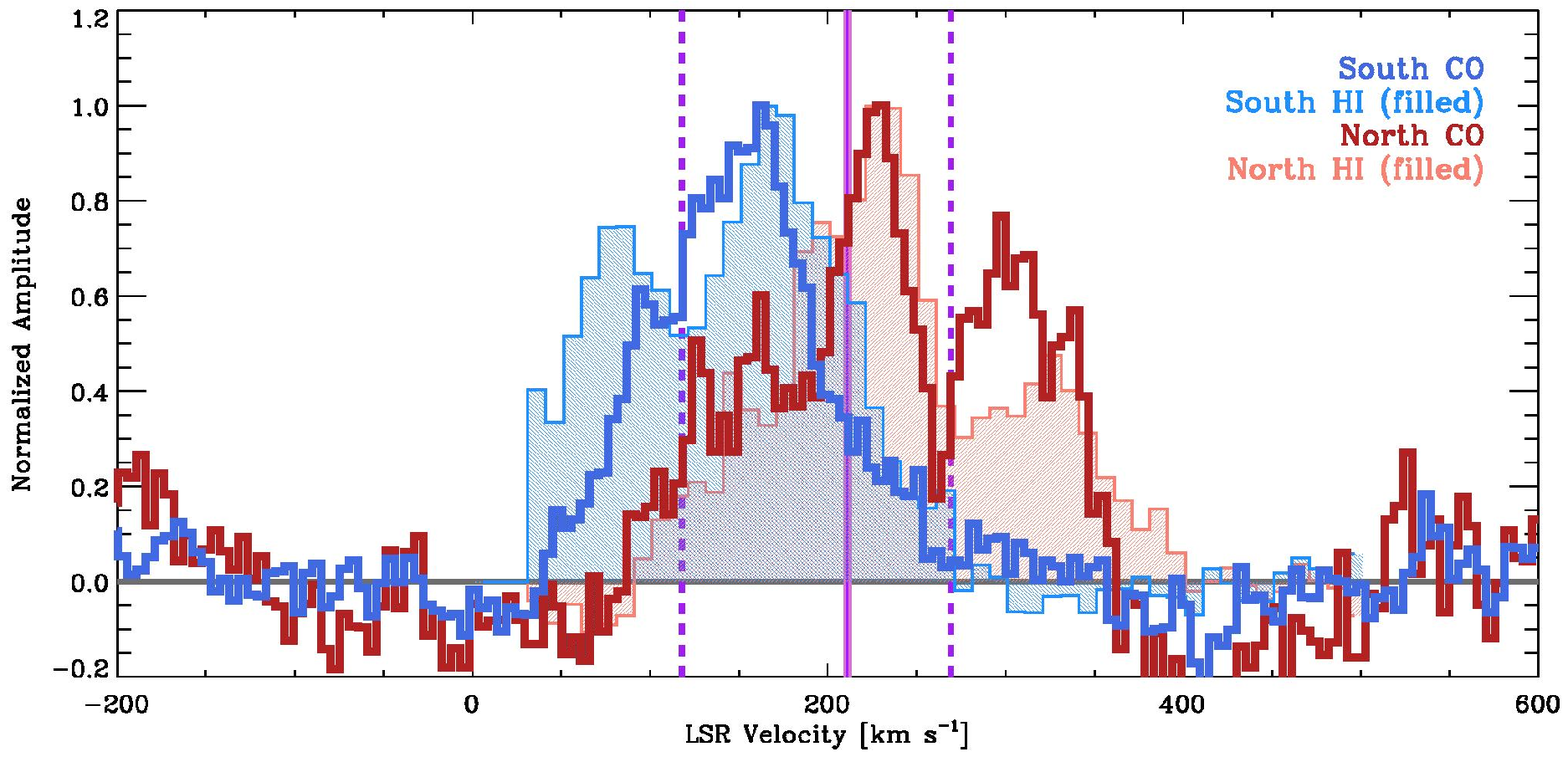}
\caption{Integrated and normalized spectra from $1.5$--$2.5$~kpc above and below the disk in CO (thick line) and {\sc Hi} (filled histogram). Both the  northern and southern outflow region show double peaked line profiles, indicating a conical geometry with cold gas bounding the hot outflow. The mixture of molecular and atomic gas in the profile varies with region. In the south the two-peaked shape is more clearly visible in {\sc Hi} but is not clearly visible in the CO. In the north the double peak appears clearly in CO and {\sc Hi}. Purple lines indicate the systemic velocity (solid)  and the mean velocity of molecular gas more than 1.5~kpc above and below the disk (dashed).
\label{fig:outspec} }
\end{figure*}

The lines in Figure \ref{fig:posvel} are very broad and especially in the northern part of the galaxy the position-velocity diagram shows multiple components at a given vertical displacement. This can be seen more clearly from direct plots of the peak and integrated spectra along the minor axis in Figure \ref{fig:spec}. ``Line splitting,'' that is observing two components separated in velocity along each line of sight, has often been sought and invoked as a signature of a emission coming from the edges of a cone or cyclinder, particularly considering H$\alpha$ \citep[e.g., see][]{MCKEITH95,SHOPBELL98}. In this sketch,  emission arises from two sides of a tilted cone of emission being expelled from the galaxy. The different inclinations of the two sides of the cone yield distinct velocities,  one at $v = v_{\rm out}~\sin \left( i + 0.5~\theta \right)$ and one at $v = v_{\rm out}~\sin \left( i - 0.5~\theta \right)$, where $\theta$ is the cone opening angle. The difference, $\Delta v$, depends on the inclination, the opening angle of the cone and the outflow velocity via:

\begin{equation}
\label{eq:cone}
\Delta v = 2~v_{\rm out}~\cos i~\sin \frac{\theta}{2}~.
\end{equation}

\noindent For the case of M82, $\cos i \approx 1$ and $\Delta v \approx 2~v_{\rm out}~\sin \left( 0.5~\theta \right)$.  Even for an outflow that fills a cone or covers the edges patchily, an approximate form of Equation \ref{eq:cone} should still hold. That is, the line width should be is proportional to the opening angle and velocity of the outflow for a wide range of geometries.

Figure \ref{fig:outspec} accumulates all of the CO and {\sc Hi} spectra between $1.5$~kpc and $2.5$~kpc from the plane in the outflow region. That is, it shows the integrated spectra of the area that we consider most likely to reflect only cold material associated with the outflow. The two-component profiles are clearly visible to both the north and the south, though the details differ. In the south the double-peaked profile is most clearly visible in {\sc Hi}, while to the north the CO shows clearer line splitting. We have already seen a changing phase as a function of height, here we see evidence from the spectral profile that the material around the outflow is a mixture of H$_2$ and {\sc Hi} with the composition varying around the galaxy.

Figure \ref{fig:spec} shows line widths measured at quarter of maximum for peak and integrated spectra along the outflow. We report typical high latitude ($|z| > 1.5$~kpc) line widths for the CO in Table \ref{tab:vel}, which are $\sim 150$~km~s$^{-1}$. For an outflow velocity,  of $\sim 400$~km~s$^{-1}$ and this $\Delta v$, Equation \ref{eq:cone} suggests an opening angle $\sim 20\arcdeg$. The peaks of the two components in the integrated profiles are somewhat closer together, $\sim 100$~km~s$^{-1}$, which would argue for $\theta \sim 13\arcdeg$. These are somewhat smaller opening angles than those suggested by H$\alpha$ emission \citep[e.g.,][]{MCKEITH95,SHOPBELL98}, which range from $\approx 30$--$60\arcdeg$. A lower outflow velocity (e.g., due to a higher correct inclination) could remedy the discrepancy or our focus on high latitude material may find a more columnated outflow. In either case, the double peak profiles reinforce that the cold gas at high vertical displacement is associated with the wind seen at other wavelengths, likely confining the hot wind.

\section{Slowing Rotation with Height}

\begin{figure*}
\epsscale{1.1}
\plottwo{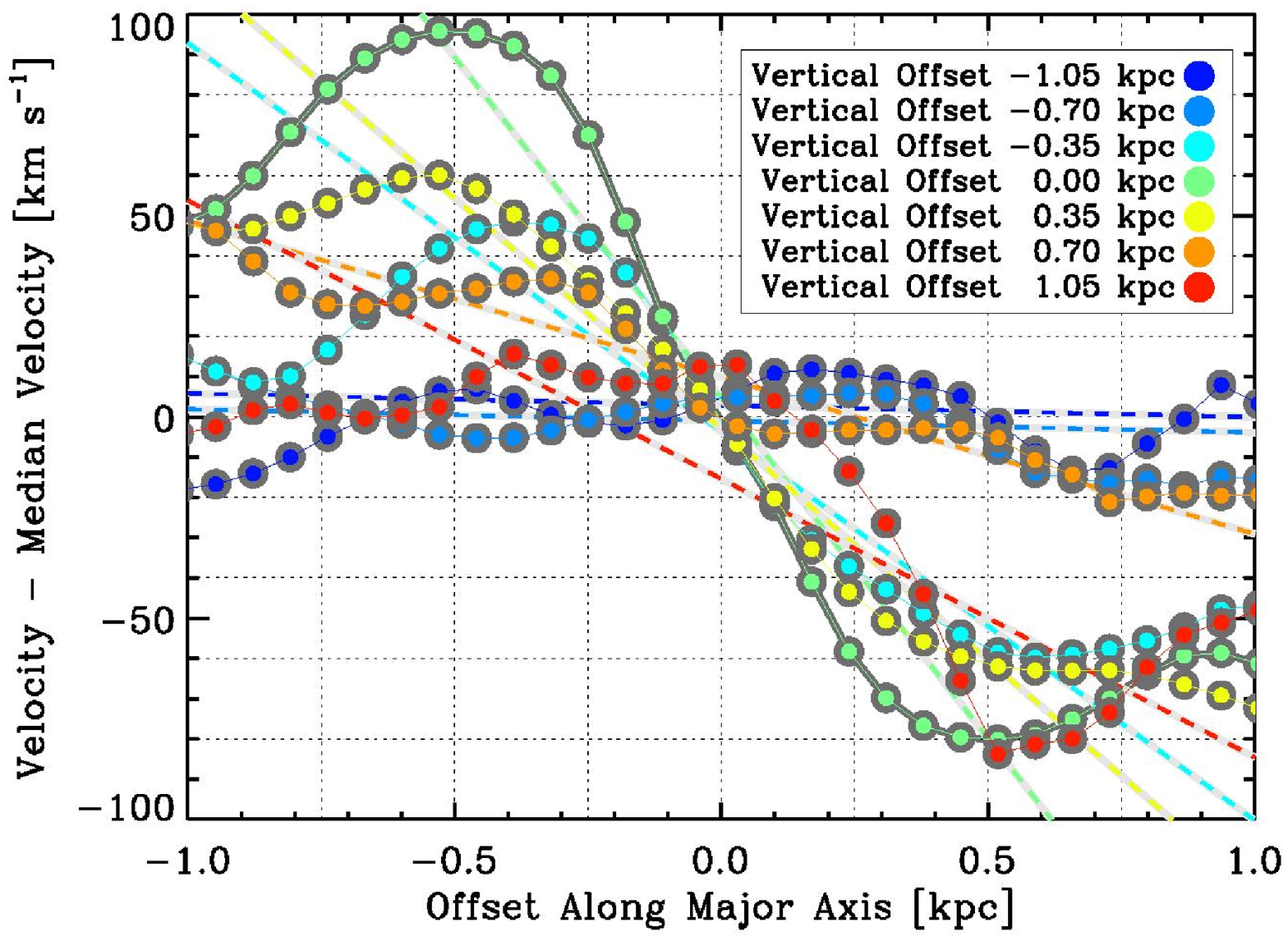}{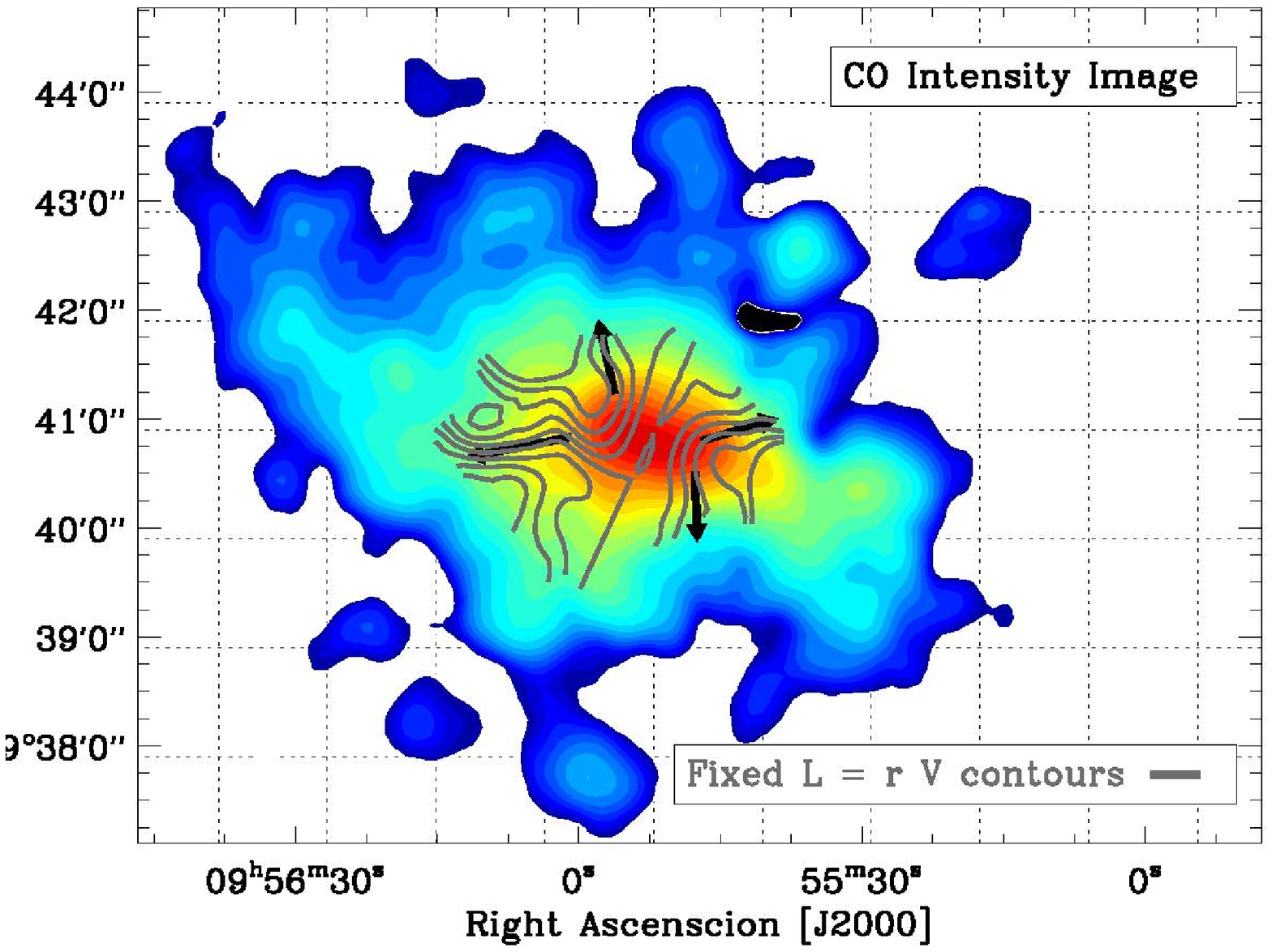}
\caption{
\label{fig:angmom}
({\em left}) Rotation of the disk, traced by mean velocity along the line of sight as a function of offset along the major axis, for emission in the midplane and in cuts parallel to the major axis but above and below the midplane. Rotation remains visible but decreases in amplitude with increasing height  until it is largely imperceptible by $z \sim 1.3$~kpc ($80\arcsec$). Note that for edge-on systems like M82, the true rotation curve requires careful modeling of the kinematics. Our purpose with this figure is to show that rotation diminishes with vertical displacement. ({\em right}) Contours of equal angular momentum, $L = r_\perp v_\parallel$, for regions of bright CO ($I > 10$~K~km~s$^{-1}$) plotted over the CO intensity image. The gray contours show lines along which angular momentum should be approximately conserved and so plausible lines along which material kicked out of the starburst might flow \citep{SEAQUIST01,WALTER02}. Black arrows illustrate the implied wide angle flow out around the starburst. Expulsion of cold material in many directions helps explain the amorphous appearance of the CO and dust maps.}
\end{figure*}

The CO (2-1) line is wider and smoother closer to the disk and the velocity field near the disk shows rotation (Figure \ref{fig:cube}). This is the other component visible in Figure \ref{fig:posvel}, a ``puffed up'' molecular disk around the starburst. If material is ejected from the central, rapidly rotating  starburst to larger radii, then conservation of angular momentum will cause it to slow down. This slowing rotation with vertical displacement has been noted by \citet[][]{WALTER02} and \citet{SEAQUIST01} \citep[see also][]{SOFUE92}.  The left panel in Figure \ref{fig:angmom} shows that this effect is  present in our data along the minor axis out to $\approx 1$~kpc, beyond which rotation is no longer visible. 

Our data have coarse resolution compared to changes in the rotation curve of M82 \citep{SOFUE98} and so are not ideal to measure structure near the disk. Still, we can quantify the decline in rotation to some degree. In our data,  the slope of the best-fit velocity gradient along the inner $2$~kpc parallel to the major axis drops by as much as $100$~km~s$^{-1}$~kpc$^{-1}$ from the midplane to $\pm 1$~kpc. This is larger than gradients seen in edge-on galaxies with extended {\sc Hi} or ionized gas distributions \citep[such as NGC 891][]{OOSTERLOO07}, where the magnitude of slowing rotation is typically $\approx 15$ to $25$~km~s$^{-1}$ per  scale height \citep{RAND08}. Taking the scale height as $\sim 0.5$~kpc, we find a drop of $\sim 30$ to 50~km~s$^{-1}$ per scale height in M82. 

For rotation to slow so quickly with increasing height, material must be moving to larger radius very quickly. This measurement argues that  material is being violently flung out of M82 at a wide opening angle, not only along the visible hot outflow. Following \citet{SEAQUIST01} and \citet{WALTER02}, we  also consider a more direct version of angular momentum conservation to gauge the flow of material in the immediate vicinity of the starburst. They suggest that the quantity $L = r_\perp v_\parallel$ offers a reasonable observational analog of the angular momentum. Here  $r_\perp$ the projected radius along the major axis (the axis of rotation) and $v_\parallel$ the intensity-weighted mean velocity along the line of sight. As discussed by \citet{SEAQUIST01} this assumes the observed $v$ to map to the tangential velocity and so is is an approximation. Still the calculation provides useful insight and we plot the resulting angular momentum isocontours in the right panel of Figure \ref{fig:angmom}.

The shape of the iso-$L$ curves in Figure \ref{fig:angmom} argues that the starburst expels cold material over a wide range of angles, not only  coincident with the hot wind along the minor axis. This provides a natural explanation for the round, extended appearance of the CO and dust maps  compared to the columnated hot outflow. The hot outflow cavity is very good at ``lighting up'' associated material but material is expelled over a broad range of angles.

\section{Flow of Material, Momentum, and Energy Along the Minor Axis}
\label{sec:outflow}

\begin{figure}
\epsscale{1.2}
\plotone{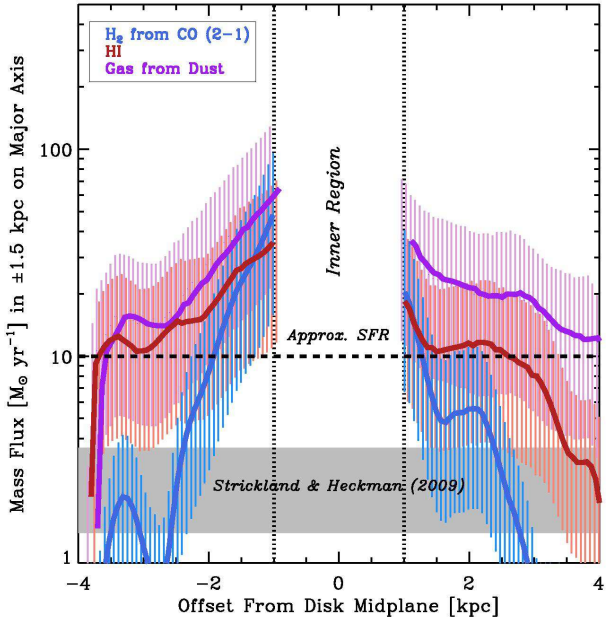}
\caption{\label{fig:flow}
Mass flow through a plane parallel to the major axis, in M$_\odot$~yr$^{-1}$, calculated from multiplying the local mass surface density of H$_2$ (from CO), {\sc Hi}, or $\Sigma_{\rm gas} = \Sigma_{\rm dust} / \delta_{\rm DGR}$ by the mean outflow velocity. The error bars indicate the effect of changing the inclination by $\pm 5\arcdeg$. For comparison, we plot the outflow rate calculated based on observations of hot gas by \citet{STRICKLAND09} and a rough estimate of the star formation rate of the system. The mass flow rate corresponds to an energy or momentum outflow rate via Equations \ref{eq:energy} and \ref{eq:momentum} and the outflow may account for $\sim 20\%$ of the energy and momentum produced in stars. Taking dust as a tracer of all cool gas, the mass outflow rate declines with an $e$-folding length of $\approx 1.4$~kpc; for an outflow velocity of $450$~km~s$^{-1}$ this corresponds to a timescale of  $\approx 3$~Myr.}
\end{figure}

To understand the amount and fate of the molecular gas associated with the M82 outflow, we combine what we learn from the kinematics  and the mass profiles. If we consider an outflow moving straight up along the minor axis and adopt an outflow velocity, then the mass profiles in Figure \ref{fig:vertprof} imply a mass flux through each successive surface. Figure \ref{fig:flow}, shows the mass flux through progressive surfaces along the minor axis for a fixed outflow speed of $450$~km~s$^{-1}$. We also experimented with using the local  velocity field to calculate a mass flux. This yields similar results; the approach to calculate the velocity appears to  matter less than the large systematic uncertainty introduced by our imperfect knowledge of the inclination. At our adopted speed, gas will travel 1 kpc in $2.3$~Myr. Error bars indicate the effect of changing the inclination of the outflow by $\pm 5\arcdeg$. Along with the uncertainty in the mass estimates already discussed, this renders the outflow rate uncertain by a factor of $\approx 2$--$3$. However, the internal consistency is likely substantially better than this. For comparison, the Figure also shows the approximate star formation rate of the galaxy \citep{WESTMOQUETTE07} and the mass outflow rate inferred for the hot wind by \citet{STRICKLAND09}.

Figure \ref{fig:flow} can be recast in terms as a rate of momentum or energy flow away from the disk. For a fixed outflow velocity, the corresponding energy rate is

\begin{equation}
\label{eq:energy}
\frac{dE}{dt} = 2 \times 10^{49}~{\rm erg~yr}^{-1}~\left( \frac{\dot{M}}{10~M_\odot~{\rm yr}^{-1}} \right)~\left( \frac{v}{450~{\rm km~s}^{-1}} \right)^2
\end{equation}

\noindent and the momentum flux is

\begin{equation}
\label{eq:momentum}
\frac{dp}{dt} = 4500~M_\odot~{\rm km~s}^{-1}~{\rm yr}^{-1}~\left( \frac{\dot{M}}{10~M_\odot~{\rm yr}^{-1}} \right)~\left( \frac{v}{450~{\rm km~s}^{-1}} \right)~.
\end{equation}

Figure \ref{fig:flow} and its momentum and energy analogs show first that cold material seems to be leaving the outflow as it moves away from the disk. That is, the profiles decline. Second, the outflowing cold gas is changing phase to become more atomic and less molecular; the CO (blue) profile declines fastest. Third, the amount of cold gas near the disk is so large that the inferred rates dwarf the hot gas outflow rates near the disk.  However, the outflow rate of molecular gas, at least, is less than or comparable to the hot gas outflow rate by $\approx 4$~kpc. Thus despite the visible molecular outflow and large amount of CO emission, the molecular wind is not a major factor on large scales. Fourth, mass flux rates around the disk are of the same order or even larger than the star formation rate in the galaxy. Thus the redistribution of material due to stellar feedback will be just as important to the evolution of the starburst as the depletion of gas into stars. Finally, the figure shows that distinguishing outflowing from  tidal {\sc Hi} and dust at large distances is critical to understanding the mass budget in the outflow.

Following \citet{OSTRIKER11}, a reasonable momentum output due to supernova is $\approx 3000$~M$_\odot$~km~s$^{-1}$ per $1$~M$_\odot$ formed and a reasonable kinetic energy injection is $\approx 10^{49}$~erg per 1~M$_\odot$ formed. For a star formation rate of $10$~M$_\odot$~yr$^{-1}$ there should then  be $\approx 30,000$~M$_\odot$~km~s$^{-1}$~yr$^{-1}$ and $10^{50}$~erg~yr$^{-1}$ released into the ISM. Above about $1.5$~kpc (where we are fairly confident that the outflow dominates), the  momentum associated with the mass outflow rate of 10--20~M$_\odot$~yr$^{-1}$ represents $\approx 10$--$30$~\% of the momentum available from star formation each year  and the energy available represents a similar fraction of the available energy. Both the energy and momentum injection and the outflow numbers are uncertain, but  this first order comparison shows plausible agreement.

Again, we note that this translation of distance from the disk into evolution of the outflow with time assumes a roughly constant mass outflow rate over the last $\sim 10$~Myr. A useful future avenue of investigation, but one beyond the scope of this paper, would be to attempt to combine these measurements with information about the recent  star formation history in M82's burst \citep[e.g., see][]{DEGRIJS01,FORSTERSCHREIBER03} along with plausible lag times for launching of a wind.

\section{Synthesis}

\begin{figure*}
\plotone{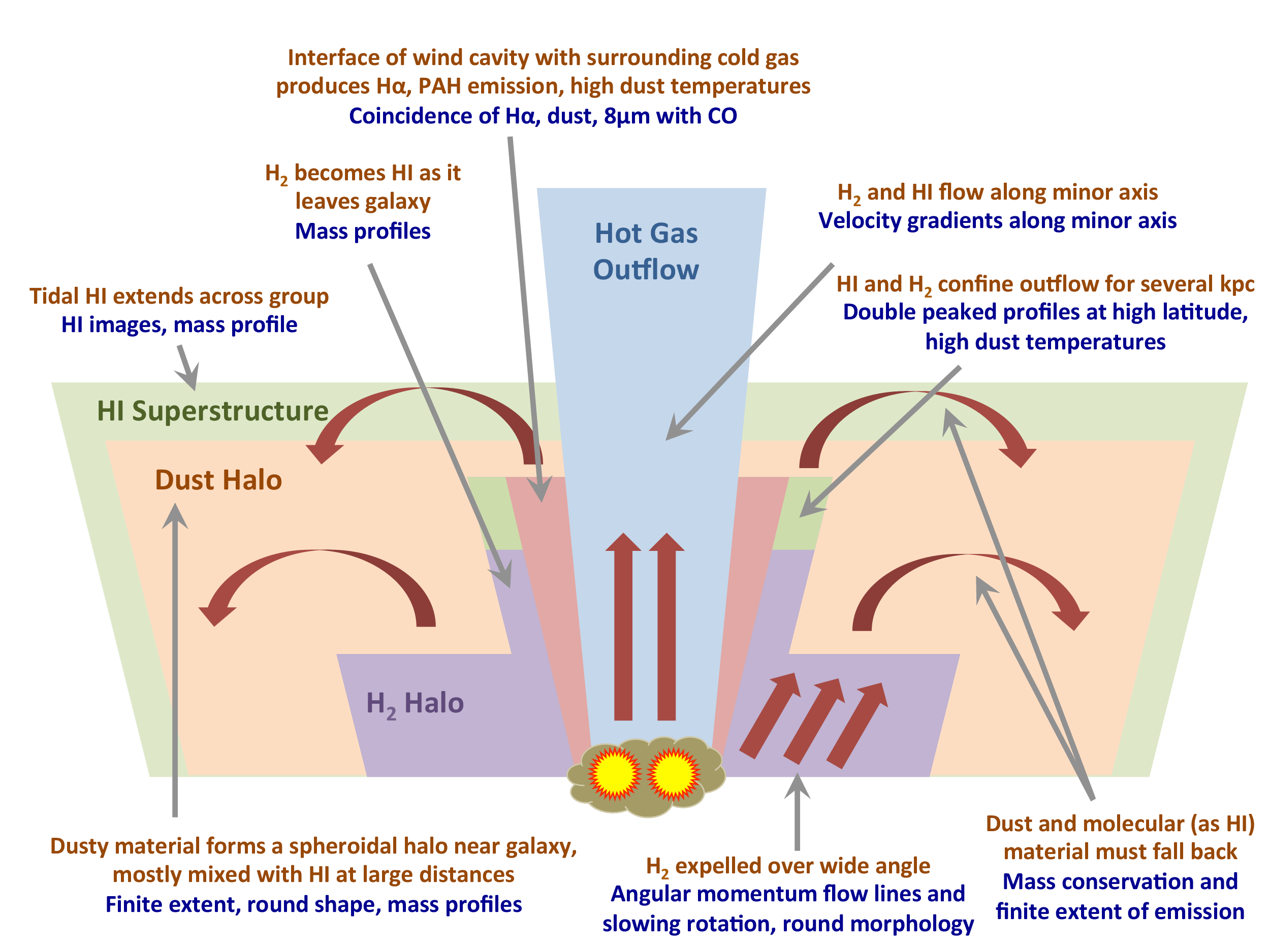}
\caption{\label{fig:sketch} A schematic of the view of cool gas and dust around M82, not to scale. Blue notes cite evidence for the red statements.
}
\end{figure*}

We present the most sensitive wide-field map of the CO emission around M82 to date, which traces the distribution and kinematics of molecular gas in the sprawling gaseous halo that surrounds the famous starburst. This includes extended CO emission coincident with other tracers of M82's well-known galactic wind, with clear detections at distances $> 3$~kpc above the starburst disk and well outside the stellar disk. By combining these new CO data with maps of  dust emission \citep{ROUSSEL10,ENGELBRACHT06,KANEDA10}, {\sc Hi} \citep{YUN93,YUN94}, tracers of the hot outflow \citep{LEE09,STRICKLAND09,HOOPES05}, starlight \citep{JARRETT03}, and other CO transitions \citep[CO 1--0 and CO 3--2][]{WALTER02,WILSON12} we attempt to build a complete  picture of the cool gas and dust halo around M82. 

Figure \ref{fig:sketch} illustrates our conclusions schematically. We find CO associated with the hot outflow. Both CO and {\sc Hi} play an important role confining the outflow.  The interface between the hot wind cavity, with its associated intense radiation field, and the surrounding cold material gives rise to many of the most striking aspects of the outflow: the H$\alpha$ and PAH emission, the hot dust, and scattered FUV light. The association of the cool gas with the outflow is clearest outside $\approx \pm 1.5$~kpc from the disk. In these extended regions double peaked line profiles indicative of conical geometry and a minor axis velocity gradient both link the gas to the outflow kinematically. The coincidence of CO, hot dust, and H$\alpha$ along the minor axis also support this picture.

However, although cold material is associated with the hot outflow, it does not appear to travel far into the halo. Mass profiles of dust, CO, and {\sc Hi} all decline along the minor axis without any corresponding evidence for acceleration. Thus basic mass continuity arguments suggest that cold material is leaving the outflow, stalling out or falling back to the disk. Dust should remain mixed with any low temperature ionized gas, so material vanishing into an ionized phase does not appear to offer a plausible alternative. There is, however, a clear phase change associated with the molecular gas, which tends to become more atomic as it flows out of the galaxy.

A wide field view of dust, gas, and stars around M82 supports the idea of a cold ``fountain'' rather than a cold outflow. Profiles of dust emission  over a wide field do not diverge. That is, their profiles appear significantly steeper than a projected $\rho \propto r^{-2}$ profile, which would be required for mass continuity and a fixed outflow velocity. IR colors suggest that a changing dust temperature does not offer a plausible alternative scenario, the dust seems to stay warm enough for {\em Herschel}'s SPIRE instrument to detect it to large distances. Most dust, and thus presumably most of the cold material ejected from the galaxy, is  concentrated within a few kpc of the galaxy. Its morphology at low intensity is approximately circular in projection, suggesting material is not just ejected along the hot outflow cavity.

Expulsion of material in all directions is supported by the CO kinematics closer to the disk. Slowing rotation with height and a direct examination of lines of fixed angular  momentum show ``flow lines'' with much wider opening angles than the hot outflow. This is consistent with cold material being ejected in many directions but preferentially lit up along the hot wind cavity.

Thus we argue that the data support a highly dynamic cold gas fountain around M82. Material is hurled out as molecular gas but becomes atomic over a timescale of a few Myr. At the same time, material stalls out or falls back on a slightly longer timescale. Doing so it must change phase and position around the galaxy, because fallback is not readily evident from the CO minor axis position-velocity diagram.

The mass flux along the minor axis changes accordingly. It is much higher than the outflow rate associated with the hot wind or the star formation rate near the galaxy, but it drops with increasing distance. By $\approx 4$~kpc from the disk, the mass flux of molecular material is not an important contribution compared to the hot wind. The situation with the total or atomic gas is less clear. The mass flux also drops, but in future work it will be important to distinguish tidal features from outflow features in the atomic gas to cleanly measure the outflow component. Still, the implied trend, especially considering the dust to be ejected material, is that the cold gas  outflow does not reach far out into the halo, unlike the hot material. The motion of this cold material may account for $\approx 10$--$30\%$ of the energy and momentum produced by  star formation each year.

These results sound something of a cautionary note for outflows detected in unresolved or marginally resolved spectra. We show in the appendix that if viewed at a different inclination, M82 could produce a dramatic example of the wide CO line wings commonly used to identify outflows. However the evidence we present argues that the molecular material, in particular, is not a true outflow. Large scale feedback effects on the molecular gas are clearly critical to the evolution of the system. They will remove gas from the starburst region and affect the structure of the galaxy but we do not see good evidence that molecular gas makes it far out into the halo of M82.

Finally, we note several areas for future investigation that will help shed further light on this key example of a molecular wind. We have assumed a constant recent mass outflow rate, but in fact it will be productive to jointly model the flow of material out of the galaxy and the recent star formation history of the galaxy. Also, although we present good arguments that radiative and not collisional heating dominates the dust energy budget, the role of shocks and hot gas heating dust at the interface of the hot superwind and cold material has yet to be explored in detail. Last, the edge-on orientation that makes M82 so tractable to this sort of study also leaves the inclination of the galaxy uncertain; any more stringent constraints on this quantity will have important implications for how we think about the energetics, momentum, and possible escape of the material leaving the burst.

\acknowledgments We thank the anonymous referee for a careful and constructive report that improved this paper. This work benefited from helpful discussions and feedback from many colleagues including  Todd Thompson, Eve Ostriker, Jeffrey Mangum, John Hibbard, and the University of Virginia/NRAO star formation group (Kelsey Johnson, Amanda Kepley, and Scott Schnee).  This paper is based on observations carried out with the IRAM 30-m telescope. IRAM is supported by INSU/CNRS (France), MPG (Germany) and IGN (Spain). The {\sc Hi} data comes from the Very Large Array, operated by the National Radio Astronomy Observatory. The National Radio Astronomy Observatory is a facility of the National Science Foundation operated under cooperative agreement by Associated Universities, Inc. We made use of the NASA/IPAC Extragalactic Database (NED), the NASA's
Astrophysics Data System, the IDL Astronomy User's Library, and the Coyote IDL graphics library.

\begin{appendix}

\section{Prospects for Detecting Outflows Similar to M82 in Unresolved Spectra}

\begin{figure*}
\plottwo{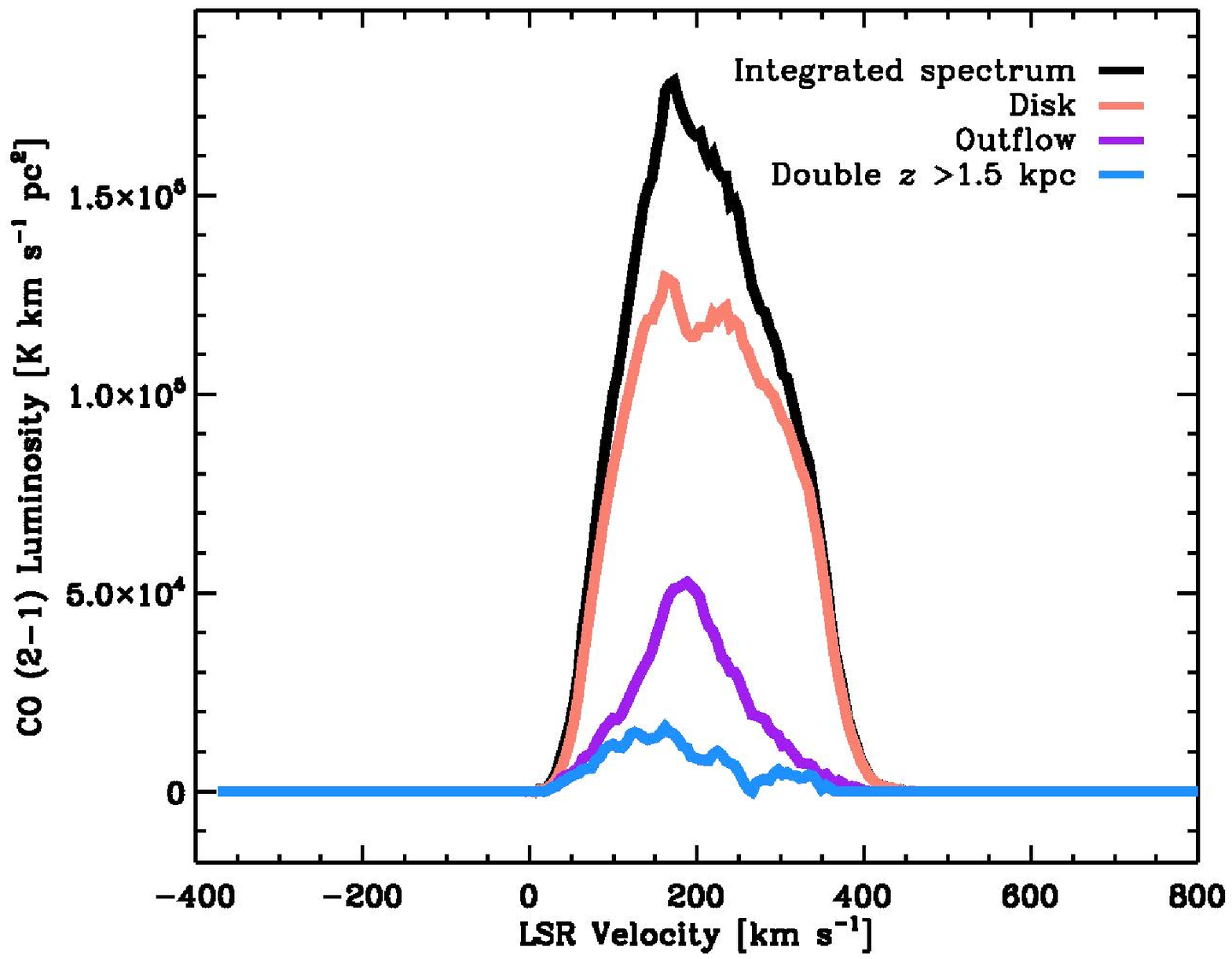}{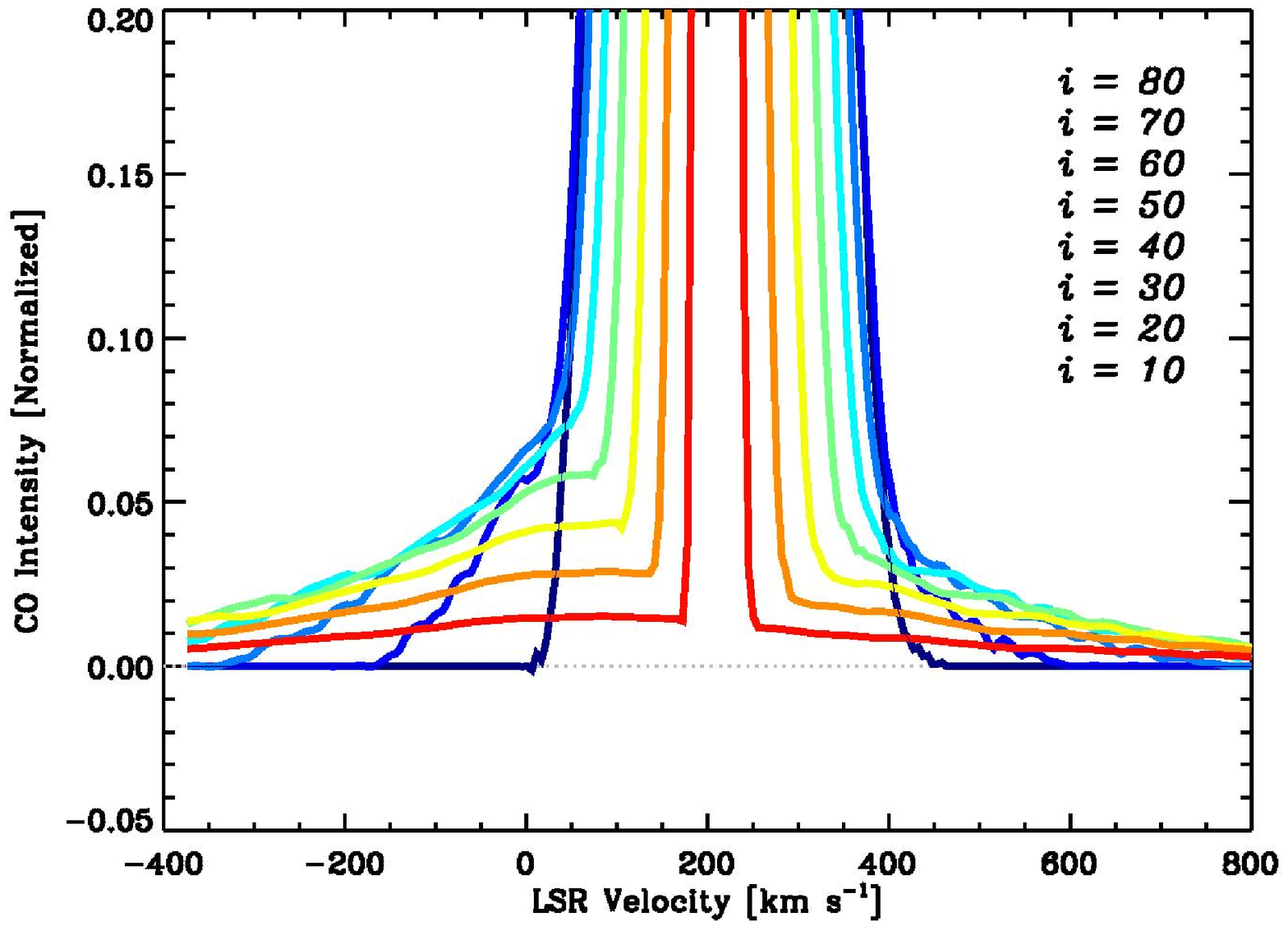}
\plottwo{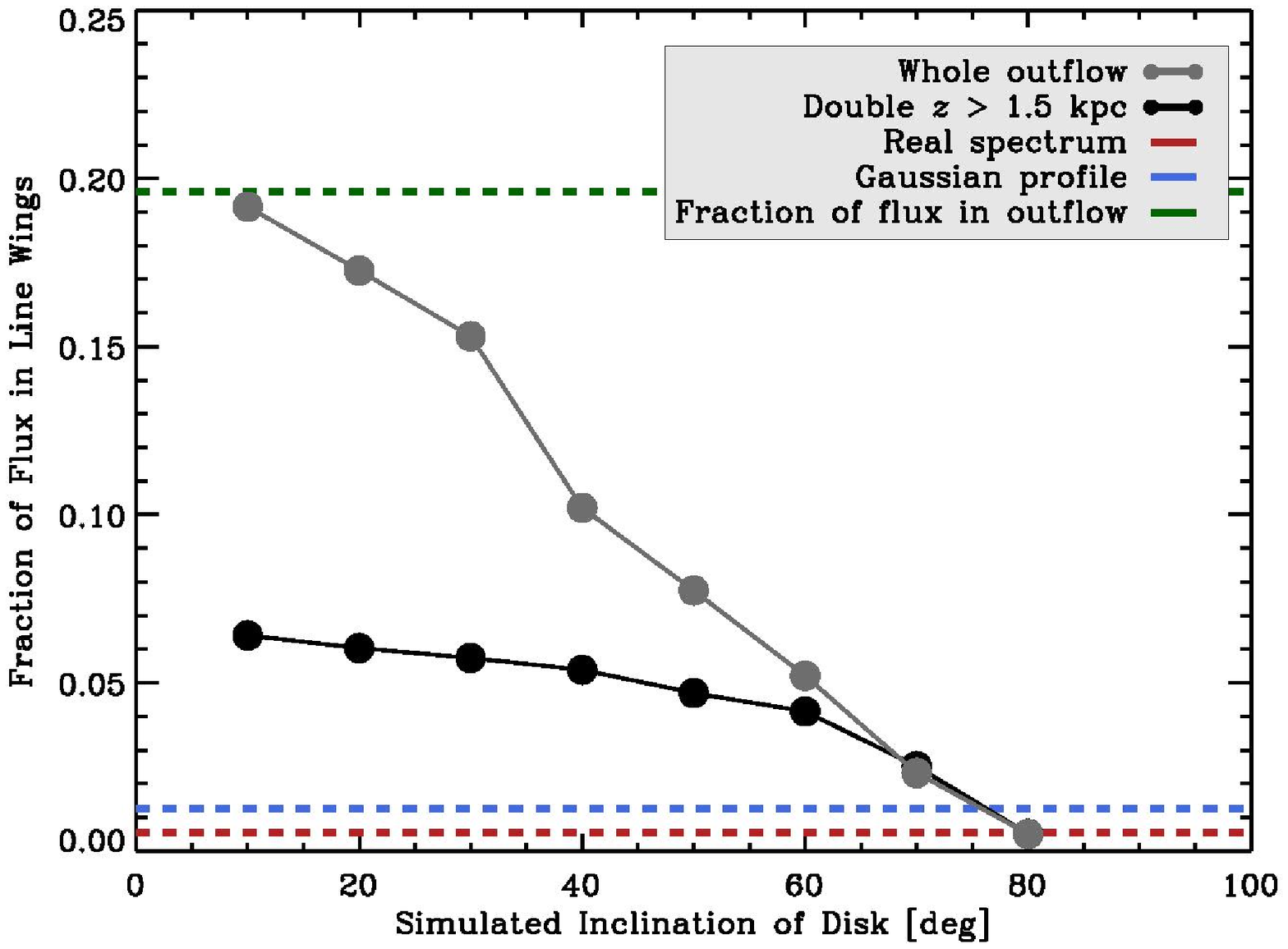}{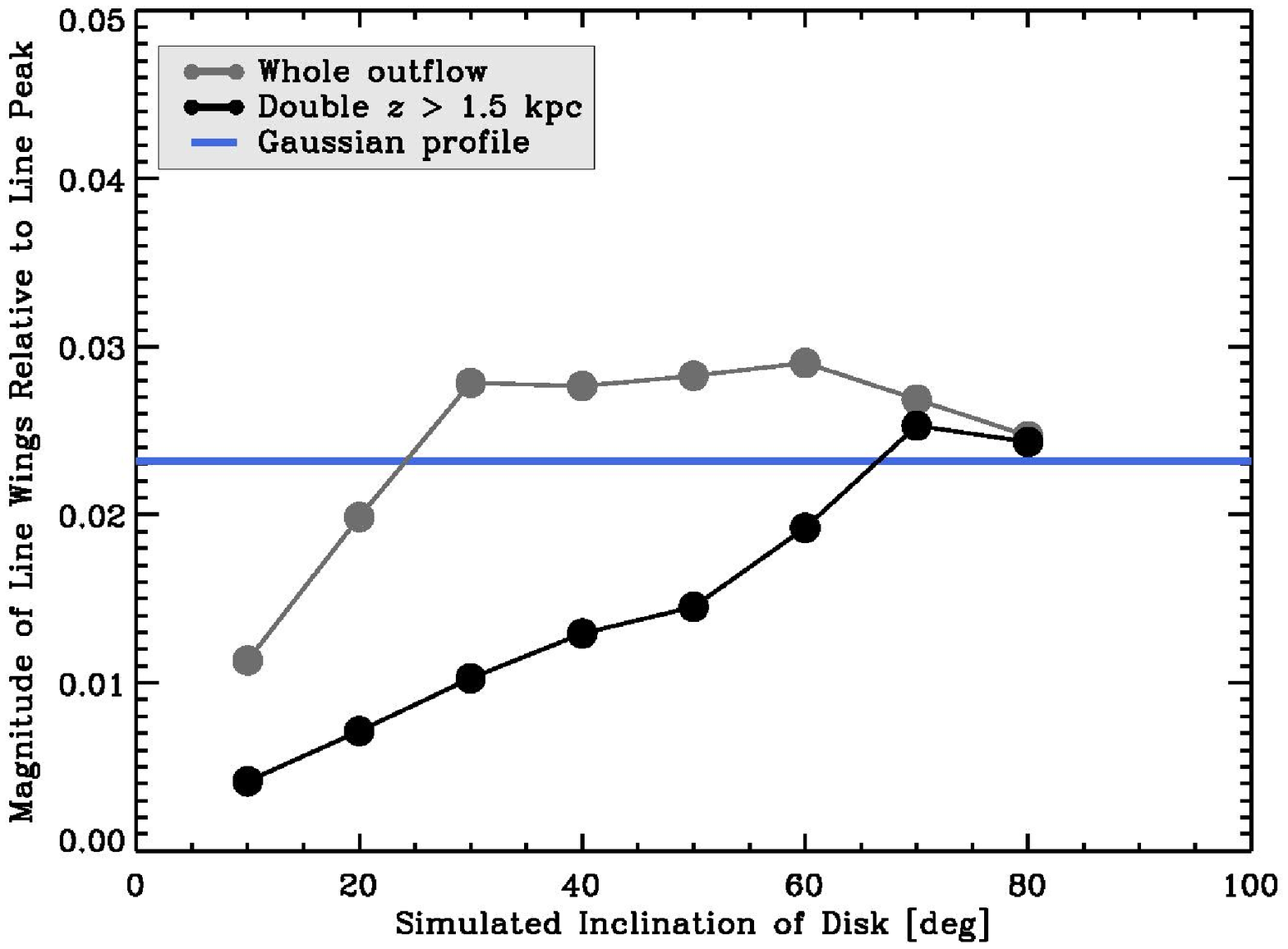}
\caption{
\label{fig:intspec} 
Prospects for detecting M82-like outflows from unresolved CO spectroscopy. ({\em top left}) The integrated spectrum of M82 (black) from our masked data cube with the spectra of the individual regions --- the starburst disk (red), the whole outflow (purple), the outflow outside $z \pm 1.5$~kpc multiplied by two (blue; see text), and the extended disk (gray). Conserving flux and applying varying corrections for inclination, we estimate how the integrated spectrum would appear if we viewed M82 from other angles ({\em top right}). The bottom left panel shows the fraction of flux in the ``wings'' of the integrated spectrum, with wings defined as channels with $< 5\%$ of the peak flux density. We plot the values for a Gaussian line profile and the real integrated spectrum as horizontal lines for comparison. The bottom right panel shows the amplitude of the line wings in the simulated spectrum; here amplitude is defined as the flux weighted flux density in parts of the spectrum with flux density $< 5\%$ of the peak value. Results for only the high latitude material suggest a conservative lower limit of $\approx 5\%$ flux in line wings with magnitude $\sim 1\%$ of the peak flux density for a moderately inclined version of M82.}
\end{figure*}

In both the radio and the optical, unresolved spectra remain a main tool for identifying candidate outflows. Though it may be the best-known  galactic wind, the M82 outflow would be difficult to identify from integrated spectroscopy because of its high inclination. As Figure \ref{fig:intspec} shows, projection effects render the wind almost invisible in the global spectrum of the galaxy. By measuring individual spectra for our sub-regions (Figure \ref{fig:wind}), we simulate the effects of seeing M82 from other angles. To do this, we assume the outflow to be orthogonal to the disk and that the true inclination of the galaxy (outflow) is $i=80\arcdeg$ ($10\arcdeg$). For a series of other  possible inclination angles, $i_{\rm sim}$, we adjust the width of the disk spectrum by a factor of $\sin i / \sin i_{\rm sim}$ while conserving its integrated flux. We carry out a similar operation on the outflow and add the two to construct a new, artificially inclined spectrum. Mechanically, we achieve the adjustment for inclination by simply interpolating the old spectrum, with the abscissa rescaled, on to our fiducial velocity grid and rescaling the interpolated spectrum to match the flux of the original version. We expect this approach to fail at inclinations below $i_{\rm sim} \approx 20\arcdeg$ because the intrinsic velocity dispersion within the starburst, which we do not model, will become important. Therefore we consider $i_{\rm sim}$ over the range $20$ to $80\arcdeg$.

The top left panel in Figure \ref{fig:intspec} shows the integrated spectrum of M82 as we observe it. The top right panel shows the integrated spectrum simulated for different inclinations. The spectral signature of the outflow in a face-on version of M82 is a broad, low component \citep[similar, e.g., to that in Markarian 231,][]{FERUGLIO10}. We quantify the flux and amplitude of these faint line wings in the bottom panels of Figure \ref{fig:intspec}. For these calculations, we define the ``line wings'' to be all channels that have flux density less than 5\% of the peak value \citep[similar to][]{LEROY15B}. For a Gaussian line profile, just over $1\%$ of the flux emerges from the line wings defined in this manner. For the flux in these wings, we sum all channels that meet this criteria. To measure the average amplitude, we calculate $\sum F_{\nu}^2 / \sum F_{\nu}$ where $F_\nu < 0.05~F_\nu^{\rm max}$; that is, we derive the flux-weighted flux density in part of the spectrum with amplitude below 5\% of the peak value.

Figure \ref{fig:intspec} shows results for two cases: (1) the entire outflow region and (2) a spectrum constructed by doubling the emission outside $\pm 1.5$~kpc. At these high latitudes, we expect to avoid foreshortened disk material and gas ejected at a range of angles, observing only material flowing away from the disk along the minor axis. We double the spectrum because from mass continuity arguments (see main text), we expect that at least an approximately equal amount of material must flow out along the minor axis inside $\pm 1.5$~kpc. Therefore, doubling the spectrum represents a realistic, but still conservative (because it accounts for no phase change or fall-back) lower limit for this exercise.This doubled $|z| > \pm 1.5$~kpc curve represents our conservative lower limit and the main result of this exercise: in a face-on version of M82, faint line wings with amplitude $\lesssim 1_{-0.5}^{+1}\%$ of the peak are likely to contain at least $\approx 5$\% percent of the total line flux. The results for the whole outflow region suggest that the true signature of a face-on M82 could be even stronger than this and could be as high as $\gtrsim 10$--$20\%$ of the flux in line wings with flux density $\approx 3\%$ the maximum. These more optimistic results carry the caveat that they may include blended disk emission and emission from ejecta at a wide range of inclinations. For these ejecta a a wide range of angles, the effect will be to move some of the flux to a narrower component with higher amplitudes, one that may be harder to pick out from the disk in an integrated spectrum.

\section{The CO-to-H$_2$ Conversion Factor in the Outflow}

The procedure to convert our observed CO luminosity into a molecular gas mass is complicated by the fact that physical conditions in the outflow may differ substantially from those in the disk galaxy giant molecular clouds usually used to calibrate $\alpha_{\rm CO}$ \citep[e.g.,][]{DONOVANMEYER13}.  In this section, we step through what we can infer about the conversion factor in the M82 outflow based on comparing different CO transitions and tracers of different ISM phases. We refer the reader to \citet{BOLATTO13B} for a thorough discussion of all of the issues involved in estimating the CO-to-H$_2$ conversion factor.

\subsection{Optically Thin Limit}

Following the treatment of CO (1-0) in the NGC 253 wind by \citet{BOLATTO13A}, we consider the case of an optically thin line in local thermodynamic equilibrium (LTE)  as a lower limit. Using Einstein A coefficients and other data from the Leiden Atomic Molecular Database \citep{SCHOIER05}, we calculate the optically thin $\alpha_{\rm CO}$ for the three lowest $J$ CO transitions:

\begin{eqnarray}
\label{eq:thin}
\alpha_{\rm CO}^{1-0} &\approx& 0.34~\left(\frac{T_{\rm ex}}{30~{\rm K}}\right)~\exp \left(\frac{5.53}{T_{\rm ex}}\right) \acounits \\
\alpha_{\rm CO}^{2-1} &\approx& 0.12~\left(\frac{T_{\rm ex}}{30~{\rm K}}\right)~\exp \left(\frac{16.6}{T_{\rm ex}}\right) \acounits \\
\alpha_{\rm CO}^{3-2} &\approx& 0.095~\left(\frac{T_{\rm ex}}{30~{\rm K}}\right)~\exp \left(\frac{33.2}{T_{\rm ex}}\right) \acounits~. \\
\end{eqnarray}

\noindent Here we have assumed local thermodynamic equilibrium (LTE) and neglected CMB effects; see \citet[][]{BOLATTO13A} and \citet{LEROY15A} for an explicit list of the equations used in the LTE calculation. The implied line ratios in the optically thin case, $R_{21} \equiv I_{21} / I_{10} \approx 2.8$ and $R_{32} \equiv I_{32} / I_{21} \approx 1.3$ (both in brightness temperature units) for $T_{\rm ex} = 30$~K match expectations from RADEX \citep{VANDERTAK07} in the limit of large line width, low column, and high volume density with $T_{\rm kin} = 30$~K. For the LTE calculation, varying $T_{\rm ex}$ from 20--40~K  makes only a $\sim 20\%$ difference on $\alpha_{\rm CO}^{2-1}$. The impact is larger for a non-LTE treatment, as we see below.

\subsection{Comparing Low-$J$ CO Transitions}

\begin{figure*}
\plottwo{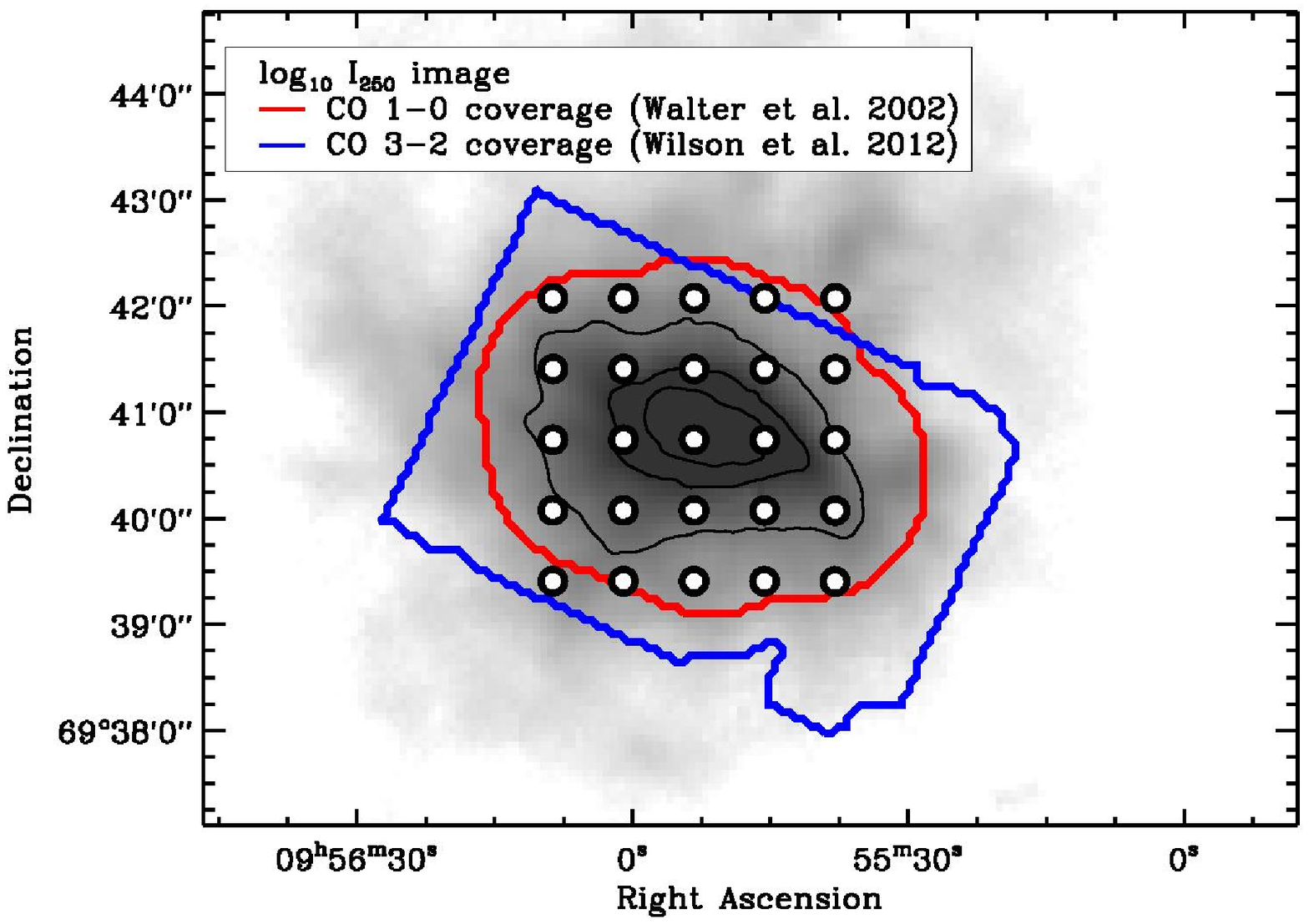}{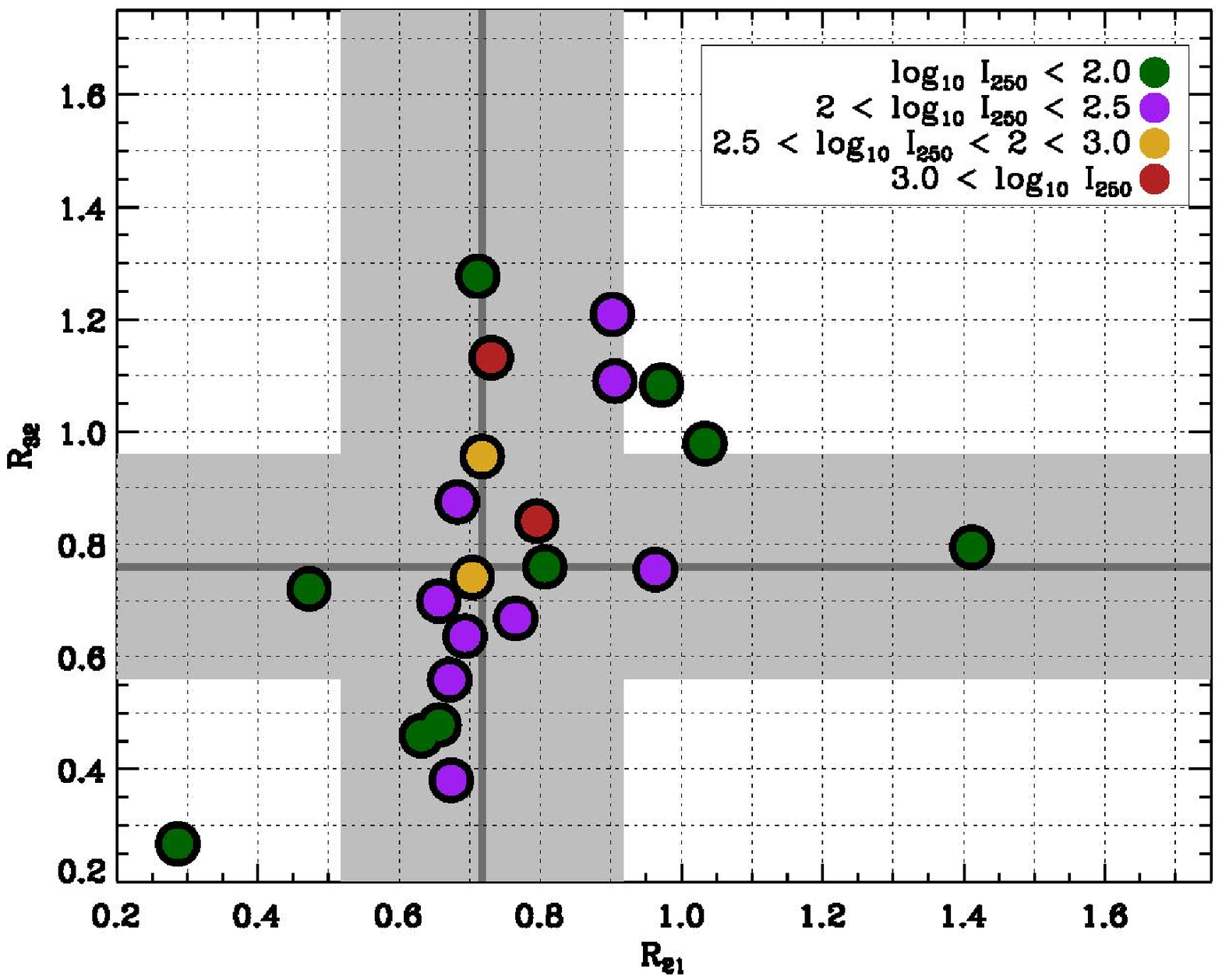}
\plotone{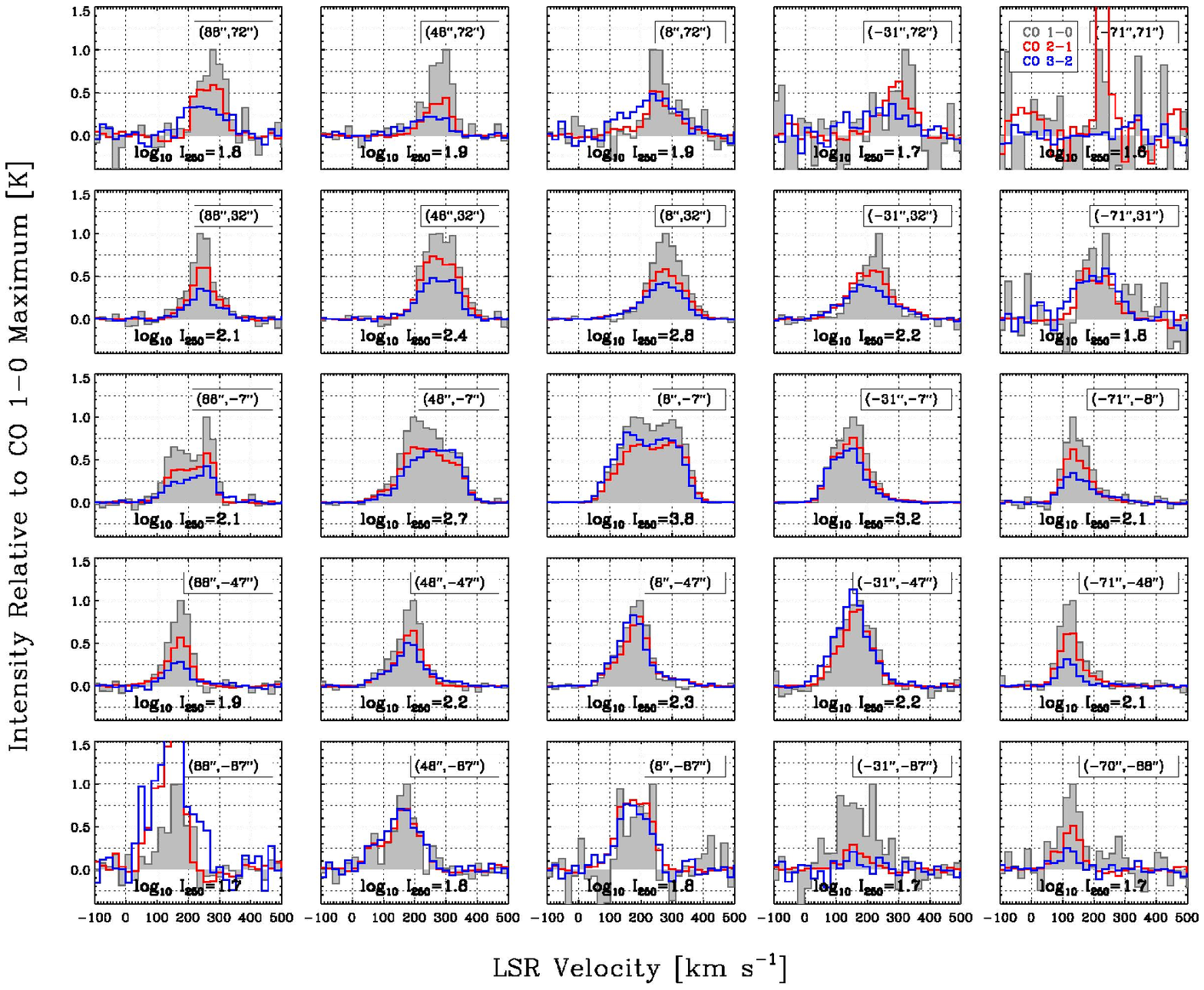}
\caption{
\label{fig:linerat} Comparison of different CO transitions in M82.  ({\em top left}) SPIRE 250$\mu$m map with contours indicating the coverage of the CO (1--0) OVRO+FCRAO map \citep{WALTER02} and the JCMT CO (3--2) map \citep{WILSON12}. Dots indicate the central positions for the heavily binned spectra in the bottom panel. ({\em top right}) The three line ratios as a function of SPIRE intensity, which roughly corresponds to displacement from the central starburst. Here each point is an individual line of sight at the matched $20\arcsec$ resolution of the HERA map. ({\em bottom}) Spectra for all three low-$J$ CO transitions binned to $\approx 40\arcsec$ independent pixels and $\approx 20$~km~s$^{-1}$ channel width to increase the  signal to noise. The grid cells are centered at the dots indicated in the top left panel and the average $250\mu$m intensity in each bin is noted in that panel.}
\end{figure*}

The line ratios implied by Equation \ref{eq:thin} are high, with both $R_{21}$ and $R_{32}$ greater than unity; indeed $R_{21} \sim 3$. M82 has been studied  extensively in molecular gas,  so that we have some knowledge of the true line ratios. Across the whole system, comparing our luminosity to that of \citet{SALAK13} yields $R_{21} \approx 0.8$. Though the apertures are imperfectly matched, dividing our outflow luminosity by their high-latitude luminosity yields  $R_{21} \gtrsim 0.3$ but certainly not above unity.  More concretely, we have in-hand CO (1--0) and CO (3--2) maps from \citet{WALTER02} and \citet{WILSON12}. We use these for a direct calculation of line ratios, although note that this comparison is limited by the coverage and sensitivity of the other data (see Figure \ref{fig:linerat}).

\begin{figure*}
\plottwo{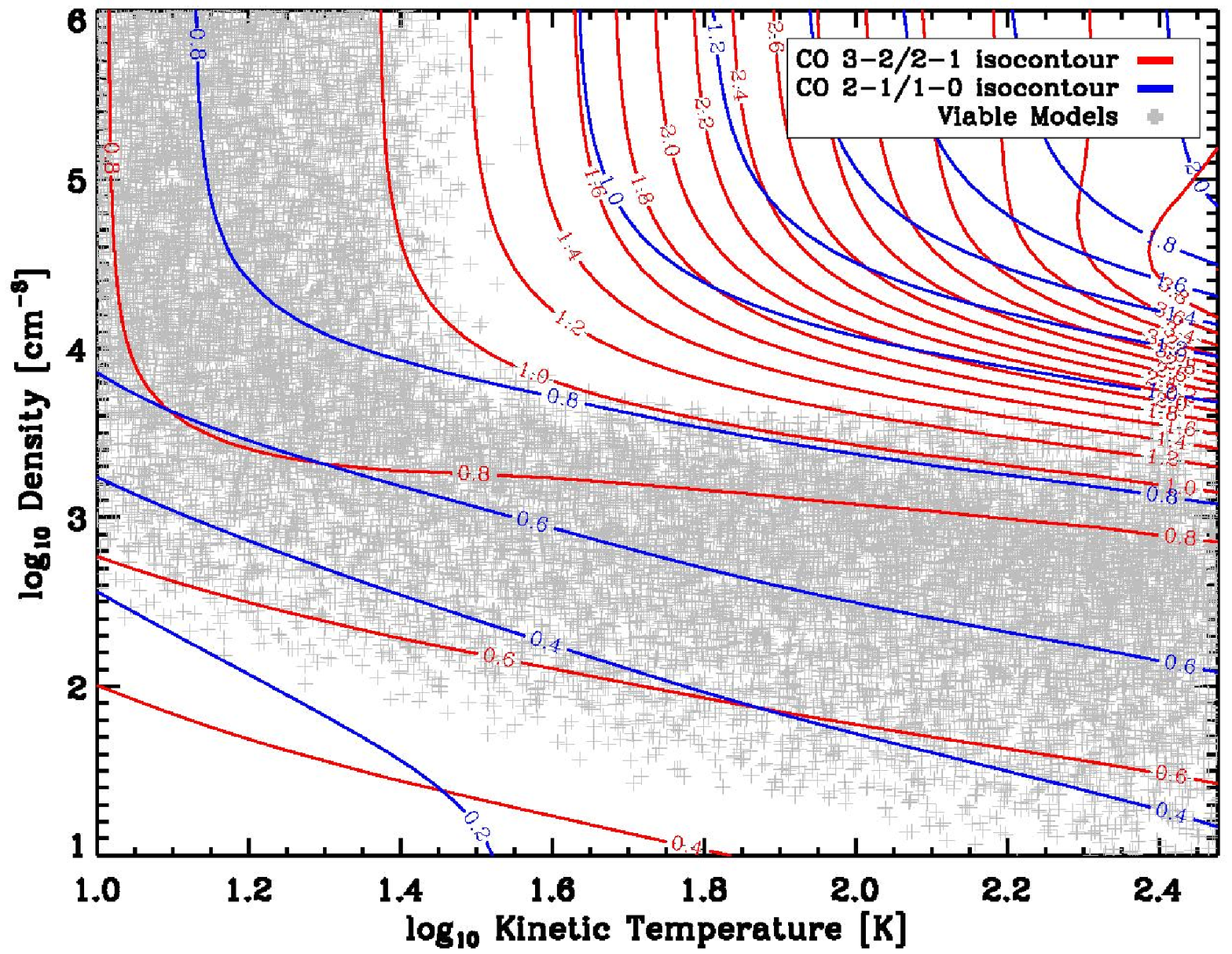}{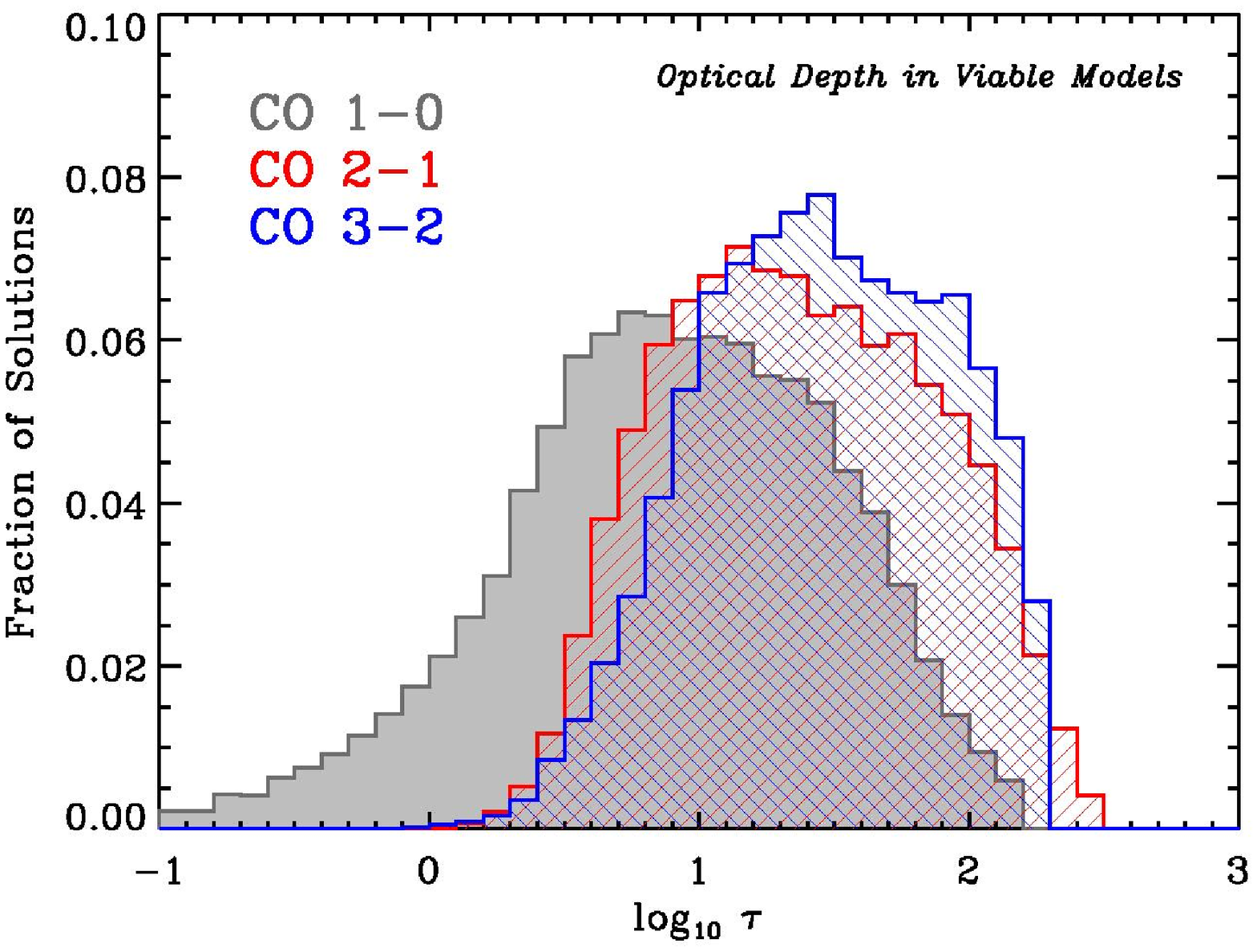}
\caption{
\label{fig:radex}  Results for RADEX modeling of $^{12}$CO line ratios around M82. We run a large grid of RADEX calculations covering a range of kinetic temperature ($T_{\rm kin}$), collider density ($n_{\rm H2}$), and column per unit line width ($N_{\rm CO} / \Delta v$). These two panels show our main results: ({\em left}) the observed line ratios are consistent with models that have either low temperatures ($T_{\rm kin} \lesssim 30$~K) and high densities ($\log_{10} n_{\rm H2} \gtrsim 3$~cm$^{-3}$) or high temperatures ($T_{\rm kin} \gtrsim 30$~K) and low densities ($\log_{10} n_{\rm H2} \sim 2$--$3$~cm$^{-3}$). Red and blue lines show isocontours of the CO (3--2) / (2--1) (blue) and (2--1) / (1--0) (red) ratios for a $N_{CO} / \Delta v \approx 10^{17}$~cm$^{-2}$~(km~s$^{-1}$)$^{-1}$, a reasonable value for a molecular cloud. Gray points show all combinations of $T$ and $n$ in our grid that match the typical observed line ratios within 0.25 (we add a small amount of noise to the data for plotting). ({\em right}) Optical depth for all grid solutions that match the observed line ratios. The calculations suggest optically thick emission in all three lines, consistent with the comparison to LTE. Work combining $^{12}$CO lines with optically thin transitions by \citet{WEISS05} suggests the high $T_{\rm kin}$, low $n_{\rm H2}$, optically thick solutions as most likely, which in turn suggest conversion factors $\sim 2$--$4$ times lower than Galactic.
}
\end{figure*}

To compare our CO (2--1) map with the other transitions, place all three CO data cubes on to a matched grid with coarse $\approx 40\arcsec \times 40\arcsec \times 20$~km~s$^{-1}$ pixels. The centers of the independent pixels appear as circles in Figure \ref{fig:linerat} along with the coverage of the CO (1--0) and CO (3--2) maps. In Figure \ref{fig:linerat}, we plot the spectra for transition at each pixel with intensities (in Kelvin) normalized to the peak of the CO (1--0) spectrum. We also plot the two line ratios $R_{21}$ and $R_{32}$, color coded by the 250$\mu$m intensity in the pixel, which serves as a rough indicator of distance from the starburst.

These spectra show that the overall line profile shape remains the same among the transitions, although the signal to noise of the other data are lower than our CO (2--1) map (this is particularly an issue before the heavy rebinning used to make the figure). This consistency among all three maps  offer some check on the accuracy of our calibration and error beam correction scheme. In M82, the intensities of the CO (2--1) and (3--2) lines tend to be lower than that of the CO (1--0) line, but not by a large factor. The exceptions to this general agreement are the corners of the grid, where the finite extent of the other maps make the comparison unreliable. Future comparison to the wide field map of \citet{SALAK13} will allow a check on whether the variations reflect limitations of the data or variations in physical conditions in regions of faint, extended emission. Line ratios near unity but not above agree with \citet{WEISS05}, who studied multiple CO transitions for individual pointings across the streamer/outflow area as defined in \citet{WALTER02} and found average $R_{21} = 1.0 \pm 0.1$ and $R_{31} = 0.8 \pm 0.2$. For comparison, we find median $R_{21} = 0.7$ with $\approx 0.15$ scatter and median $R_{32} \approx 0.75$ with $0.3$ scatter. Figure \ref{fig:linerat} shows that the ratios are reasonably indicative of areas across the overlap between the maps, though the central burst may be slightly more excited.

These ratios are not what one would expect for optically thin gas in LTE. To investigate what physical conditions could produce the observed ratios, we compare these ratios to a large grid of {\em Radex} calculations. {\em Radex} \citep{VANDERTAK07} is a non-LTE code that models line emission using an escape probability formalism for a single set of physical conditions (density, temperature, and column density per line width). We calculate expected $^{12}$CO emission for a large grid of these physical parameters and compare our observed line ratios to these calculations in Figure \ref{fig:radex}. Our interest here is broad results, not a point-by-point comparison, so we identify points in the grid that match our median $R_{32}$ and $R_{21}$ with a tolerance of $\pm 0.2$. This does a good job of capturing most of the observed ratios (Figure \ref{fig:linerat}). Figure \ref{fig:radex} shows our two main results for the calculation. First, almost all models that match the observed ratios are optically thick in all three low-$J$ CO lines, further arguing against the optically thin case. Second, the observed line ratios can be produced by either cold ($T_{\rm kin} \lesssim 20$~K) and dense ($\log_{10} n_{\rm H2} {\rm [cm}^{-3}] \gtrsim 3.5$ gas or hot ($T_{\rm kin} \gtrsim 30$~K) but rarer ($\log_{10} n_{\rm H2} {\rm [cm}^{-3}] \sim 2$ gas. These results agree with \citet{WEISS05}, who found significant opacity in the $^{12}$CO lines considering individual pointings in the same area we study here. By including $^{13}$CO they were able to further constrain $\log_{10} n_{\rm H2} {\rm [cm}^{-3}] \sim 3$~cm$^{-3}$, towards the lower end of viable densities. This also led \citet{WEISS05} to argue for a higher temperature, $T_{\rm kin} \gtrsim 30$~K. Note, also that this temperature cannot be arbitrarily high: in addition to the constraints from \citet{WEISS05}, direct observations of the H$_2$ rotational lines by \citet{BEIRAO15} constrain the bulk H$_2$ temperature to be less than a few hundred K and even this ``cooler'' component of warm H$_2$ represents only a modest fraction of the total molecular gas mass in the system.
 
What does opaque, warm gas mean for the CO-to-H$_2$ conversion factor? \citet{WEISS01} used analysis of a more limited set of CO lines, but still including optically thin tracers, to demonstrate a reasonable correspondence of $\alpha_{\rm CO}$ with the $T_{\rm kin}^{-1}~n^{0.5}$ expected for an opaque, virialized cloud. Following this logic, a $T_{\rm kin} \approx 2$--$3$ times higher than Milky Way clouds, with a similar density \citep[see][]{HEYER09} implies a conversion factor $\sim 1/3$--$1/2$ the  ``disk galaxy'' value. This roughly matches the values found by \citet{WEISS01} in positions that were vertically offset from the starburst disk,  $\approx 5$--$10 \times 10^{19}$~\xcounits , or $\approx 0.25$--$0.5$ times the Milky Way value. It also matches the value of $\approx 10^{20}$~\xcounits shown by  \citet{KETO05} to produce approximate virial equilibrium in the resolved cloud population in the M82 starburst.

\subsection{Comparing Dust and Gas}

\begin{figure*}
\plottwo{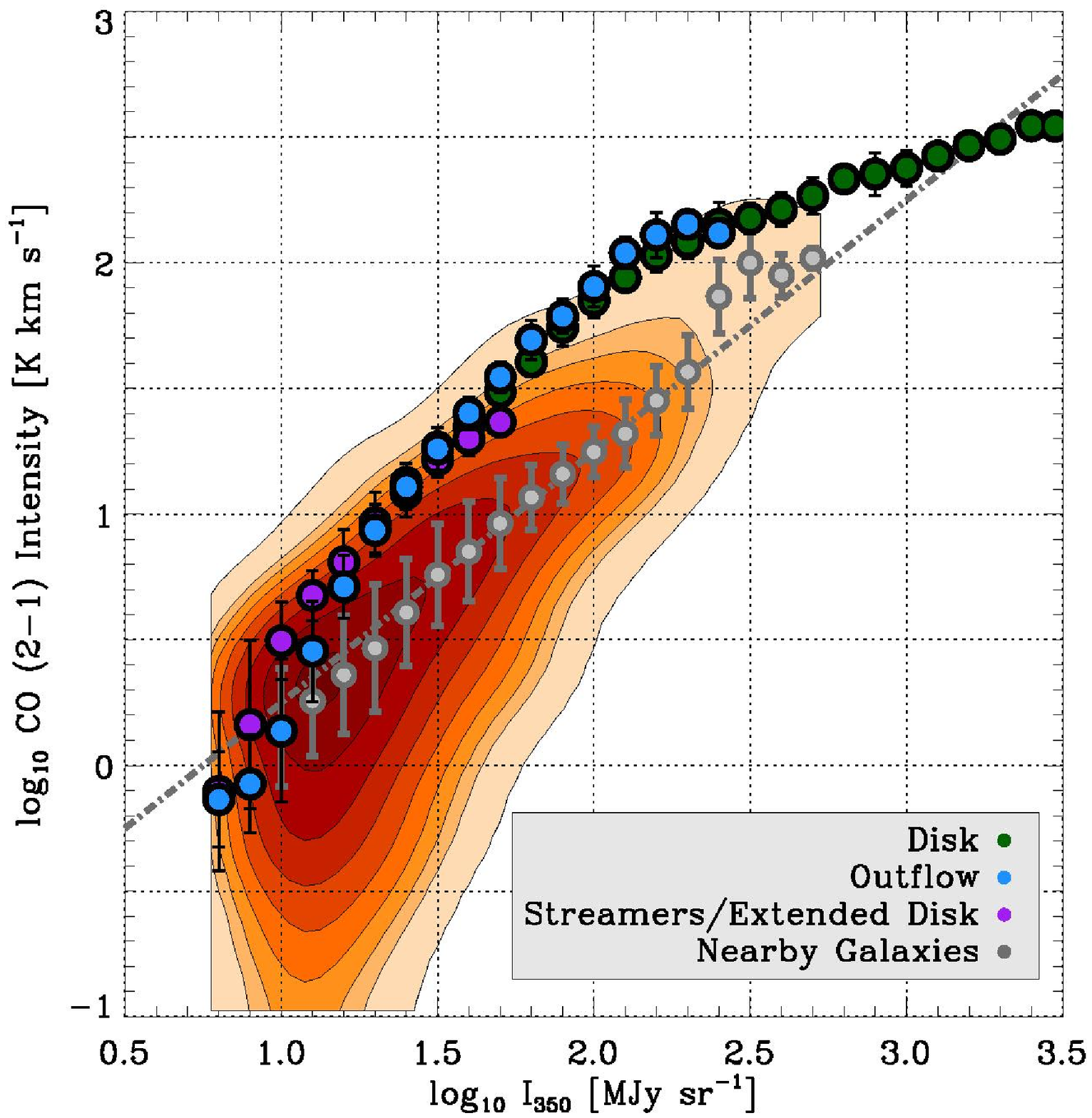}{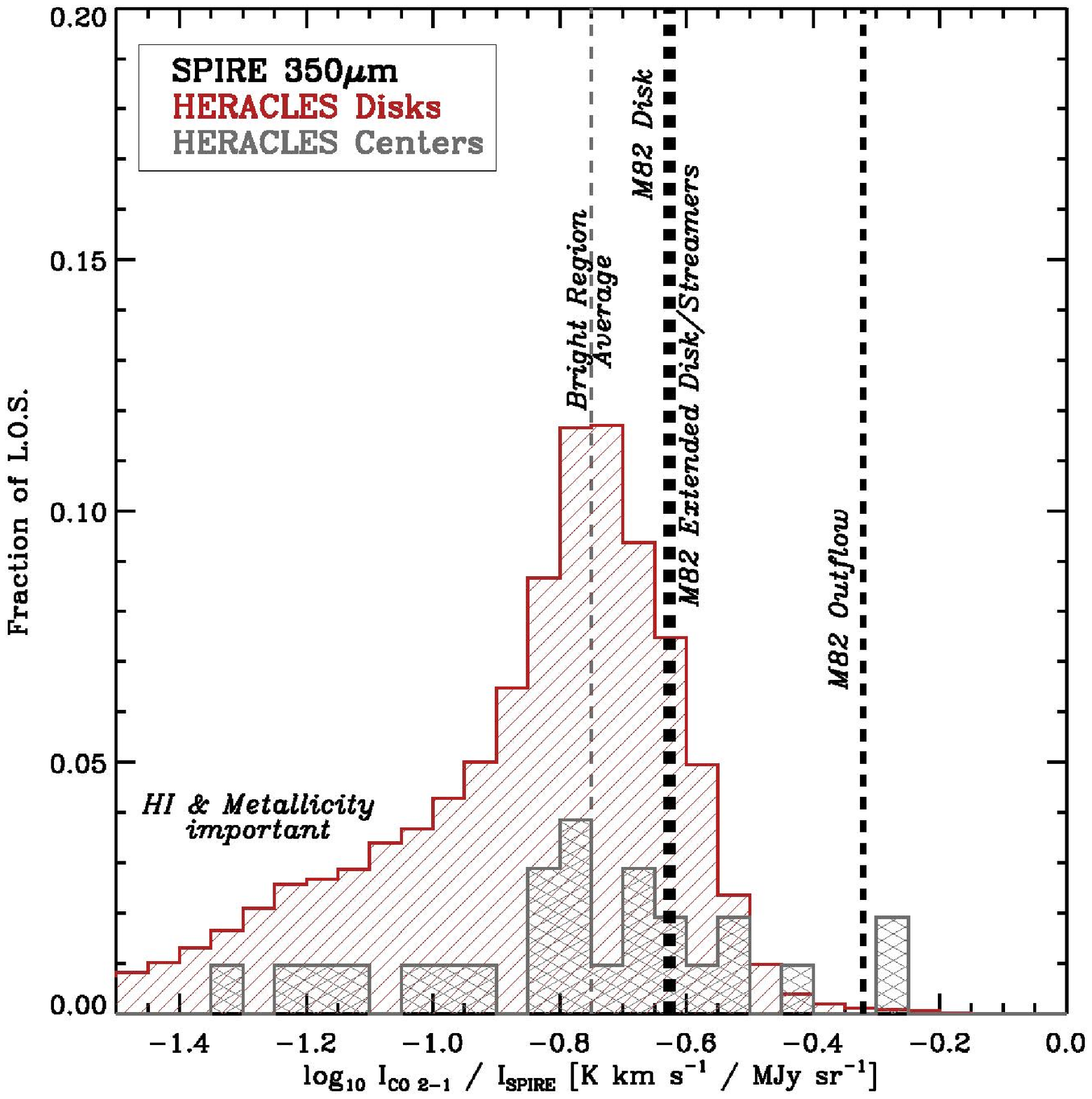}
\caption{
\label{fig:spire} CO (2--1) emission ($I_{\rm CO}$) as a function of far-infrared intensity at 350$\mu$m ($I_{350}$) measured with SPIRE on {\em Herschel} for a reference sample compiled from the HERACLES and KINGFISH surveys (contour and histogram, taking all lines of sight with $I_{350} > 10$~MJy~sr$^{-1}$) and the three regions that we use to analyze M82. Though the effects of {\sc Hi} and metallicity are clearly visible at low intensities and as a tail in the histogram, the reference sample suggests a typical ratio of $\log_{10} I_{\rm CO} / I_{350} = -0.75 \pm 0.2$ for the molecule-dominated parts of disk galaxies. M82 shows a systematically higher amount of CO emission per unit IR emission, $\log_{10} I_{\rm CO} / I_{350} = -0.35$. The translation of $I_{350}$ to dust mass is not trivial, but the simplest explanation for the difference is a conversion factor, $\alpha_{\rm CO}$ that is $\approx 2.5$ times smaller in M82 than a typical disk galaxy.
}
\end{figure*}

Dust emission offers an powerful, orthogonal constraint on the conversion factor. Following \citet{ISRAEL97}, the dust mass derived from the infrared spectral energy distribution has been used as a tracer of the total neutral gas column. This technique has been used to estimate H$_2$ mass without the use of CO in a variety of systems. Indeed, versions of this approach have been applied to the inner region of M82 as early as \citet{WILD92}. Details of using dust to solve for $\alpha_{\rm CO}$  are discussed in \citet{SANDSTROM13}, \citet{LEROY11}, and references therein. Here we first compare the  correlation between CO and long wavelength emission measured by {\em Herschel} in the outflow region to that observed across a wide set of nearby galaxies. We demonstrate that CO emission is bright compared to $250\mu$m emission in M82, a strong empirical indication of a different $\alpha_{\rm CO}$. Then we compare the dust surface density derived from SED modeling, $\Sigma_{\rm dust}$, to {\sc Hi} and CO emission to estimate $\alpha_{\rm CO}$. We solve for the dust-to-gas ratio, $\delta_{\rm DGR}$, in the extended molecular disk and then, assuming this value to be approximately constant, apply it to constrain $\alpha_{\rm CO}$ in the outflow region.

\subsubsection{Submillimeter Intensity and CO Emission}

Figure \ref{fig:spire} $I_{\rm CO}$ as a function of 350$\mu$m intensity measured by {\em Herschel} \footnote{Note that the results are essentially the same if we instead use 250$\mu$m emission; we prefer $I_{350}$ only because it lies closer to the Rayleigh-Jeans regime.}. The 350$\mu$m intensity, though not entirely on the Rayleigh-Jeans tail, will be approximately proportional to $T_{\rm dust} \times \Sigma_{\rm dust}$ and for a constant dust-to-gas ratio, $I_{350} \propto T_{\rm dust} \times \Sigma_{\rm gas}$ \citep[for caveats see][]{SANDSTROM13,LEROY11}. Similarly, in the limit where molecular gas dominates the gas reservoir, $\Sigma_{\rm gas} \sim \alpha_{\rm CO} \times I_{\rm CO}$, so that the ratio $I_{\rm CO} / I_{\rm 350}$ is an observational ``color'' closely related to the CO-to-H$_2$ conversion factor. If all properties other than $\alpha_{\rm CO}$ remain fixed, then $\alpha_{\rm CO} \propto \left( I_{\rm CO} / I_{\rm 350} \right)^{-1}$ in regions where H$_2$ dominates the gas reservoir \citep[see related discussion cast in terms of luminosity in][]{GROVES15}.

As a benchmark, we take the entire HERACLES sample \citep[see][]{LEROY13} and the {\em Herschel} images from KINGFISH \citet{KENNICUTT11}. We convolve HERACLES to the resolution of the 350$\mu$m data \citep[following][]{ANIANO11} and extract matched intensities for each line of sight where $I_{350} > 10$~MJy~sr$^{-1}$. Figure \ref{fig:spire} shows the resulting correlation appears as a data density plot in $I_{\rm CO}$-$I_{350}$ space and a histogram. The figure also motivates the adopted intensity cutoff; below $I_{350}$~MJy~sr$^{-1}$ the effects of metallicity and {\sc Hi} on the ratio  $I_{\rm CO} / I_{350}$ become substantial. That is, the dust-to-gas ratio begins changing and {\sc Hi} constitutes a non-negligible part of the gas reservoir \citep[][]{SANDSTROM13}. Indeed, the Figure shows that these effects are already visible, though not yet dominant, below $I_{350} \sim 30$~MJy~sr$^{-1}$. From either the ridge line (median $I_{\rm CO}$ at  a given $I_{350}$) in the left panel or the histogram in the right panel, we calculate the ratio for bright emission in the HERACLES and KINGFISH overlap to be

\begin{equation}
\log_{10} \frac{I_{\rm CO 2-1}}{I_{350}}  = -0.75 \pm 0.2~{\rm where }~I_{350} > 30~{\rm MJy~sr}^{-1}
\end{equation}

\noindent with $0.2$~dex the rms scatter among the data, not the uncertainty in the mean and the relationship calculated for mostly star-forming disk galaxies at $\approx 1$--$2$~kpc resolution.

Figure \ref{fig:spire} also plots our measurements of M82. We find a good correlation between $I_{350}$ and $I_{\rm CO}$ in M82 but also that the normalization of the M82 correlation differs from the reference sample. Outside the highest and lowest intensity regions, M82 exhibits systematically higher $I_{\rm CO} / I_{350}$ than normal disk galaxies. This is exactly the sense expected for a low $\alpha_{\rm CO}$. Ignoring for the moment the low and high intensity deviations, the median $\log_{10} I_{\rm CO 2-1} / I_{350}$ for M82 is $-0.35$, which is $0.4$~dex or a factor of $2.5$ times lower than the median for the reference sample. 

Taken at face value, $I_{\rm CO 2-1} / I_{350}$ implies that $\alpha_{\rm CO}^{2-1}$ in M82, is $\approx 2.5$ times  lower than in a typical disk galaxy. Effects other than $\alpha_{\rm CO}$ can alter this ratio, however. The main effects should temperature variations ($I_{350} \propto T_{\rm dust}$, the contribution of {\sc Hi} to the total gas reservoir, and potential differences in the  dust-to-gas ratio between M82 and the reference sample. The sense of a high $T_{\rm dust}$ in M82 would be to make $I_{350}$ higher, which would only lower the inferred $\alpha_{\rm CO}$.  Similarly, any additional gas component would render this simple estimate an upper limit. Only the dust-to-gas ratio represents a plausible way to recover $\alpha_{\rm CO}$ nearer to the  overall reference sample. However, the weaker shielding associated lower dust-to-gas ratio also typically suppresses CO emission \citep[e.g.,][and references therein]{LEROY11,LEE15}, which would act in the direction of again lowering $I_{\rm CO 2-1} / I_{350}$. 

CO (2--1) emission thus appears about $2.5$ times brighter relative to $350\mu$m emission in M82 than in the whole HERACLES sample. Though this is an empirical indicator with caveats, we argue that the most straightforward interpretation of this high ratio is a lower CO-to-H$_2$ conversion factor in M82 compared to most of the bright regions in HERACLES. At very low and very high $I_{350}$ this conclusion appears less secure, likely due to the importance of {\sc Hi} at the low intensity end and the complexity of emission from immediate starburst region (which is not the topic of this paper) at the high intensity end.

\subsubsection{Results Using Dust Spectral Energy Distribution Fitting}

\begin{figure*}
\plottwo{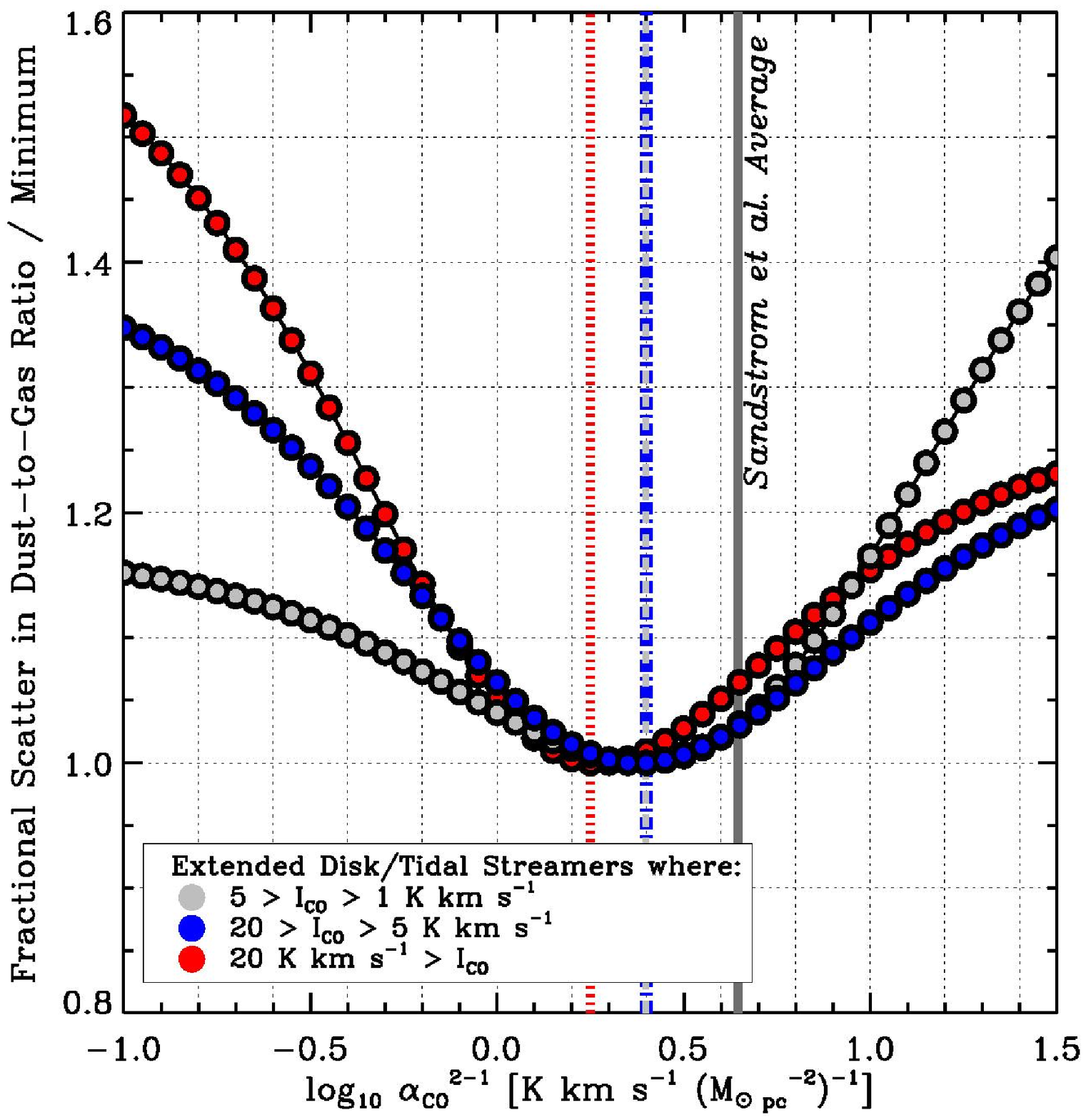}{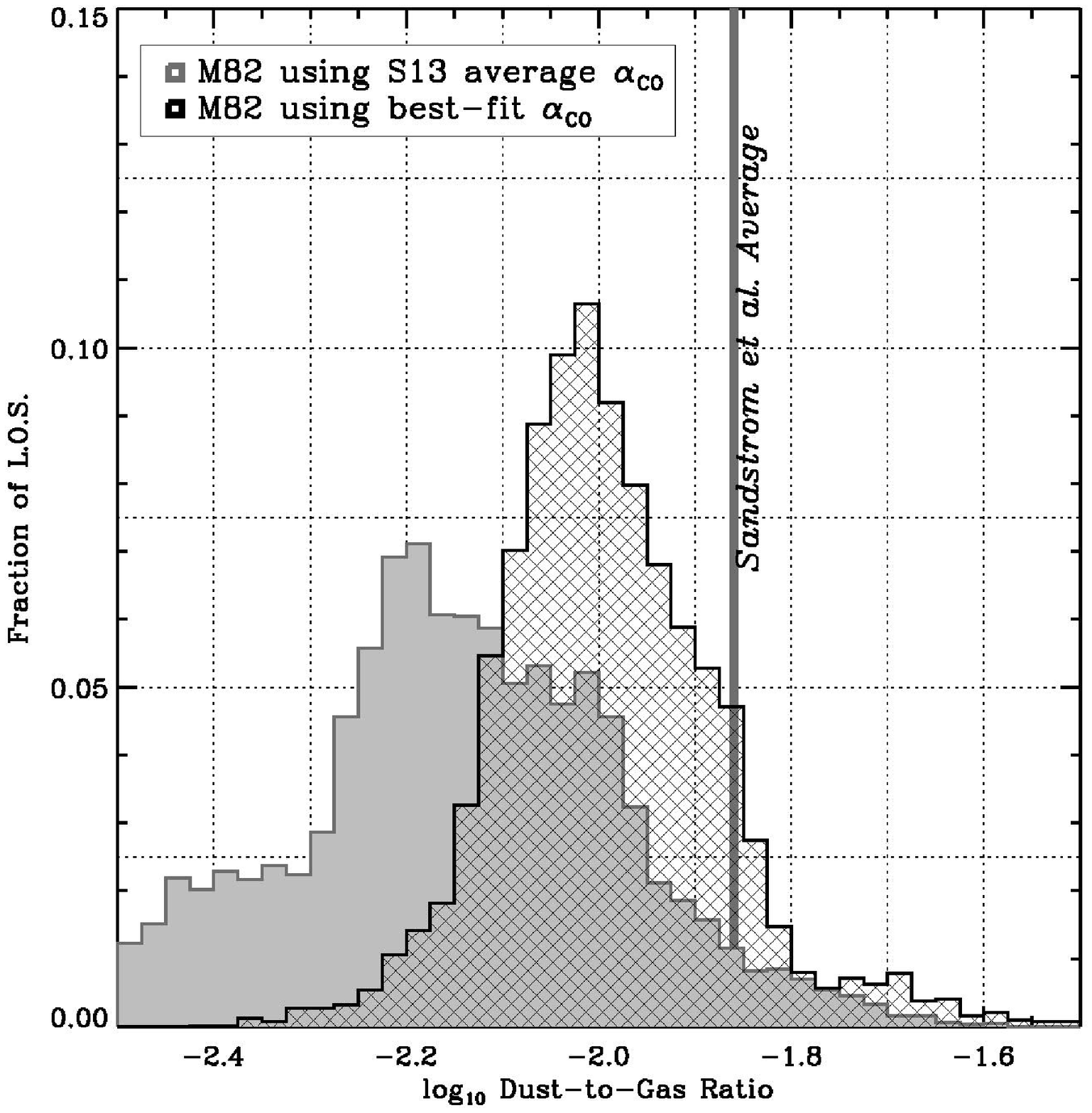}
\plottwo{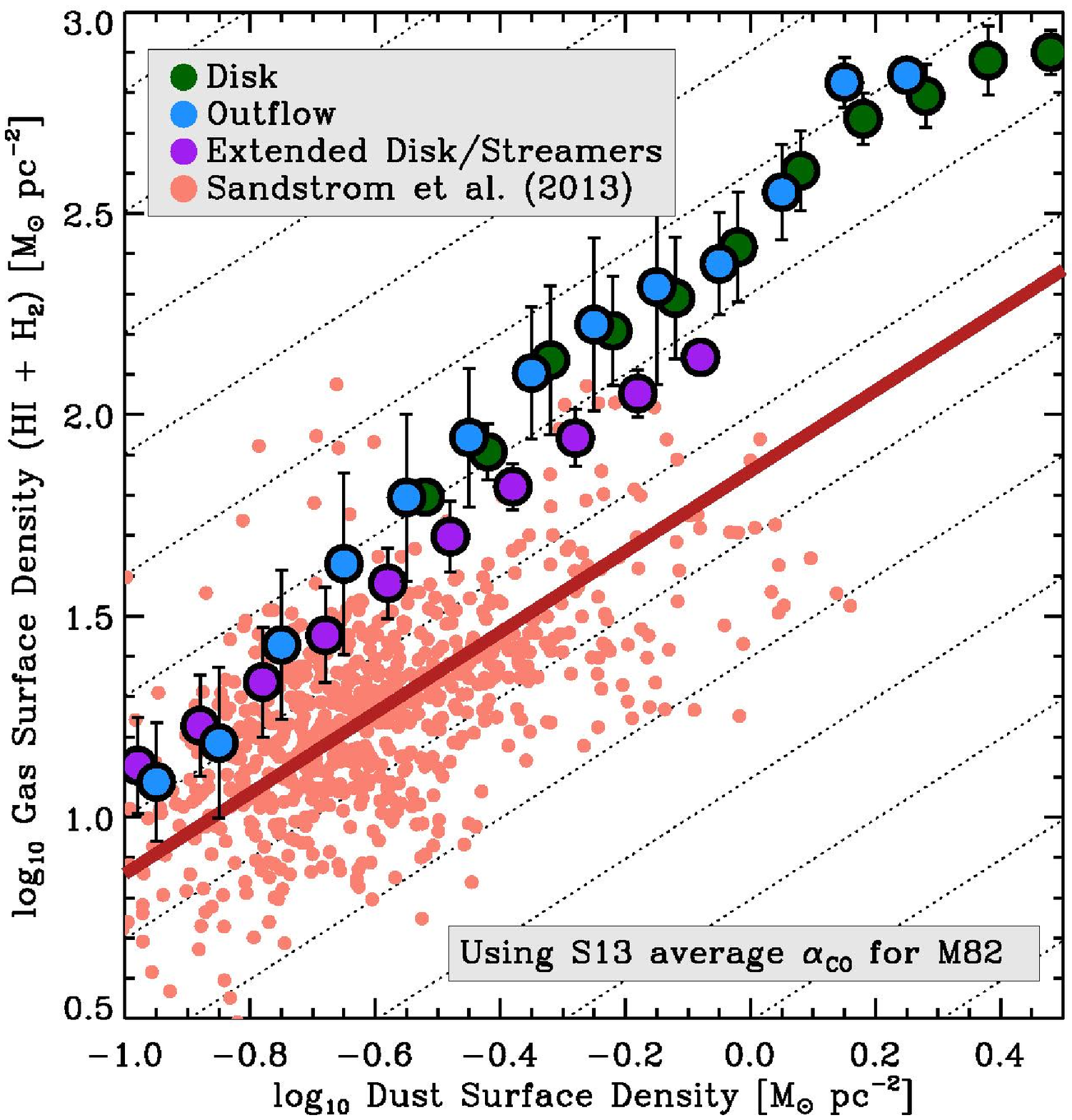}{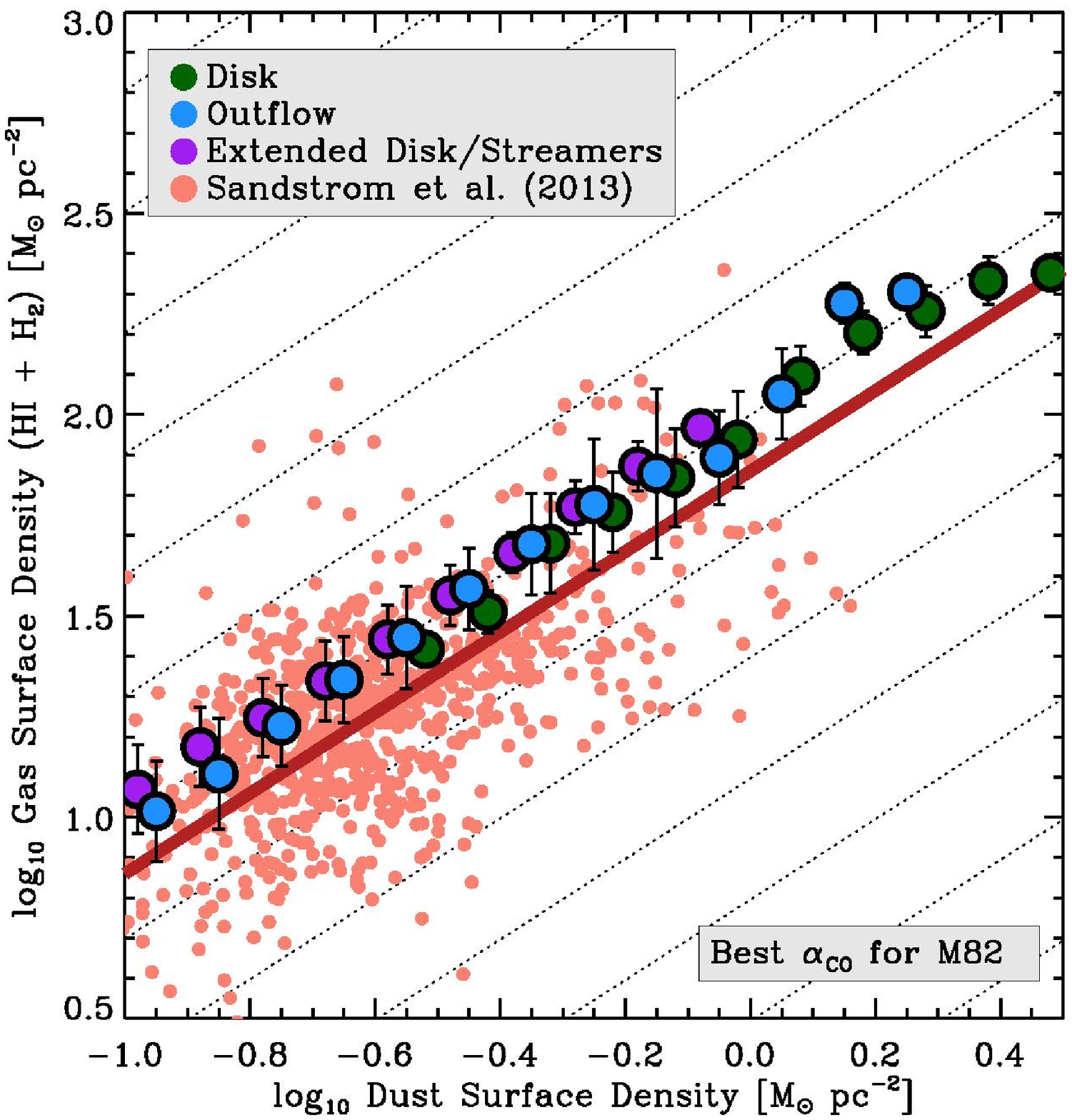}
\caption{
\label{fig:dustaco} Constraints on $\alpha_{\rm CO}$ from comparison of CO emission to dust and {\sc Hi}. ({\em top left}) Scatter in the dust-to-gas ratio as a function of our adopted $\alpha_{\rm CO}$ for the extended disk, where we have sufficient dynamic range in H$_2$/{\sc Hi} to solve for $\alpha_{\rm CO}$ following \citet{SANDSTROM13} and \citet{LEROY11}. Our solution in the extended disk is stable across several regimes in CO brightness (different colors). The implied dust-to-gas ratio ({\em top right}) using our best prescription (black histogram) shows less scatter and is more consistent with a single dust-to-gas ratio than if we applied the typical DGR from disk galaxies using the same dust models\citep{SANDSTROM13}. The implied dust-to-gas ratio in M82 is somewhat lower than in more massive disk galaxies, perhaps consistent with its lower mass. Bootstrapping this dust-to-gas ratio to the rest of the galaxy, we constrain $\alpha_{\rm CO}^{2-1}$ in the area around the starburst and the outflow. The resulting scaling between dust and total neutral gas ({\sc Hi}+H$_2$) appears in the ({\em bottom right}) panel, which uses a two-component prescription for $\alpha_{\rm CO}$ (Equation \ref{eq:brokenaco}). Outside of the brightest part of the starburst, this prescription yields a reasonably fixed dust-to-gas ratio across the system, achieving the same sort of linear scaling of dust and gas seen by \citet{SANDSTROM13} for a large set of disk galaxies (red points and line). If a single dust-to-gas ratio is expected for M82, this approach is a clear improvement over applying only a typical $\alpha_{\rm CO}$ ({\em bottom left}), which yields larger scatter and a non-linear relation between gas and dust.}
\end{figure*}

Following \citet{SANDSTROM13} and \citet{LEROY11}, we adopt the more physical --- but also more model-dependent --- approach of comparing the $\Sigma_{\rm dust}$ inferred from theoretical  models to $I_{\rm CO}$ and $\Sigma_{\rm HI}$. As described in the main text, after cleaning the infrared maps we fit the models of \citet{DRAINE07A} to the infrared SED point-by-point at the resolution of the SPIRE 350$\mu$m map. We compile matched vectors of $I_{\rm CO}$, $\Sigma_{\rm HI}$, and $\Sigma_{\rm dust}$ at this resolution and attempt to solve for $\alpha_{\rm CO}$ by minimizing the scatter in the dust-to-gas ratio as a function of $\alpha_{\rm CO}$. 

As discussed in \citet{SANDSTROM13}, the method requires both {\sc Hi} and H$_2$ to contribute meaningfully to the ISM and a single dust-to-gas ratio to describe the system. In M82, this condition is best met in the extended disk and tidal streamers (the gray region in Figure \ref{fig:wind}). Figure \ref{fig:dustaco} shows the dust-to-gas ratio as a function of tested $\alpha_{\rm CO}$ in this region. For several  subset of the data for this region, we obtain $\alpha_{\rm CO}^{2-1} \approx 2.5$~\acounits , just over half the average value found by \citet{SANDSTROM13} using the same CO line and the same dust models for a large sample of disk galaxies, $\alpha_{\rm CO}^{2-1} = 4.4$~\acounits .

Thus in the extended disk of M82, scatter minimization in the dust-to-gas ratio implies  $\alpha_{\rm CO}$ about half the value found in normal disk galaxies. This this is approximately the same result returned by comparing infrared intensity at 350$\mu$m to CO emission and so reinforces that simple result. More, the fit yields a dust-to-gas ratio for M82 that we expect to also hold in the cold gas entrained in the wind and likely the starburst. The dust-to-gas ratio implied by our best-fit $\alpha_{\rm CO}$ is $\delta_{\rm DGR} \approx 0.009$, or about 110 times more gas than dust by mass. For comparison, \citet{SANDSTROM13} found $\log_{10} \delta_{DGR} = -1.86$, or 1-to-72, for a large set of disk galaxies using the same line and models; that study focused on the CO-bright inner parts of galaxies, so this average value may be somewhat high compared to, say, the Solar Neighborhood or integrated galaxies. Considering integrated galaxy SEDs, \citet{DRAINE07B} found $\log_{10} \delta_{DGR} \approx -2$, or 1-to-100 using CO (1-0) and the same dust models. Thus the extended disk of M82 shows a slightly lower dust-to-gas ratio than metal rich disk galaxies, but well within the range of normal values seen in previous studies using similar data and perhaps consistent with M82's low stellar mass. We emphasize comparative statements because several lines of evidence suggest that the  normalization of the dust masses produced by the \citet{DRAINE07A} models may be as much as a factor of $\sim 2$ times too high \citep[e.g.,][]{FANCIULLO15}. Thus our best estimate for the true value of the dust-to-gas ratio in M82 would be $\sim 1$-to-$220$, slightly lower than $\delta_{DGR}$ inferred in the Solar Neighborhood from depletions \citep[$\approx 1$-to-$150$, e.g., see synthesis in][]{DESERT90}.

Applying the best first $\delta_{\rm DGR}$ from the extended disk, we can place constraints on the molecular mass, and so $\alpha_{\rm CO}^{2-1}$, in the outflow and disk. We calculate $\alpha_{\rm CO}$ in three ways:

\begin{enumerate}

\item We take $\delta_{\rm DGR} = 0.009$ and estimate $\alpha_{\rm CO}$ via the equation:
\begin{equation}
\label{eq:fixdgr}
\alpha_{\rm CO} = \frac{\sum \delta_{\rm DGR} \times \Sigma_{\rm dust} - \sum \Sigma_{\rm HI}}{\sum I_{\rm CO}} 
\end{equation}
with the sum over regions of interest: the outflow region, the starburst disk, and the extended disk. We also subdivide those regions to explore $\alpha_{\rm CO}$ variations within them.

\item We also place a conservative limit on  $\alpha_{\rm CO}$ by neglecting any contribution from {\sc Hi}:
\begin{equation}
\alpha_{\rm CO} < \frac{\sum \delta_{DGR} \times \Sigma_{\rm dust}}{\sum I_{\rm CO}} 
\end{equation}
which avoids uncertainty about the relative geometry of {\sc Hi} and CO and does not rely on the uncertain $\Sigma_{\rm HI}$ closer to the starburst disk, where the line goes into absorption.

\item We apply Equation \ref{eq:fixdgr} but with $\delta_{\rm DGR} = 0.01$ \citep{DRAINE07B} as a check on the impact of the adopted dust-to-gas ratio.
\end{enumerate}

\begin{deluxetable}{lcccc}[h]
\tabletypesize{\scriptsize}
\tablecaption{Results for $\alpha_{\rm CO}^{2-1}$ from Dust Comparison \label{tab:aco}}
\tablewidth{0pt}
\tablehead{
\colhead{Region} & 
\colhead{for $\delta_{\rm DGR} = 0.009$} & 
\colhead{for $\delta_{\rm DGR} = 0.01$} & 
\colhead{for $\delta_{\rm DGR} = 0.009$, no {\sc Hi}} & 
\colhead{based on $I_{\rm CO}/I_{\rm 350}$\tablenotemark{a}} \\
\colhead{} & 
\colhead{} & 
\colhead{} & 
\colhead{(upper limit)} & 
\colhead{} 
}
\startdata
Extended disk & $2.7$ & $2.2$ & $<6.2$ & 2.5 \\
Outflow \\
\ldots whole region & $1.3$ & $1.0$ & $< 2.6$ & 1.6 \\
\ldots faint ($I_{\rm CO} < 5$~K~km~s$^{-1}$) & $3.4$ & $2.7$ & $< 8.2$ & 4.4 \\
\ldots bright ($I_{\rm CO} > 5$~K~km~s$^{-1}$) & $0.9$ & $0.8$ & $< 1.8$ & 1.1 \\
Starburst Disk \\
\ldots whole region\tablenotemark{b} & $3.2$ & $2.9$ & $< 3.6$ & 3.2 \\
\ldots faint ($I_{\rm CO} < 50$~K~km~s$^{-1}$) & $1.4$ & $1.2$ & $< 2.3$ & 1.3 \\
\ldots bright ($I_{\rm CO} > 50$~K~km~s$^{-1}$)\tablenotemark{b} & $3.4$ & $3.1$ & $< 3.7$ & 3.5 \\
\enddata
\tablenotetext{a}{Assuming that $\alpha_{\rm CO}^{2-1} = 4.4$~M$_\odot$~pc$^{-2}$ in the reference sample following \citet{SANDSTROM13} and scaling by the difference
between $I_{\rm CO}/I_{350}$ in the region and the reference sample.}
\tablenotetext{b}{We expect that in the bright part of the starburst disk our modeling fails. But the general result of $I_{\rm CO 2-1}/I_{350}$ similar to a disk galaxy holds.}
\tablecomments{Units of \acounits throughout.}
\end{deluxetable}

Table \ref{tab:aco} summarizes these calculations. Neglecting the very bright center of the galaxy we distill the findings as follows: dust emission suggests that the conversion factor throughout M82 is likely about half the disk galaxy value in the extended disk and faint parts of the wind. It appears to be lower, by about an additional factor of $2$, in the bright regions near the starburst. In the very brightest parts of the starburst itself the ratio $I_{\rm CO 2-1}/I_{\rm 350}$ actually resembles that found in the disks of normal galaxies; however, this region is not the subject of the paper. In agreement with the multiline comparison, even the lowest plausible $\alpha_{\rm CO}$ values from the dust treatment are still many times the optically thin value.

\subsubsection{Synthesis}

All of the arguments considered lead us to expect a moderately low conversion factor throughout M82. As a conservative single number, we take the $\alpha_{\rm CO}^{2-1} \approx 2$~\acounits , or about $1/2$ the disk galaxy value argued by \citet{SANDSTROM13} and about $1/3$ the value adopted by other papers studying HERA maps of galaxies \citep[e.g.,][]{LEROY13}. This single value agrees within a factor of $\approx 2$ with almost all results: line ratios, comparison to sub-mm intensity, comparison to dust model, and cloud virial masses in the starburst \citep{KETO05}. It implies optically thick emission, which is consistent with the observed line ratio, but likely requires a low $^{13}$C in the active region given the low observed ratios of $^{13}$CO to $^{12}$CO intensity in \citet{WEISS05}.

A more realistic, but more complex, prescription adopts $\alpha_{\rm CO}^{2-1} \approx 2.5$~\acounits\ in the extended disk and the faint part of the outflow and takes $\alpha_{\rm CO}^{2-1} \approx 1$~\acounits\  in starburst region and the bright part of the outflow, which is still very near the disk. A convenient cut for these purposes is $I_{\rm CO} \approx 20$~K~km~s$^{-1}$ in the outflow region:

\begin{equation}
\label{eq:brokenaco}
 \alpha_{\rm CO}^{2-1} = \begin{cases}
    2.5~\text{\acounits }, & \text{extended disk, elsewhere if $I_{CO} < 5$~K~km~s$^{-1}$}.\\
    1.0~\text{\acounits }, & \text{starburst, outflow where $I_{CO} > 5$~K~km~s$^{-1}$}.
  \end{cases}
\end{equation}

\noindent This approach posits that CO is a more effective emitter in the bright starburst region than the cooler, more quiescent extended disk. This agrees with the sub-mm and dust fitting results well. However, with coarse resolution and only a single line for the very extended part of the outflow, we lack concrete  knowledge of what the CO emitting structures in the wind look like in detail. Bear in mind that this two-regime prescription is specific to our data because we expect the intensity to be a strong function of angular resolution.

Figure \ref{fig:dustaco} shows that Equation \ref{eq:brokenaco} yields results that are consistent with a single, linear relation between dust and gas over a large area in M82 and that this approach represents a substantial improvement over using only the typical value for disk galaxies found by \citet{SANDSTROM13}. The figures show that the histogram of dust to gas ratios over and individual line of sight and the binned scaling relation between dust and gas surface density both become more consistent with a single dust-to-gas ratio using this approach. Note that the bright starburst itself does not appear to follow this prescription. Here, the conversion factor may become even more complicated as it convolves many distinct sets of physical conditions and opacity becomes important; however, the burst itself is not the topic of this paper. For the rest of galaxy, Equation \ref{eq:brokenaco} results in single linear scaling between dust and gas.

\section{Tests Related to the Dust Treatment}

\begin{figure*}
\plottwo{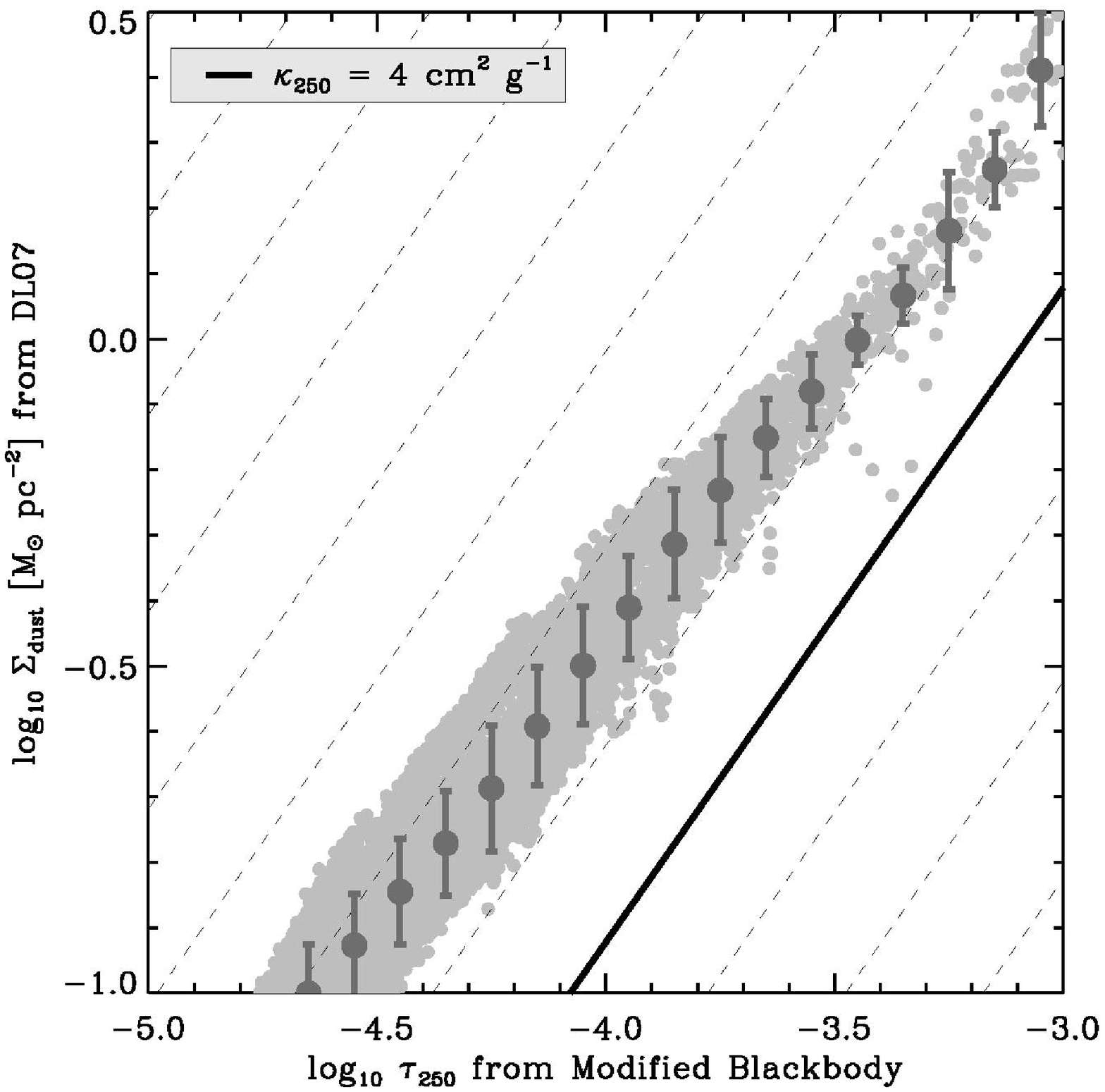}{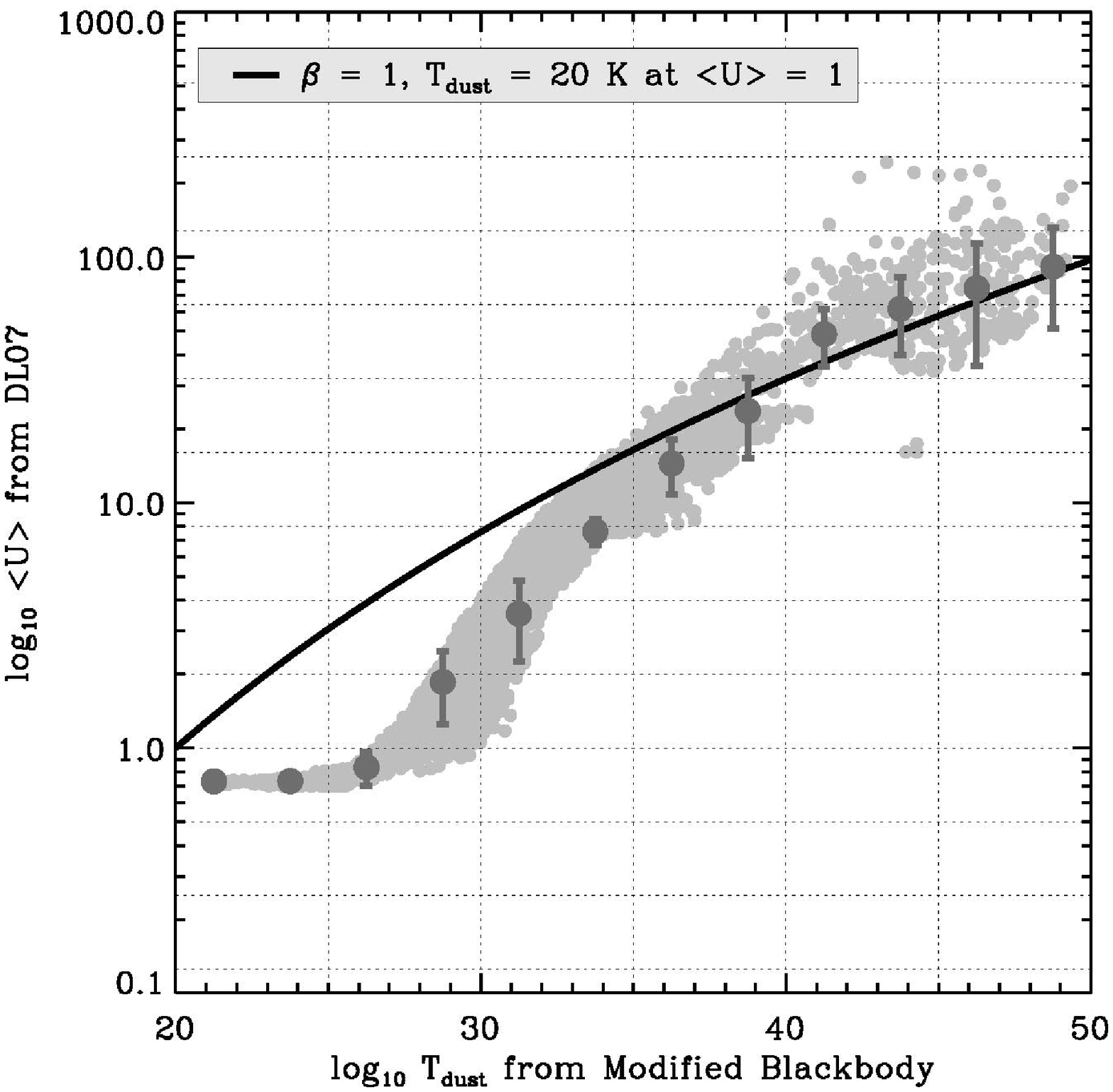}
\plottwo{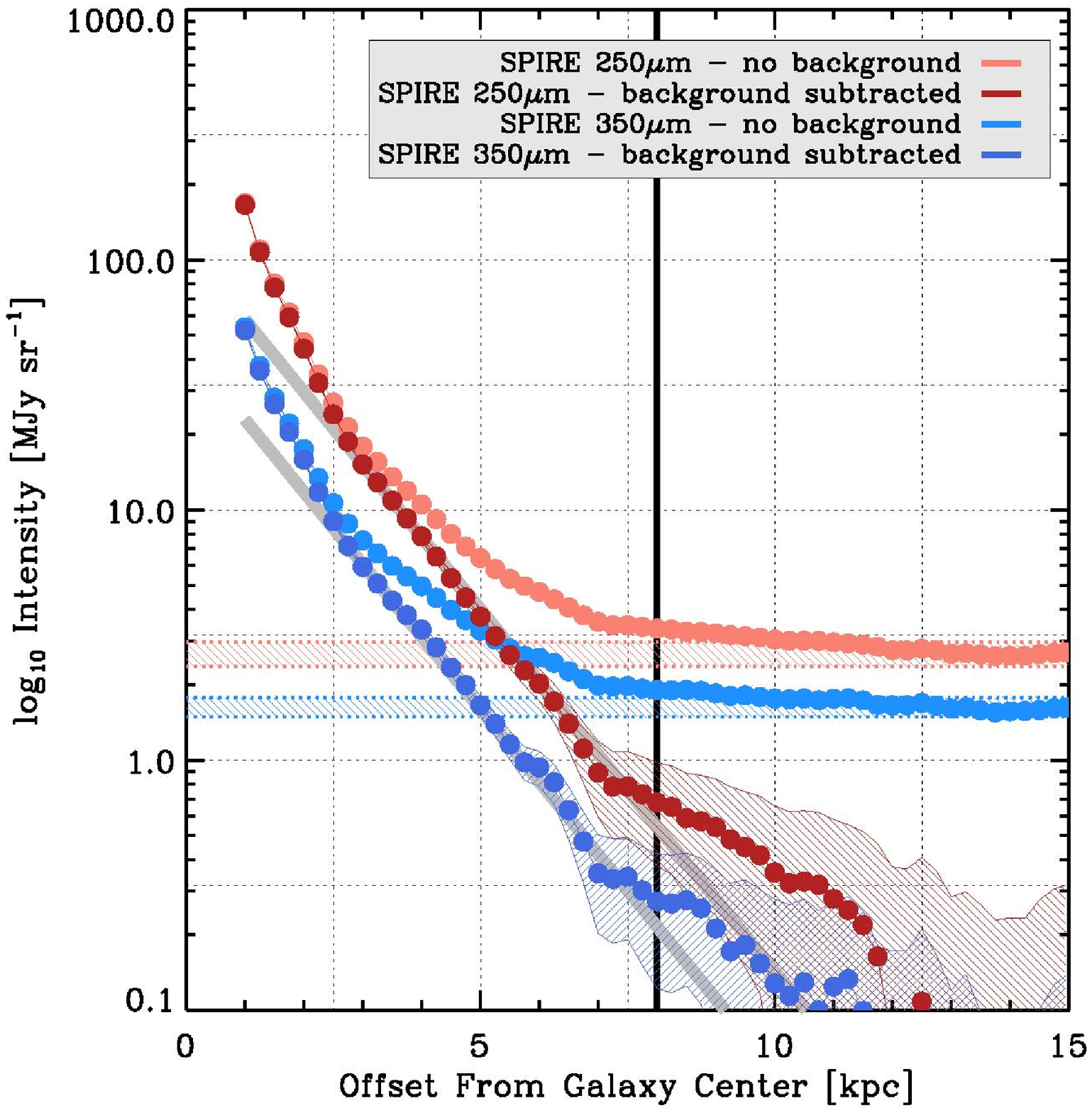}{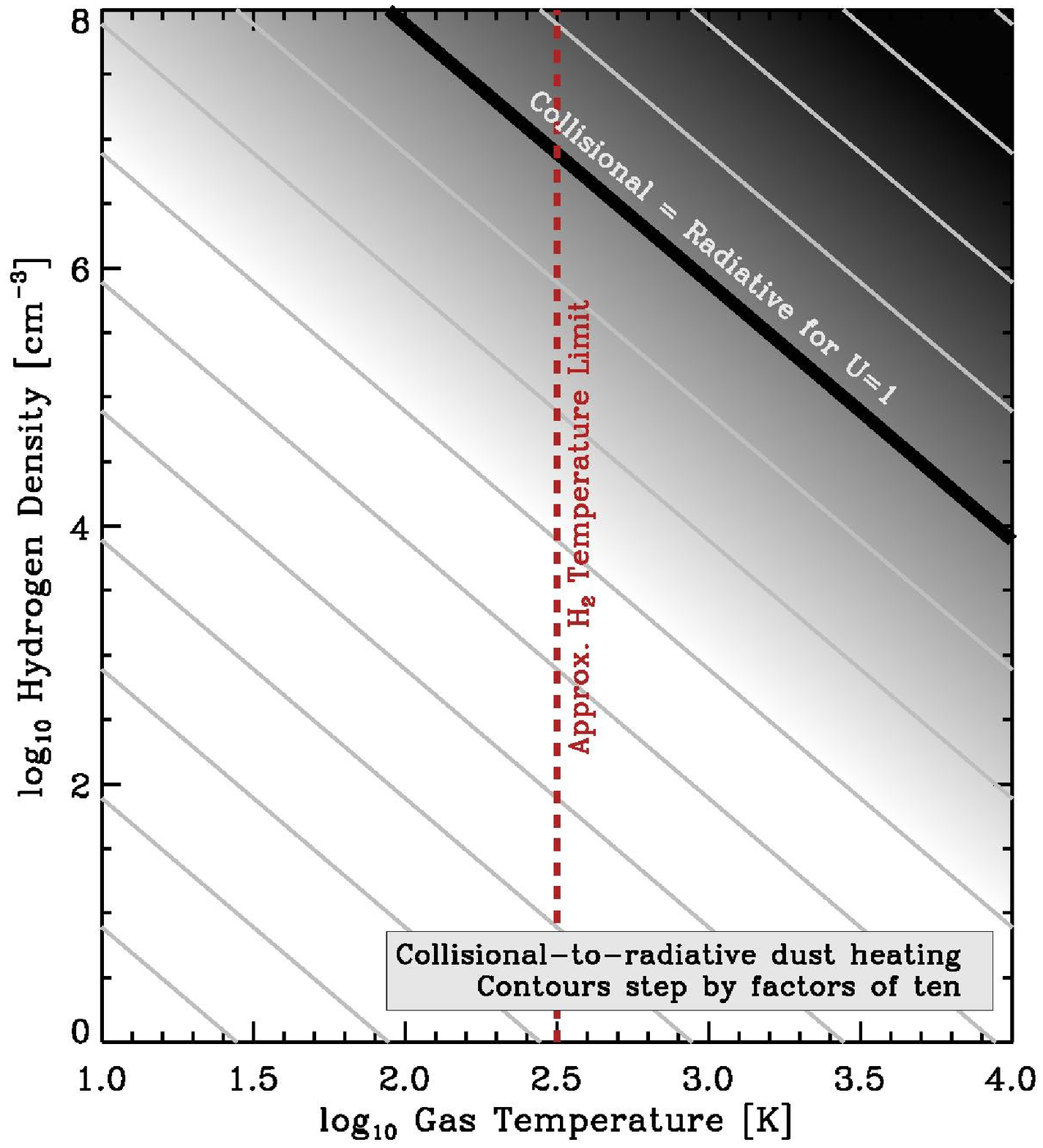}
\caption{
\label{fig:dusttests} Calculations related to our dust treatment. Top row: comparisons of modeling using the \citet{DRAINE07A,DRAINE07B} models to fitting a single modified blackbody to the infrared SED. ({\em top left}) The dust surface density from the \citet{DRAINE07A} modeling as a function of modified blackbody optical depth, a tracer of dust column. The two track one another well modulo a normalization factor (the dashed lines are separated by a factor of two with slope unity). ({\em top right}) Mean radiation field from the \citet{DRAINE07A} modeling as a function of dust temperature from a blackbody fit with a simple expectation as a solid black line. The two converge in the high-temperature, high radiation field regime. They track one another almost monotonically at lower temperatures but differ because of the \citet{DRAINE07A} treatment of an unresolved population. The comparisons show that the results of the paper would not change substantially if we shifted our dust treatment. ({\em bottom left}) The need for and effect of the subtraction of a median background from the SPIRE maps. We show averaged annular profiles around M82 at 250 and 350$\mu$m before and after subtraction of a median background. We also indicate a plausible range of zero-point values (light rectangular shaded regions) and the effect of varying the background within this range on the subtracted profiles (darker shaded region). The effect is minimal within the region that we study in the paper (out to the black line). ({\em bottom right}) Simple calculation of the balance of collisional and radiative heating following \citet{DRAINE11} for a mean radiation field $U=1$. Radiative heating dominates collisional heating for the plausible range of bulk gas temperatures and densities, supporting the interpretation of the dust temperature in terms of illuminating radiation field, though more detailed modeling of the interface of cold, dense and hot, rare material is needed.}
\end{figure*}

Figure \ref{fig:dusttests} illustrates four tests related to our treatment of IR emission around M82. The top two panels show the effect of our choice of the \citet{DRAINE07A} model by comparing these to the result of a simple modified blackbody fit to the IR SED between 70 and 350$\mu$m. The top left panel shows our modeled $\Sigma_{\rm dust}$ as a function of the optical depth at 250$\mu$m from the blackbody fit with each pixel as a single gray point. The two track each other closely, with about a factor of two scatter and a similar systematic trend across the range of dust surface densities in the galaxy. The normalization needed to equate the two would correspond to a mass absorption coefficient at $250\mu$m of $\kappa \approx 1$--$2$~cm$^2$~g$^{-1}$. This is moderately low compared, e.g., to the theoretical value \citep[e.g.,][]{DRAINE07A} because the \citet{DRAINE07A} models model a population of dust and so tend to yield somewhat higher dust mass surface densities than single population fits. The treatment of heating also differs and matters to the derived surface density (the right panel). Given that most of the results in this paper are comparative and do not depend on the normalization of the dust models Figure \ref{fig:dusttests} shows that our adopted model has only a modest impact. The right panel compares the derived temperature from the modified blackbody fit and the mean radiation field from the \citet{DRAINE07A} model fit. For comparison the line expected if $T = 20$~K corresponds to $U = 1$ and $\beta = 1$ appears as a black line. The range of radiation fields modeled by the \citet{DRAINE07A} model lead to a lower mean $U$ at low dust temperatures, which is a goal of the model and a more realistic treatment of the SED for low resolution observations of other galaxies \citep[see][]{DALE02,DRAINE07B}. At high temperature, high radiation field regions, the two treatments tend to converge. The two quantities do track one another, though not linearly. Given that use of $\left< U \right>$ is mostly qualitative, the comparison shows that our results for $\left< U \right>$ could mostly be recast in terms of dust temperature if one prefers to avoid the assumptions about unresolved populations underlying the \citet{DRAINE07A} model.

The bottom left panel illustrates the effect of subtracting a median background from the SPIRE maps when constructing the large-scale profiles. We plot the profiles with and without background subtraction, illustrating that the wide maps around M82 do converge to a non-zero level at large radii. As seen in the main text, this emission does not appear to be highly structured or to resemble emission at other wavelengths and the {\em Herschel} data processing is not intended to provide a reliable zero-point intensity. The figure illustrates a plausible range of background levels via the shaded regions and then shows the effect of varying the subtraction within this range on the subtracted profiles. Within the $\approx 8$~kpc from the galaxy center this treatment appears fairly robust. The nature of faint, extended emission and the presence of any faint, smooth emission on very large scales around M82 will have to be addressed with facilities or data processing optimized for that purpose.

Finally, the bottom right panel of the  plot the expected ratio of collision to radiative heating for a simple treatment following \citet{DRAINE11}. They provide an estimate of this ratio for atomic hydrogen, their Equation 24.7. We show the results of this calculation for a range of densities and temperatures, a solar neighborhood radiation field ($U=1$), which is already very low given our results on M82. We adopt a grain size $0.1\mu$m and an equal mix of graphite and silicate grains following their treatment; larger grains would only increase radiative heating. The plot shows that for any plausible set of conditions in the bulk of the molecular or atomic gas, radiative heating is expected to dominate over collisional heating in M82. Shocks have been shown to be important in a variety of ways to both the dust and the gas around M82 \citep[e.g.,][]{KANEDA10,BEIRAO15} but we do not expect that they push the bulk of the gas mass into the extreme regime needed for collisional heating to dominate radiative heating in the presence of the strong radiation fields around M82. In individual shocked regions or the dense, well-shielded parts of the starburst, the picture may be different. For the large-scale view we study, where neutral gas and dust show good correspondence, we know that the bulk H$_2$ temperature is likely a few times 10 K and constrained by IR spectroscopy to be not more than a few times 100 K. Meanwhile, the hot gas at extreme temperatures is expected to have low densities \citep{STRICKLAND09}, even though its temperature is extreme. More detailed study is needed to understand dust heating in the interface of the hot wind material and the cold material, but this first order calculation supports the interpretation of dust temperature as a tracer of the radiation field around M82.

\section{The Effects of the IRAM 30-m Error Beam}

M82 represents a prime example of the general astronomical problem of faint emission in the presence of a bright source. Although the problem is far less several in mm-wave line emission than at other wavelengths, stray light pickup can still be a problem. The main beam efficiency of the IRAM 30-m, $\eta_{\rm mb}$, is measured to be $\approx 58\%$ at 230~GHz. Much of the rest of the response to emission on the sky is in an ``error beam'' that has been characterized as a series of Gaussians much larger in size than the main beam. As a result, our map could include low-level pickup of extended structures. In fact, the large size of the error beam and the fact that it operates ``by channel'' (i.e., at a specific frequency) combine to make this a small effect in M82 and indeed throughout the HERACLES survey. This appendix describes the quantitative test that we use to arrive at this conclusion.

Consider a data cube that has been reduced and converted to an image on the sky under the assumption the error beam is free of astronomical signal, so that the cube in $T_A^*$ units has been converted to $T_{\rm mb}$ units by multiplying by a factor $\eta_{\rm mb}^{-1}$. This cube represents an upper limit to the true sky emission because some emission observed with the error beam may be included in the cube. Under the assumption that the error beam effect is a perturbation on the overall emission, we can estimate the magnitude of such contamination by simply convolving the observed cube with the error beam plane by plane and then scaling the result by $\eta_{\rm mb}^{-1}$ (we scale to account for the fact that this multiplication by $\eta_{\rm mb}^{-1}$ was already applied to the contamination in the original cube). This now represents an estimate of the error beam contamination in the $T_{\rm mb}$ cube that can be subtracted from the original cube. Often this calculation will be enough to demonstrate the minimal impact of the error beam. In the case where the contamination is significant, the calculation may need to be refined because this first estimate of contamination will be an overestimate (recall that the original cube itself was an overestimate because it includes the contamination). To refine the estimate, one subtracts the first estimate of error beam contamination from the original cube and then convolves the new, contamination-corrected cube with the error beam. This is a refined estimate of contamination by the error beam that can be subtracted from the original cube to yield an improved estimate of the true intensity distribution. As long as the error beam represents a modest correction to the overall emission, iteration should yield a corrected estimate of the true emission.

Procedurally, the correction scheme is:

\begin{enumerate}
\item Generate a cube in $T_{\rm mb}$ units assuming no contamination. 
\item Convolve the cube with the known error beam plane by plane. Scale by $\eta_{\rm mb}^{-1}$. This represents an estimate of error beam contamination.
\item Subtract the estimated contamination from the original cube. This is an estimate of the true, error-beam corrected sky distribution in $T_{\rm mb}$ units.
\item Return to step \#2 and use the corrected cube to generate the contamination estimate.
\item Repeat steps \#2--4 until the change between successive iterations is small.
\end{enumerate}

In practice, even this iterative approach often yields an overestimate of the effect of the error beam on single dish maps of other galaxies because the procedure of baseline fitting will tend to suppress stray light pickup in the original cube. The error beam is often comparable to the size of the galaxy, so that its effect is to place a heavily reduced version of the spectrum of the galaxy into individual spectra. For a highly resolved galaxy, this spectrum will often be quite extended compared to the line profile at a given location within the galaxy. Most baseline fitting procedures \citep[for example the local {\sc Hi} and CO windowing function in][]{LEROY09} will remove such low, broad emission. A corollary of this, the absence of broad line wings in stacked HERACLES spectra, is discussed in \citet{SCHRUBA11}.

In an IRAM memo, Kramer, Pe\~nalver, and Greve (2013) provide the latest characterization of the IRAM 30-m beam, the forward efficiency, and forward spillover efficiency term that is the difference between the main beam and forward efficiencies. Adopting their prescription for the error beam, we iteratively model the error beam for M82 (and all of the other HERACLES) targets following the procedure above. We then re-fit spectral baselines in the cubes following the same robust polynomial fitting procedure used to produce the original cubes. We iterate a total of five times for M82, though the last three iterations have only small effect. The result of this modeling for M82, which should be among the worst cases for the HERACLES galaxies, is a decrease of $\sim 10\%$ in the overall flux of the cube. Locally, the effect can be more extreme, with $\approx 20$--$30$\% differences between the corrected and uncorrected maps in the area immediately around the bright starburst disk.

\end{appendix}

\end{document}